\RequirePackage{fix-cm}
\documentclass[pdftex,epjc3,twocolumn,smallextended]{svjour3}
\RequirePackage[T1]{fontenc}
\smartqed  % flush right qed marks, e.g. at end of proof
\RequirePackage{graphicx}

\RequirePackage{flushend}
\RequirePackage{mathptmx}      % use Times fonts if available on your TeX system
\RequirePackage{latexsym}
\RequirePackage[numbers,sort&compress]{natbib}
\RequirePackage[colorlinks,citecolor=blue,urlcolor=blue,linkcolor=blue]{hyperref}
\usepackage{hyperref}
\usepackage{mathtools}
\usepackage{amsmath}
\usepackage{cuted}
\usepackage{mathtools,times,cuted}
  \newcommand{\Qa}{\mathcal{Q}}
  \newcommand{\MB}{$\mathcal{MB}$}
\newcommand{\Pa}{\mathcal{P}}
 \usepackage[latin1]{inputenc}
\usepackage{color}
\usepackage{url}
\usepackage{graphicx}
\hypersetup{linkcolor=red,citecolor=blue,filecolor=cyan,urlcolor=magenta}
\usepackage{float}
\usepackage{amsmath}
\usepackage{tikz}
\usepackage{graphicx}
\usepackage{latexsym,color}
\usepackage{amssymb}
\usepackage{xcolor}
\usepackage{amsfonts,dsfont}
\usepackage[T1]{fontenc}
\usepackage{ae,aecompl,latexsym}

\definecolor{lightgray}{rgb}{0.75,0.75,0.75}

\def\be{\begin{equation}}
\def\ee{\end{equation}}
\def\bea{\begin{eqnarray}}

\newcommand{\apj}{APJ}

\def\eea{\end{eqnarray}}

\newcommand{\la}{\mathcal{A}}
\newcommand{\laa}{\mathcal{L}}
\usepackage{physics}

\newcommand{\pp}{\textbf{()}}

\newcommand{\il}{~}

\usepackage[varg]{txfonts}

\journalname{Eur. Phys. J. C}
\begin{document}

\title{On light surfaces in  black hole  thermodynamics
}


\author{D.  Pugliese\thanksref{e1,addr1}
        \and
        H. Quevedo\thanksref{e2,addr2,addr3} %etc.
}

\thankstext{e1}{e-mail: dany.pugliese@gmail.com}
\thankstext{e2}{e-mail: quevedo@nucleares.unam.mx}


\institute{Research Centre for Theoretical Physics and Astrophysics,
Institute of Physics,
  Silesian University in Opava,
 Bezru\v{c}ovo n\'{a}m\v{e}st\'{i} 13, CZ-74601 Opava, Czech Republic \label{addr1}
           \and
 Instituto de Ciencias Nucleares, Universidad Nacional Aut\'onoma de M\'exico,  AP 70543, M\'exico, DF 04510, Mexico \label{addr2}\and
	Dipartimento di Fisica and ICRA, Universit\`a di Roma ``La Sapienza", I-00185 Roma, Italy \label{addr3}
}

\date{Received: date / Accepted: date}

\maketitle

\begin{abstract}
 We discuss the fundamentals of classical  black hole (\textbf{BH})  thermodynamics    in  a  new framework determined by light surfaces and their frequencies. This new approach allows us to study {\bf BH} transitions inside the Kerr geometry.  In the case of {\bf BH}s, we introduce a new parametrization of the metric in terms of   the maximum extractable rotational energy or,   correspondingly, the irreducible mass, which is an alternative to the spin parametrization.
 It turns out that \textbf{BH} spacetimes with spins
$a/M= \sqrt {8/9}$ and $a/M=1/\sqrt{2}$ show anomalies in the rotational energy extraction and surface gravity whereas the case $a/M=\sqrt{3}/2$ is of particular relevance to study the variations of the horizon area. We find the general conditions under which {\bf BH} transitions can occur and express them in terms of the masses of the initial and final states. This shows that {\bf BH} transitions in the Kerr geometry are not arbitrary but depend on the relationship between the mass and spin of the initial and final states.
From an observational point of view, we argue that near the {\bf BH} poles it is possible to detect photon orbits with frequencies that characterize the light surfaces analyzed in this work.
\end{abstract}

\section{Introduction}
A black hole  (\textbf{BH}) singularity is   mainly  an  "active"  gravitational    object in the sense that it can interact with its environment constituted usually by matter and fields.
A \textbf{BH}  transition from a state to another, accompanied by a change of the
characteristic parameters,  is regulated by the  laws of \textbf{BH}  thermodynamics. These changes are also crucial   in the analysis of  \textbf{BH}  energy extraction detectable,  for example,  by observing   jet   emissions.
 The progenitor  collapse into a \textbf{BH} appears as an  irreversible process of
information degeneration, where  the information   becomes inaccessible  to the distant observers and the \textbf{BH} has no observable
hair. A (stationary) \textbf{BH}  macro-state  is defined and determined only by the mass parameter $M $ mass, spin $J$, and electric charge $Q$.
However,  the \textbf{BH}  micro-states  number can be enormously large,  providing eventually
a very high \textbf{BH} entropy.
The  number of \textbf{BH} micro-states increases with the \textbf{BH} size and the  \textbf{BH}  size, function of  outer horizon  area,   becomes a measure of its  entropy.
The horizon area cannot decrease
 in any process with suitable conditions on energy and  no repulsive gravity. Hawking's area theorem, states that the total
horizon area of classical \textbf{BHs} cannot decrease over time -\cite{Hawking71,Hawking74,Hawking75}.

In the processes of  \textbf{BH} energy and rotational energy  extraction, underlying for instance  jet emission processes, this result is maintained.
  The
Penrose  process is, for example, a classical method for extracting energy from a Kerr \textbf{BH} using its frame-dragging effects in the ergoregion.
 A small body  may have  in this region  negative energy as measured by an observer at infinity.
 If the body is fragmented into two parts, and    the negative-energy part is swallowed by the \textbf{BH}, decreasing its (total) mass,
the other part can be   ejected. In this process $J$  decreases, consequently, its area increases (as determined by the outer horizon)--\cite{Penrose69}.
The balance of the mass increase (which increases the \textbf{BH} area) and spin variations in this process is maintained within the constraint that
the event horizon can increase or
remains constant.
For example, the recently observational confirmation of Hawking's black-hole area theorem  has been presented based on data from GW150914 \cite{Isi:2020tac}.

Black holes are classical solutions of the Einstein field equations, representing  spacetime  regions bounded by the event horizon and  causally disconnected from
observers at  infinity.
  Bekenstein first suggested the  idea that the  event horizon\footnote{It should be noted that the inner horizon has been
shown to be unstable    and it has been conjectured it could
end  into a singularity
\cite{Simpenrose,DafermosLuk}.} defines the \textbf{BH}  entropy--\cite{Bekenstein73,Bekenstein75}.  However,  \textbf{BH} thermodynamic properties   seem to be purely intrinsically geometric characteristics, defined on the basis of the outer \textbf{BH}  (event) horizon  alone. Probably,
an unknown   quantum mechanism would explain and complete \textbf{BH} classical thermodynamics, leaving  however  open complex problems   (for instance, the well-known information paradox as a consequence of Hawking  thermal radiation),  and touching  the limit of applicability of quantum gravity.

However, if   \textbf{BHs} have an entropy, we can expect them to have also a temperature.
This implies that  a \textbf{BH} must also radiate. This radiation was explained by effects of quantum fluctuation  of matter fields  in the vacuum close to the singularity as
Hawking radiation.
Fluctuations of the  (matter) field  vacuum  produce   couples of  particles and
antiparticles in the close vicinity of a  black hole. A particle  is captured by the singularity  with  negative energy, the other particle can escape to
infinity  with  positive energy, constituting observable radiation  with  thermal profile.

In this semi-classical scenario,  the \textbf{BH} would evaporate emitting only thermal Hawking radiation with non-vanishing  (non constant) entropy; however, such an evolution  appears not to be fully
 described by quantum mechanics
 (raising the so called  information paradox).
The derivation of Hawking radiation is indeed  semi-classical  and yet does not establish the \textbf{BH} situation at the Planck size,  where quantum gravity is expected to play a major role. Specifically, it does not establish  if, for example, the \textbf{BH}
 evaporates completely or  viceversa some kind of remnants is left. A further  question is how  the emitted radiation, reaching infinity, may be  entangled with such  remnant, and what is its entropy.
The \textbf{BH} may also continue the evaporation process completely, leaving  no remnants.
One could expect that  (Hawking) radiation  from the collapsing matter into a \textbf{BH}  would carry out the information of the matter, the \textbf{BH} would  evaporate away as a result of thermal emission and the  process preserves the  unitarity  foreseen
in quantum mechanics. This expectation, however,  fails with
the difficulties  existing   in  obtaining  a unitary
description  of \textbf{BH} evaporation (and \textbf{BH}  formation).
A concrete possibility is that the final radiation of the complete evaporation does not ``bring" information on the
 initial matter state (contradicting  quantum mechanics  unitarity  realized through the action of an unitary  evolution operator).
In this case,  the quantum fluctuation would produce
a mixed state of the outgoing emitted particles with the ingoing particles;  consequently, the  outgoing radiation  would be entangled  with the final hole state (the  final state is a mixed state if the fate of the \textbf{BH} is to evaporate to nothing)--see, for example, \cite{JP16} for a review on the  \textbf{BH} information problem.
All these are aspects of still open debated issues, at the heart of the quantization of the geometric theories of gravity, of the treatment of unified theories for matter field--quantized  gravity, and very often addressed in quantum theories such as loop  quantum gravity or in the context of string theories.

 From the classical point of view the effects of a change of state in \textbf{BHs}  are constrained  by the laws of thermodynamics.
One might ask if there are privileged or not allowed state transitions  or if  transitions are singled out by some anomalous characteristics. Constrains on these processes are  generally determined by the initial state of the black hole,   its spin $J$ or by the specific mass ratios of the initial and final states.
There are strong indications that
the creation of  a  naked singularity (\textbf{NS}) through a conversion process from an extreme \textbf{BH} is forbidden and  that no accretion process can bring a \textbf{BH} into a \textbf{NS} state, while the reverse process, assuming  the  \textbf{NS} existence, is still debated in different contexts. However, it is necessary to consider possible processes of  rotational energy extraction. For example, geometrically thin accretion disks  have been studied  for  possible mechanisms  to   convert asymptotically
\textbf{NSs} into   extreme \textbf{BH} states-\cite{z02}.

In this work, we study    aspects  of   classic black hole thermodynamics reformulated  in terms of light surfaces.
We focus on Kerr  singularities.
Using some characteristics of specific null surfaces of the Kerr geometries, including the case of the horizons, we study the relation between the initial and final states of a \textbf{BH} transformation regulated by the laws of \textbf{BH}  thermodynamics. \textbf{BHs} an \textbf{NSs} geometries are therefore represented in a plane called the extended plane.
The surfaces are defined by a characteristic null frequency $\omega$, which connect all the geometries described by the Kerr solution and all the points of these geometries where a null orbit with frequency $\omega$  is defined. These geometries always include  \textbf{BHs} and in some cases naked singularities. Characteristic frequencies are frequencies of the Kerr horizons.
A special metric line parametrization, describing only \textbf{BH} spacetimes (no naked singularities), is given in terms of the maximum extractable rotational energy  $\xi$ or,   correspondingly, the inertial mass parameter $M_{irr}$,  replacing the metric spin parameter.
We focus on some particular transformations between one state and another. We identify some particular transition states,  representing the initial spin of the black hole or a specific ratio of the masses. These emerge as properties of the classical thermodynamic variables, when viewed in the extended plane. Using the representation provided by the light surfaces, we obtain an overview of the possible states. Thus,  we first introduce the  new framework, writing  the principal quantities in this new way, and the we study the laws of \textbf{BH} transformations in the extended plane in terms of the light surfaces.

\medskip

The plan of the article is as follows.
In Sec.\il(\ref{Sec:kerr}), we introduce the Kerr geometries while,   in Sec.\il(\ref{Sec:majo-sure}), we discuss \textbf{BH} Killing horizons and  the characteristic frequencies, building up   the framework for the formulation of the laws of  \textbf{BH} thermodynamics that are introduced in Sec.\il(\ref{Sec:embo}.
 In Sec.\il(\ref{Sec:nil-base-egi}), masses and thermodynamic variables
are explored  in terms of the light surfaces. The analysis continues in Sec.\il(\ref{Sec:mass-termo-smarr}), where we focus on  the laws of \textbf{BH} thermodynamics.
Rotational energy and \textbf{BH} ``rest"  mass are the focus of Sec.\il(\ref{Sec:sud-mirr-egi}) while the
\textbf{BH} irreducible mass is addressed in Sec.\il(\ref{Sec:cou.mass-irr}).  \textbf{BH} transformations in the new frame  are  investigated in
Sec.\il(\ref{Sec:furth-mirr}).
An in--deep discussion on the  inner and outer horizons in  \textbf{BH} transitions is presented in
Sec.\il(\ref{Sec:comp-inner-outer}).
The case of constant irreducible mass is described in Sec.\il(\ref{Sec:constant-Mirr}). In Sec.\il(\ref{Sec:alc-cases}), we discuss the
 the inner horizon relations in \textbf{BH} thermodynamics.
In Sec.\il(\ref{Sec:non-iner-extr-surf}), we present the metric tensor re-parameterizations  in the new framework in terms of inertial mass, extractable rotational energy, and surface gravity.
Discussion and final remarks follow in
Sec.\il(\ref{Sec:conclu}).
In the appendix Sec.\il(\ref{Sec:BHsextended}), there are further details of
black holes  in the extended plane.
Some remarkable areas in the extended plane are described in Sec.\il(\ref{Appendix:mainbundleproperties}).

\section{The Kerr geometry}\label{Sec:kerr}
 The   metric tensor of the Kerr geometry  can be
written in Boyer-Lindquist (BL)  coordinates
\( \{t,r,\theta ,\phi \}\)
as follows
\bea\label{alai}&& ds^2=-dt^2+\frac{\rho^2}{\Delta}dr^2+\rho^2
d\theta^2+(r^2+a^2)\sin^2\theta
d\phi^2+
\\&&\nonumber
\frac{2M}{\rho^2}r(dt-a\sin^2\theta d\phi)^2\\&&\nonumber
\rho^2\equiv r^2+a^2\cos\theta^2,\quad\Delta\equiv r^2-2 M r+a^2,\\
&&\nonumber
r_{\pm}\equiv M\pm\sqrt{M^2-a^2};\quad r_{\epsilon}^{\pm}\equiv M\pm\sqrt{M^2- a^2 \cos\theta^2},
%&&\nonumber
\eea
{  where $r\in[0,+\infty[$, $t\in [0,+\infty[$, $\theta\in[0,\pi]$ and $\phi\in[0,2\pi]$.}
The radii  $r_+$ and $r_-$ are the outer and inner Killing horizons, respectively; $r_{\epsilon}^+$ and $r_\epsilon^-$ represent the outer and inner  ergosurfaces, respectively.  Here $M$ is the {(ADM  and Komar)}  mass parameter and the specific angular momentum is given as $a=J/M$, where $J$ is the
total angular momentum of the gravitational source.
For simplicity, here and in the following we consider dimensionless parameters defined as $r\rightarrow r/M$ and $a\rightarrow a/M$.
Moreover, we will use the notation $\sigma \equiv \sin^2\theta$.

 It is $r_+<r_{\epsilon}^+$ on the planes  $\theta\neq0$  and  $r_{\epsilon}^+=2M$ on the equatorial plane $\theta=\pi/2$, where $r_-=0$. Moreover,  $0<r_{\epsilon}^-<r_-<r_+<r_{\epsilon}^+$ for $\sigma\neq0$ and $a\in ]0,M[$.
 For $\sigma=0$, the functions $r_{\epsilon}^\pm$ satisfy the properties  $0<r_{\epsilon}^-=r_-<r_+=r_{\epsilon}^+$. For
 $a=\pm M$  (extreme Kerr \textbf{BH}), we have that $r_\pm=M$.
 The case $a=0$ corresponds to the static spherically symmetric Schwarzschild solution, where $r_+=2M$ and there is no ergosurface. Naked singularities  are defined for $a^2>M^2$. The surfaces $r_{\epsilon}^\pm$  in  \textbf{NSs}  are well-defined for  $\cos\theta \in [-M/a, M/a] $ or, equivalently, $
    a/M\in ]1, 1/\sec^2\theta]$.
In the  {region $r\in]r_+,r_{\epsilon}^{+}$[} ({outer \em ergoregion} or simply ergoregion), it is  { $g_{tt}>0$} and the $t$-Boyer-Lindquist coordinate becomes spacelike.
This fact implies that a  static observer cannot exist inside
the ergoregion.
In the following analysis, we will refer to some properties of circular motion in these regions.

Since the metric is independent of $\phi$ and $t$, the covariant
components $p_{\phi}$ and $p_{t}$ of a particle   four--momentum are
conserved along its  geodesic.
Consequently,
$
{E} \equiv -g_{ab}\xi_{t}^{a} p^{b}$ and $L \equiv
g_{ab}\xi_{\phi}^{a}p^{b}\
$ are  constants of motion for test particle orbits, where  $\xi_{t}=\partial_{t} $  is
the Killing field representing the stationarity of the Kerr geometry and  $\xi_{\phi}=\partial_{\phi} $
is the
rotational Killing field (the vector $\xi_{t}$ becomes    spacelike in the ergoregion).
The constant $L$ may be interpreted       as the axial component of the angular momentum  of a test    particle following
timelike geodesics and $E$ as representing the total energy of the test particle
 coming from radial infinity, as measured  by  a static observer at infinity.
The motion  on the  fixed plane $\sigma=1$ is restricted to that plane ($u^\theta=0$) because the  Kerr metric is  symmetric  under reflections with respect to the   equatorial hyperplane $\theta=\pi/2$.
\subsection{Killing horizons and characteristic frequencies}\label{Sec:majo-sure}
In this section, we build   the framework for the    \textbf{BH} thermodynamics laws by using the properties of special light surfaces and the   \textbf{BH}  horizons.

\medskip

\textbf{BH horizons}

\medskip

Let us  introduce the  Killing vector $\mathcal{L}=\partial_t +\omega \partial_{\phi}$. We will consider mainly $\omega=$constant.
The quantity  $\mathbf{\mathcal{L_N}}\equiv\mathcal{L}\cdot\mathbf{\mathcal{L}}$ becomes null  for photon-like
particles with orbital  frequencies $\omega_{\pm}$.

The  Killing vector
$
\mathcal{L}_{\pm}\equiv \xi_{t}+\omega_{\pm}\xi_{\phi}
$,
where $\omega_{\pm}$ satisfy the null condition on the norm of $\mathcal{L}$,
can be interpreted as generator of  null curves ($g_{\alpha\beta}\mathcal{L}^\alpha_{\pm}\mathcal{L}^\beta_{\pm}=0$)
as the Killing vectors $\mathcal{L}_{\pm} $ are also generators of Killing event  horizons.

The Kerr  horizons are  {null} (lightlike) hypersurfaces generated by the flow of a Killing vector,
whose {null} generators coincide with the orbits of a
one-parameter group of isometries, i.e., in general,    there exists a Killing field $\mathcal{L}$, which is normal to the null surface.
More precisely, the \textbf{BH} horizon $r_+$ is  a non-degenerate (bifurcate) Killing
horizon generated by the vector field $\mathcal{L}$.
In the case $a = 0$ (where $\omega=0$),
$\xi_t$   (now generator of $r_+$) is hypersurface-orthogonal.

Thus, the null frequencies  $\omega_\pm$  evaluated on the \textbf{BH} horizons, are the \textbf{BH} inner and outer horizons frequencies $\omega_H^\mp$ respectively.

\medskip

\textbf{Light surfaces and characteristic frequencies}

The results we discuss in this work follow from the   investigation of the properties of the  null  vector $\mathcal{L}$.  The framework of \textbf{BH} thermodynamics is constructed upon the   solutions
$\mathcal{MB}: \mathbf{\mathcal{L_N}}=0$, with $\omega=$constant, where $\mathcal{MB}$ is a spin  parameter of   the set $\mathcal{MB}\in \{a,a\sqrt{\sigma}\}$.
Solutions  $\mathcal{MB}$ are functions of $(\omega,r)$,  or $(\omega,r,\sigma)$, and can be represented  as a curve called metric bundle  $\mathcal{MB}$ on a plane $a-r$ or $\la-r$ (extended plane) where $\la\equiv a\sqrt{\sigma}$--\cite{remnants,bundle-EPJC-complete,LQG-paper,bundle-ragtime,bundle-MG,Pugliese:2021aeb,GRG-lette}.

The quantity $\omega: \mathbf{\mathcal{L_N}}=0$, for the null vector  $\mathcal{L}$ will be called  characteristic  $\mathcal{MB}$ frequency.

A particular null vector  $\mathcal{L}$ is defined by $\omega=\omega(r_{\pm})\equiv \omega_H^\pm$, defining the Killing horizons of the metric.
In fact, the event horizons  of a spinning \textbf{BH}  are   Killing horizons   with respect to  the Killing field
$\mathcal{L}_H=\partial_t +\omega_H^{\pm} \partial_{\phi}$, where  $\omega_H^{\pm}$ is the angular velocity {(frequency)} of the horizons  representing   the \textbf{BH} rigid rotation. The event horizon of a stationary asymptotically flat solution with matter, satisfying suitable hyperbolic equations,  is a Killing horizon. The strong
rigidity theorem connects the event horizon with a Killing
horizon.
In the limiting case of spherically symmetric, static  spacetimes,
 the event horizons are  Killing horizons with {respect} to the  Killing vector
$\partial_t$ and  the
event, apparent, and Killing horizons  with respect to the  Killing field   $\xi_t$ coincide (we can say that $\omega_H^+=0)$.

Therefore,  in the extended plane, the curve $a_\pm\equiv \sqrt{r(2M-r)}$, is called horizon curve and it is tangent at any point to a  $\mathcal{MB}$ curve. The tangent point $(a_g,r_g)$, distinguishes  the \textbf{BH} rotation $a_g$ and its outer (inner) horizon $r_g\in [M,2M]$ ($r_g\in ]0,M]$).
The tangent  point defines the \textbf{BH} outer (inner) horizon frequency which, therefore, is the  $\mathcal{MB}$ characteristic frequency.
In the extended plane, $\mathcal{MB}$  are  made by points $(a,r)$ or ($a\sqrt{\sigma},r)$  distinguishing a geometry $a$ (which can be also a naked singularity) and an orbit (and plane $\sigma$) where there is light-like frequency $\omega$ corresponding to  the $\mathcal{MB}$ characteristic frequency.

Two important properties follow: all metric bundles are tangent  to the horizon curves and the horizon curves are the envelope surfaces of the metric bundles. Thus, all the $\mathcal{MB}$ characteristic frequencies, all the light surfaces frequencies (also in the naked singularity solutions) are  frequencies of a \textbf{BH} (inner or outer) horizons, determined by the  corresponding $\mathcal{MB}$ tangent point. The point $a_0$ ($\la_0$) at the origin $r=0$ (central singularity) is called the bundle origin.

Note that in this case we are assuming $\omega>0$ as $a>0$; however, the study of counter-rotating photon circular orbits with frequency $\omega<0$ is possible  with   $\mathcal{MB}$s in the extended plane $a\in R$ with $\omega>0$, although in this analysis this extension is not necessary--see \cite{bundle-EPJC-complete}. The  \textbf{NS} light surfaces properties are defined by the \textbf{BH} light surfaces (through the condition of tangency of the bundles with the horizon curve). Vice versa it has been shown that most bundles  $\mathcal{MB}$ with \textbf{NS} origin are tangent to the inner horizons curve. The horizon curves point $a=0, r=0$ is the central singularity, $(a=M,r=M)$ is the horizon of the extreme Kerr \textbf{BH}, while $a=0$ with point $r=2M$ is the horizon of the Schwarzschild solution.

Each line $a_{\pm}=$constant in this extended plane  is a \textbf{BH} geometry, where  the inner horizons are for  $r\in[0,M]$ on the line $a_{\pm}$,  while $r\in [M,2M]$ contains the outer horizons. The maximum of the curve $a_{\pm}$ is the point $a=M$ and $r=M$ which is the horizon in the extreme Kerr \textbf{BH}. The Schwarzschild \textbf{BH} is for $a=0$; therefore, it corresponds to the zeros of the metric bundles curves, where $r=2M$ is its horizon--and in the extended plane it coincides  also with the outer ergosurface  on the equatorial plane of all the  Kerr \textbf{BHs} and \textbf{NSs}\footnote{
 $\mathcal{MB}$s are a conformal invariants  of the metric as many other
 quantities considered in this analysis that
 inherit  some of the properties of the Killing vector $\mathcal{L}$}.

The characteristic frequency $\omega$, the bundles origin $a_0$,  the tangent spin $a_g$ and radius $r_g$  to the horizon curve in the extended plane   are related by the following  conditions
\bea&&
\omega=\frac{1}{\la_0}=\frac{1}{a_0\sqrt{\sigma}}= \frac{\sqrt{(2-r_g) r_g}}{2 r_g}= \omega_H^{\pm}(r),\\&&\nonumber
 r_g(\omega)=\frac{2}{4 \omega ^2+1}=\frac{2 \la_0^2 }{\la_0^2 +4},\\\nonumber
 &&
 a_g(a_0)=\frac{4 \la_0 }{\la_0^2 +4}=4 \sqrt{\frac{\omega^2}{\left(4 \omega^2+1\right)^2}},\quad a_0=\frac{2}{\sqrt{\sigma}}\sqrt{\frac{r_g}{2-r_g}} \ ,%=2\frac{r_g}{a_g\sqrt{\sigma}}=\frac{4\omega_H^{\pm}}{\sqrt{\sigma}}
\eea
where
$r_g\in [0,2M]$, $a_0\in [0,+\infty[$,
$a_g\in [0,M]$  (with the extension to negative values for the counter-rotating case),
$\omega\in [0,+\infty]$ and $\sigma\in [0,1]$.

The bundle is tangent to the horizon curve at one point only; therefore, this defines uniquely the \textbf{BH} with the characteristic frequency $\omega$ and, consequently, bundles can contain either only \textbf{BHs} or  \textbf{BHs} and \textbf{NSs}.
Bundles are not defined in the  region $]r_-,r_+[$  of \textbf{BH} spacetimes, but they are defined in the region $r\in [0,2M]$ of \textbf{NSs}).

$\mathcal{MB}$  curves in the extended plane connect points of different  (\textbf{BH} or \textbf{BH} and \textbf{NS}) geometries  having all the same characteristic null frequency $\omega$.
They are related to the light surface  $r_s(\omega):\mathcal{L}_{\mathcal{N}}=0$ defined in a fixed spacetime as the light surfaces $r_s$
are the collections of all points, crossing of the $\mathcal{MB}$ with a line $a=$constant on the extended plane.

\medskip

\textbf{Constrains on matter distribution}

 $\mathcal{MB}$s, defined  by light-surfaces, determine also the
 timelike (matter circular) motion  providing the limiting orbital frequencies of stationary observers.

As the solutions $\omega: \mathcal{L}_N=0$, for each point $(a,r,\theta)$, are generally two frequencies $\omega_\pm$, this implies that
there are two  bundles crossing at any point of the extended plane. The frequencies $\omega_\pm$,  at  the crossing point $r$, bound the possible stationary observers frequencies.
Specifically, the causal structure defined by timelike stationary  observers is characterized by a frequency   bounded in the range $\omega\in]\omega_-,\omega_+[$.
On the other hand, static observers are defined by the limiting condition  $\omega=0$  and  cannot exist in the ergoregion.

The limiting frequencies  $\omega_{\pm}$, which are photon orbital frequencies, solutions of the condition $\mathcal{L_{N}}=0$, determine
the frequencies $\omega_H^{\pm}$ of the Killing horizons as well as the bundles characteristic frequencies.

The vector $\laa$ appears in the description of certain \textbf{BH} evolution processes because it enters the definitions of thermodynamic variables and stationary observers.
The vector  $\mathcal{L}$, the condition $\mathcal{L_N}=0$ and \MB s are closely related to the definition of stationary observes, i. e., observers with
 a tangent vector which is  a  Killing vector, that is, whose     four-velocity  $u^\alpha$ is    a
linear combination of the two Killing vectors $\xi_{\phi}$ and $\xi_{t}$; therefore,
$
u^\alpha=\gamma\mathcal{L}^{\alpha}= \gamma (\xi_t^\alpha+\omega \xi_\phi^\alpha$),  where
 $\gamma$ is a normalization factor and $d\phi/{dt}={u^{\phi}}/{u^t}\equiv\omega$.
The dimensionless quantity $\omega$  is the orbital frequency of the stationary observer.

It is clear that $\mathcal{MB}$s relate different geometries through their light surfaces defined by the characteristic frequencies which are    \textbf{BHs} frequency. Thus, they are particularly adapted to describe \textbf{BH} state transitions,  constrained  by the properties of the null vector  $\mathcal{L}$, which builds up many of the thermodynamic properties of \textbf{BHs}.
In the next section, we shall detail the \textbf{BH} thermodynamic properties in terms of the null  vector $\mathcal{L}$ and, therefore, of  the metric bundles.

   \MB s  are structures that are related to the points of the light surfaces of different geometries and, therefore,  provide relations between geometries. This different way of rewriting the properties of the null vector  $\mathcal{L}$, which defines also the \textbf{BH} horizons, emphasizes some geometric characteristics  and properties, repeating in the same spacetime or different geometries of a \MB.   One of these properties, from the  light surfaces and \MB definitions   is the frequency  $\omega$, all  the points of a \MB    curve are the replica   of the \textbf{BH} horizon frequency  $\omega_H^+$ or $\omega_H^-$ of the  \textbf{BH} individuated by the tangent point  $(a_g,r_g)$. In some cases, these horizons (frequencies) replicas are in the same spacetime.  In \cite{Pugliese:2021aeb,GRG-lette},   the  \textbf{BH} poles $\sigma\approx 0$ have been  studied  for the observation of the photon orbits with \MB s characteristic frequency.
These are light-like (circular) orbits having the  same frequency  as the black hole  $\omega_H^\pm$.

In the context of replicas, we also introduce
the concept of  {horizon confinement}. Indeed, we say that there is a replica when in the spacetime it is possible to find at least a couple of points having the same value for the property $\Qa$. Then, we say that  there is a \emph{confinement}, when   that value is not replicated.
In the  Kerr spacetime, part of the inner horizon frequencies are {"confined"}.
The confinement analysis, which is  the study of the topology of the curves ${\omega}=$constant in the extended plane,   provides  information about  the  local properties of the spacetime replicated in regions  more accessible  to observes; for example, in the case of properties defined in the proximity of the \textbf{BH} poles or of the inner horizons.  An  observer can  register  the presence of a replica at the point $p$ of the \textbf{BH} spacetime with spin $a_p$, belonging  to a  Killing bundle.  The observer will  find the replica of the \textbf{BH} horizon frequency $\omega_H^+(a_p)$ at the point $p$; therefore,  her/his orbital stationary frequency  is  $\omega_p\in]\omega_{\bullet},\omega_{\star}[$ where one of  ($\omega_{\bullet},\omega_{\star}$) is the horizon's frequency $\omega_H^{+}$, replicated on a pair of orbits $(r_+,r_p)$.     The second  light-like frequency  $\omega_\bullet$   is the frequency of a horizon in a \textbf{BH}  spacetime.
The relation between the two  frequencies $(\omega_{\star},\omega_{\bullet})$ is determined by a characteristic ratio, which we also study \cite{Pugliese:2021aeb,GRG-lette}.
In general, we  use  in this work replicas to connect \textbf{BH} spacetimes related in a transition and governed by the thermodynamic laws   in the extended plane.
We reformulate  \textbf{BH} thermodynamics on the replicas in terms of the light surfaces, exploring the thermodynamic properties of the geometries defined by the metric bundles.
\section{Black hole thermodynamics} \label{Sec:embo}
The norm $\mathcal{L_{N}}\equiv \mathcal{L}\cdot\mathcal{L}$ is  constant on the horizon.
We  start here our analysis of \textbf{BH} thermodynamics by introducing  the   {\textit{\bf BH} surface gravity} as the  constant (acceleration)\footnote{{We note that in this work we have used the notation $ \ell$ for the surface gravity instead of the more usual $\kappa$, to emphasize the role of  acceleration relating to the null vector $ \mathcal{L}$. }} $\ell: \nabla^\alpha\mathcal{L_{N}}=-2\ell \mathcal{L}^\alpha$,
 evaluated on the {outer} horizon $r_+$, or
 equivalently,
  $\mathcal{L}^\beta\nabla_\alpha \mathcal{L}_\beta=-\ell \mathcal{L}_\alpha$ and  $L_{\mathcal{L}}\ell=0$, where $L_{\mathcal{L}}$ is the Lie derivative,-a non affine geodesic equation, i.e.,
$\ell=$constant on the orbits of $\mathcal{L}$.

The \textbf{BH} surface gravity  $\ell$, which is also a  conformal invariant of the metric {\cite{Jacobson}},
may be defined as the  {rate} at which the norm   of the Killing vector $\mathcal{L}$
vanishes from
outside (i.e. $r>r_+$). For
the Kerr spacetime it becomes $\ell_{Kerr}= (r_+^2-a^2)/(r_+^2+a^2)^2$.

The surface gravity   re-scales with the conformal Killing vector, i.e., it  is not the same on all generators but,  because of the symmetries,  it is constant along one specific generator.

The  \textbf{BH} event horizon of
stationary  solutions
has  constant surface gravity, i. e., the surface gravity is constant on the horizon of stationary black holes, which is postulated as the zeroth \textbf{BH}  law-area theorem (see for example \cite{Chrusciel:2012jk,Wald:1999xu}).

More generally, the  \textbf{BH}  horizon area
is non-decreasing, a property which is considered as the second law of
\textbf{BH}  thermodynamics, establishing  the impossibility  to achieve
with any physical process a \textbf{BH} state with zero surface gravity.

Clearly, in the extreme Kerr  spacetime ($a=M$), where  $r_{\pm}=M$, the surface gravity  is zero. This implies that the temperature  is also null ($T_H = 0$),  with a non-vanishing  entropy \cite{Chrusciel:2012jk,Wald:1999xu,WW}. A  non-extremal
\textbf{BH} cannot reach the   extremal limit in a finite number of steps, which is  implied by the third law. (This fact has consequences
also regarding the stability
 against Hawking radiation.).

 On the other hand, the condition (constance of)  $\nabla^a \laa=0$  when $\ell=0$ substantially constitutes the definition of the  degenerate Killing horizon--degenerate \textbf{BH}--, in the case of Kerr geometries only the extreme \textbf{BH} case is degenerate; therefore, in the extended plane it corresponds to the point $a=M$, $r=M$. A fundamental theorem of Boyer shows that
degenerate horizons are closed.
This fact also establishes a topological difference between black holes and extreme black holes.

Now, the
 first law of  \textbf{BH} thermodynamics,

  $\delta M = (1/8\pi)\ell_H^+ \delta A^+_{area}+ \omega^+_H \delta J$, relates the
variation of the mass $\delta M$, the (outer) horizon area $\delta A^+_{area}$, and angular momentum $ \delta J$
with the surface gravity  $\ell_H^+$ and angular velocity $\omega_H^+$ on the outer horizon.
Here all the quantities, including the surface gravity $\ell$, are evaluated on the outer Killing horizon so that the notation $(+)$ can be  omitted.
The (Hawking) temperature term is  related to the surface gravity by $T_{H}= {\hslash c\ell }/{2\pi k_{B}}$  ($k_{B}$ is the Boltzmann constant) and the horizon area $A_{area}^+$
to the entropy, {$S= k_{B} A_{area}^+/l_P^2$ ($l_P$ is the Planck length},  $\hslash$ the reduced Planck constant, and $c$ is the speed of light)\footnote{If the \textbf{BH}
temperature  is $T= \ell/(2\pi)$,  its entropy is
 $S= A_{area}/(4\hslash G)$, the  pressure-term is  $p= - \omega_H $,  where the
internal energy is $U$= GM ($M = c^2m/G $= mass, where $m$ is a mass term). The term $\omega_H \delta J$ is  interpreted as the ``work''.
 The \textbf{BH} (horizon) area $A_{area}$ is  related to the outer horizon definition,  $A_{area} = 8\pi mr_+$, while the  volume term is $V= G  J/c^2$ (where $J = amc^3/G$).
The \textbf{BH} horizon area will be considered here alternatively
in the extended plane.}.

We focus our analysis on two initial and final states for \textbf{BH}  transition.
We can interpret  the terms  $(\omega^+_H \delta J)$ in the \textbf{BH} transition from one state $(0)$ to a new state $(1)$ considering the frequency as characteristic bundle frequency, therefore, with $\omega_H^+(0)=\omega(0)$ which is the  $\mathcal{MB}$ frequency  tangent to the outer horizon of the \textbf{BH} at  the initial state $(0)$, the entropy term $(\ell(0) \delta A^+_{area})$ is also considered on the  bundle, writing  the relation in the extended plane.

All the quantities are expressed in terms of  bundles at $\omega(0)=$constant and $(\delta A_{area}^+, \delta J, \delta M)$, describing the transition from the initial to final  state.  Furthermore, as in the extended plane, bundles tangent to the inner  horizons are also relevant. We write a corresponding relation  between the quantities  $(\delta A_{area}^-, \delta J, \delta M)$ and $(\omega_H^-(0),\ell^-(0))$ evaluated on the inner horizon $r_-$.
Alternately, we express   $(\delta A_{area}^\pm, \delta J, \delta M)$ and $(\omega_H^\pm(0),\ell^\pm(0))$ and their relation in a unique form as function of a generic point $r$
 of the extended plane.

In the next sections we will consider quantities $(\ell,\omega)$ and  $(M,A_{area},J)$ and their variations in the extended plane, on the horizon curve. We express the relation between these in terms of  the  properties of the light surfaces.

\subsection{Masses and \textbf{BH} thermodynamics  }\label{Sec:nil-base-egi}
Here we explore \textbf{BH} thermodynamics in the extended plane  in terms of the light surfaces using  \MB\ curves and discuss variations of  the total mass $M$, area $A_{area}$, and momentum $J$ of the \textbf{BH} defined by the tangency property with the horizons curve in the extended plane, considering the bundle frequency $\omega$ and the acceleration $\ell$.  We emphasize some ("conformal") properties, as defined below, in these relations.

We will connect the  \textbf{BHs} states before and after a transition by expressing their characteristics and parameters through the \MB s (curves defined by the properties of the light surfaces for different geometries) and their representation in the extended plane. Particularly the condition of tangency with the curve of the horizons provides  the  constraint for \textbf{BHs} transformations. Although the extended plane represents also  \textbf{NSs} solutions as the origin of the bundles and the \MB s contain \textbf{BHs} and \textbf{BHs} and \textbf{NSs} (related to the tangency to the curve portion of the internal horizons) or depending on the $\sigma$ angle (near the poles  there are interesting properties of the $ \omega$ frequencies explored in \cite{Pugliese:2021aeb,GRG-lette}), these transformations seem to confirm  cut off of  any \textbf{BH-NS} transition.

The problem of writing the relation between main \textbf{BH} quantities before and after a transition (on the horizon curve)  can be rephrased as the problem of finding replicas in the extended plane. More precisely, let us consider two points,
$r$ and  $r_p$, along the horizon curve  $a_{\pm}>0$, which can also be  in the entire range  $r\in [0,2M]$. We relate the parameters
$\Pa_{\pm}=(\ell^{\pm}, \omega^{\pm}_H)$, regulating the \textbf{BH} transition, as evaluated on the point $r$ and $r_p$.
%
%}.

We search for a quantity $\kappa$   such that
 $\Pa_{\pm}(r)=\kappa \Pa^*_{\pm}(r_p)$ (conformal property). Note that it can  also be
$ \ell^{\pm}$ or $\pm\ell^{\pm}$, connecting, therefore, properties defined on   an outer horizon with  properties defined on an  inner horizon; for this reason the notation $(*)$ represents a change in sign in one or more components of the couple.
While the frequency $\omega$ is always  positive  (according to the sign of  $J$), the surface gravity $\ell$ can be negative, when evaluated on a point of the inner horizon; therefore, it can be
  $\ell_H^{\pm}(r_p)=\ell_H^-<0$.

\begin{figure*}
\centering
   % \includegraphics[width=5.6cm]{nucleveropro}
   \includegraphics[width=6cm]{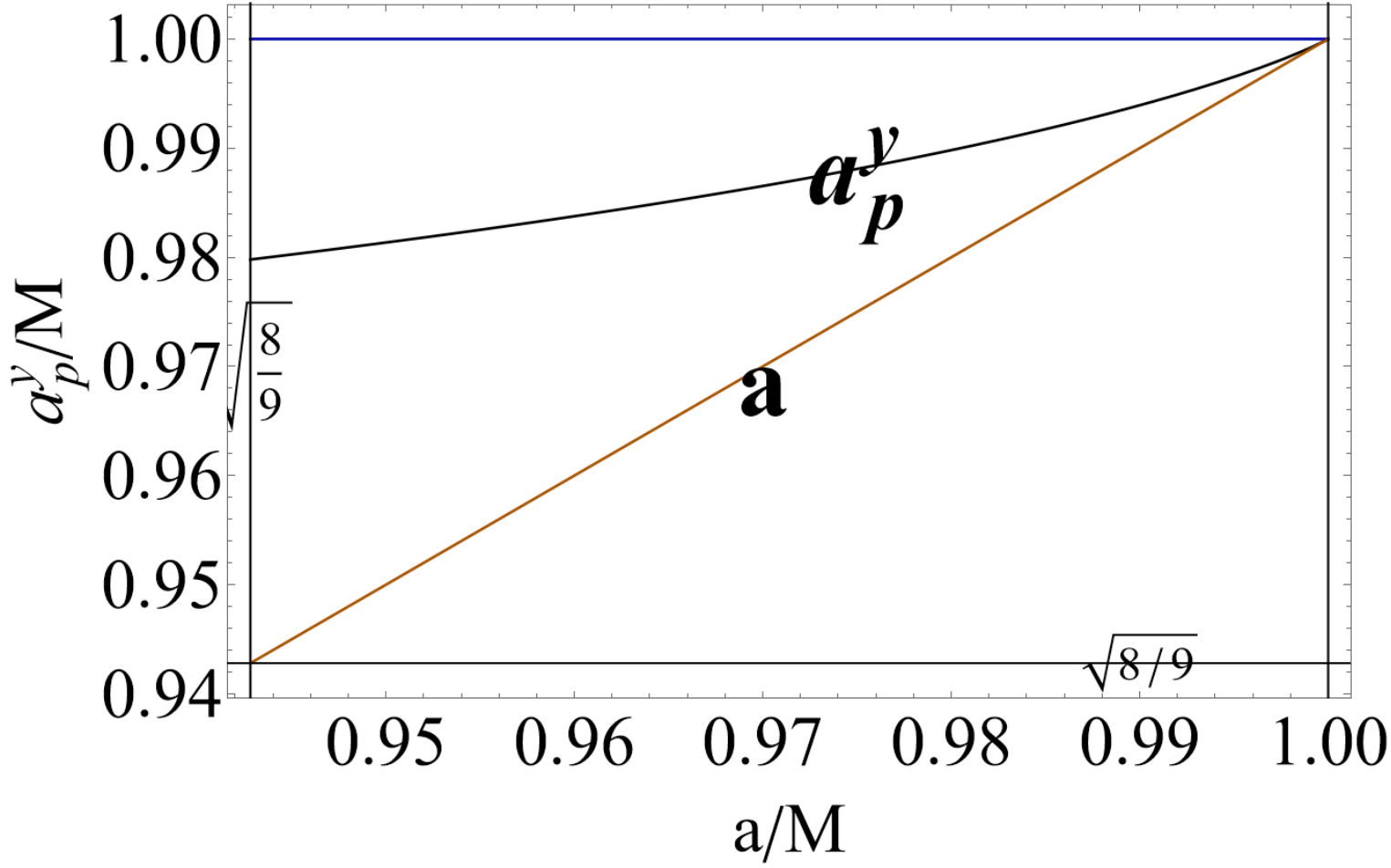}
           \includegraphics[width=6cm]{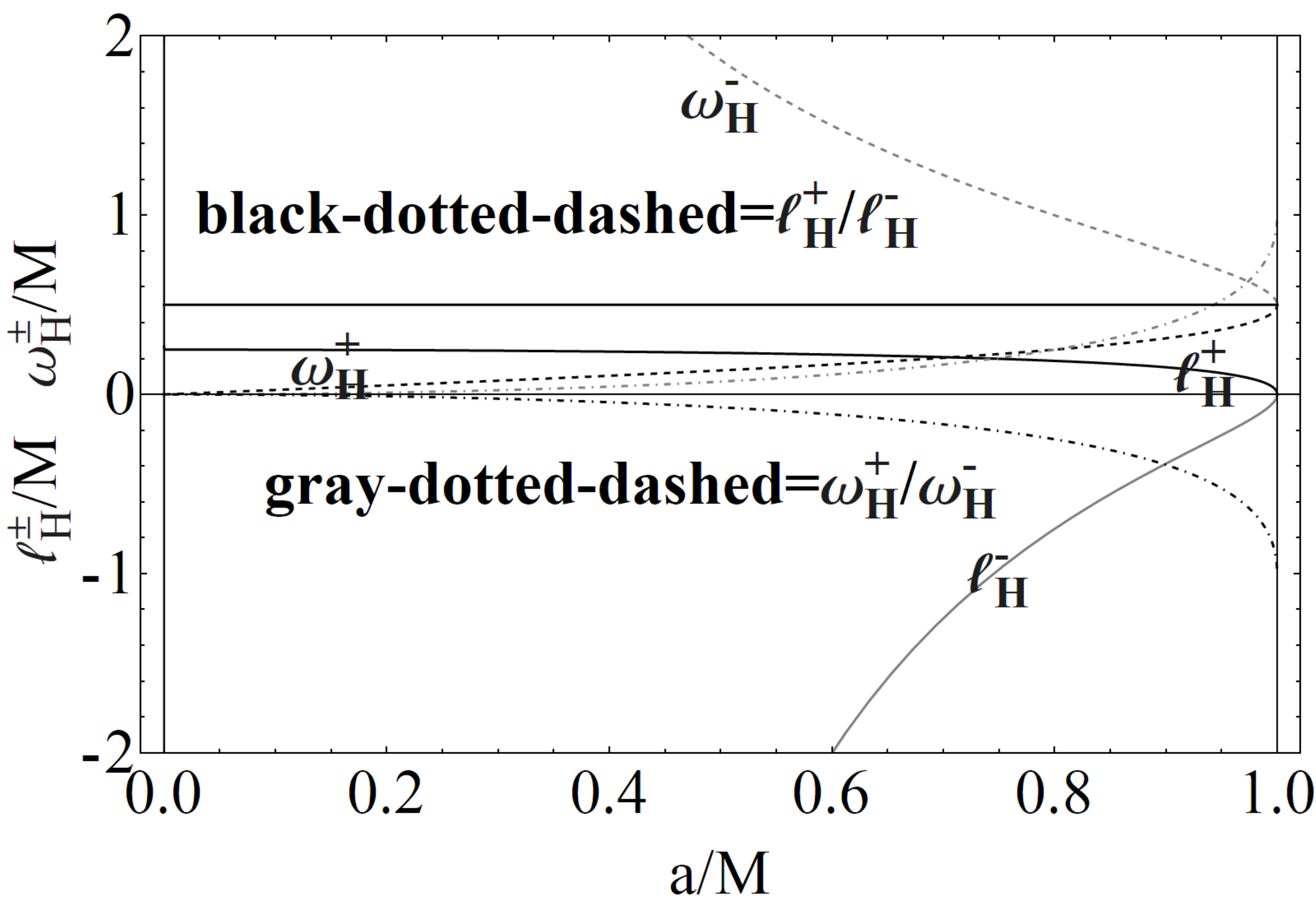}\\
            \includegraphics[width=6cm]{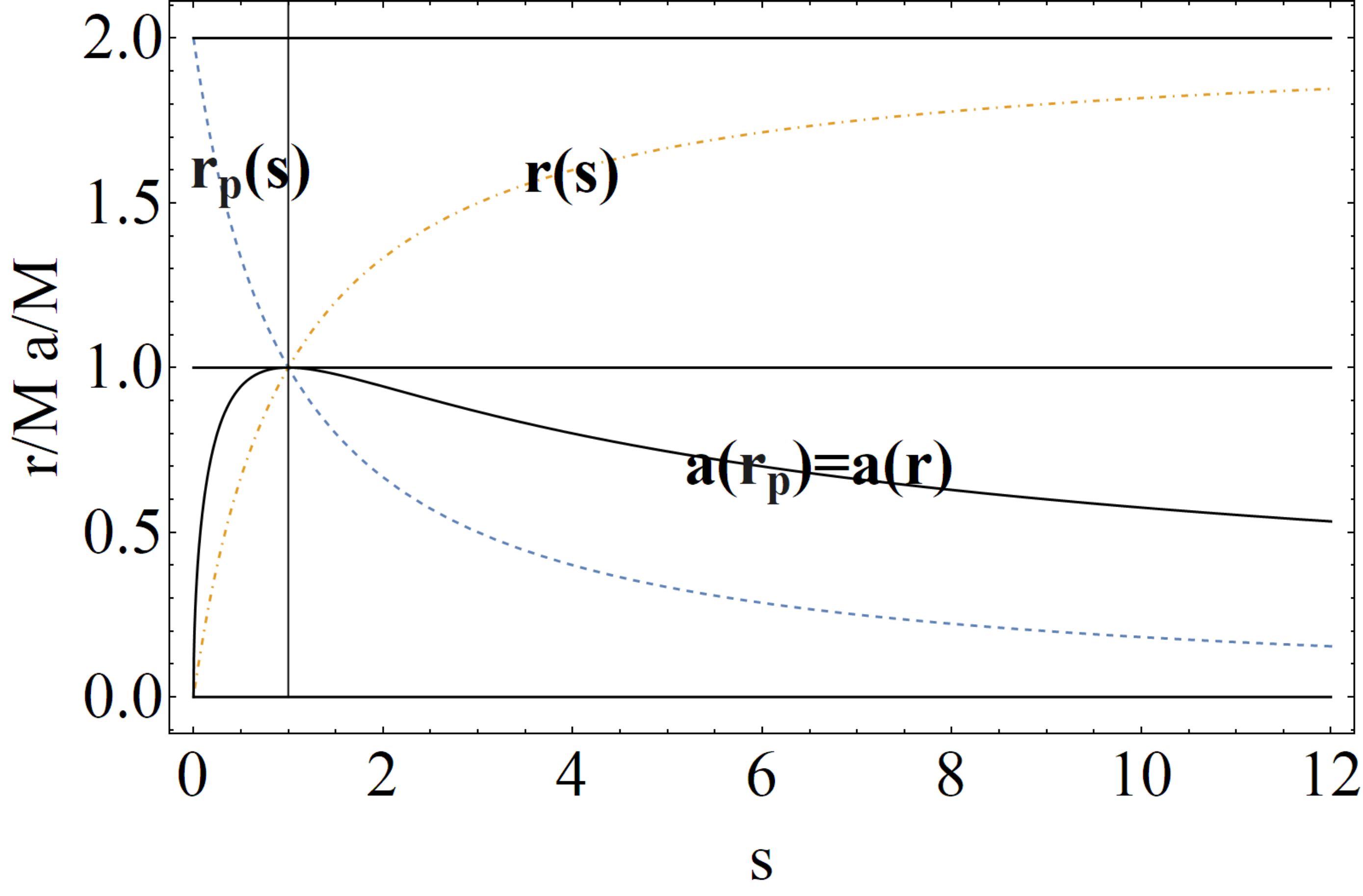}
   \includegraphics[width=6cm]{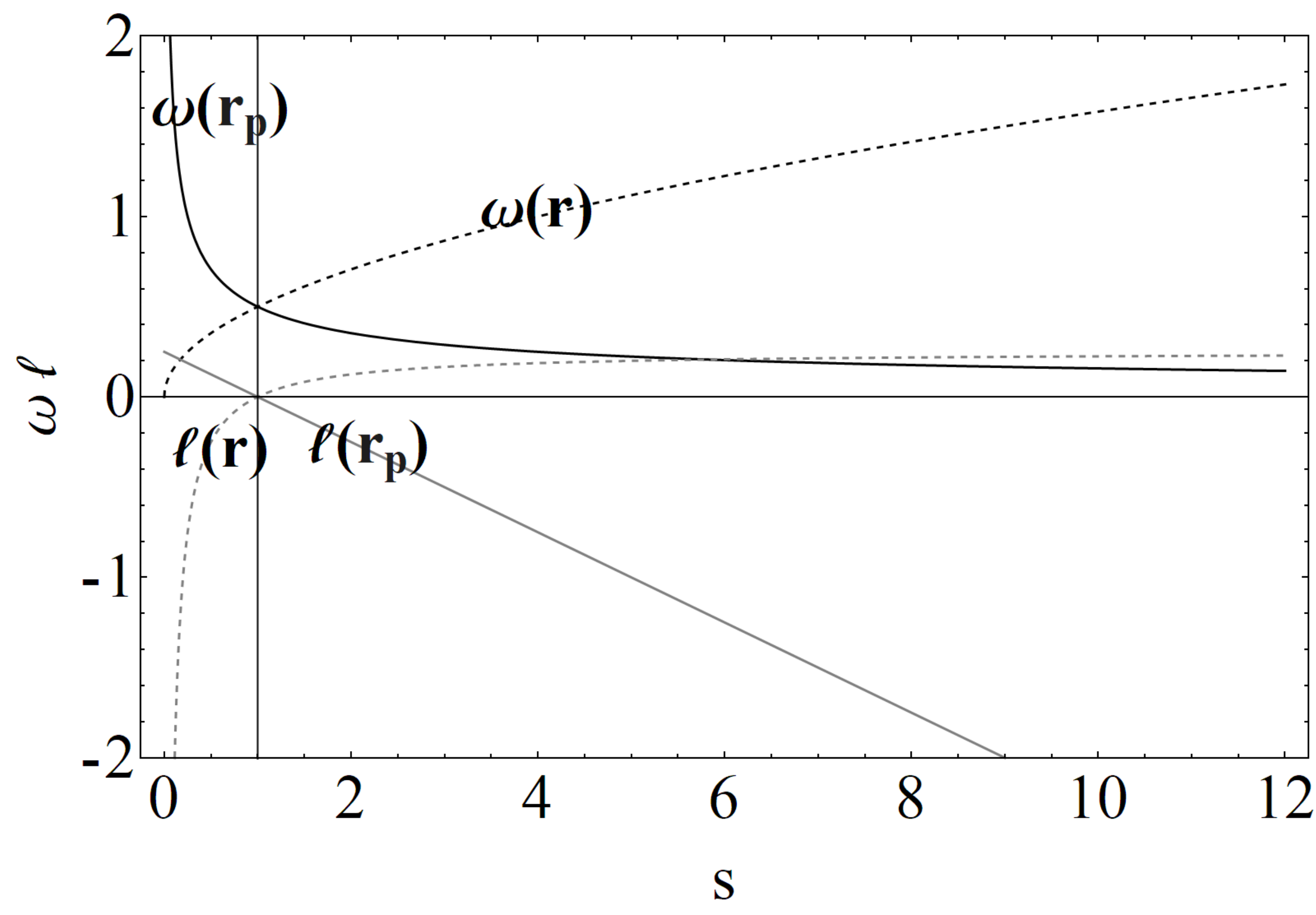}
  \caption{Upper left panel:  The \textbf{BH} spin $a_p=a_p^y$  defined in Eq.\il(\ref{Eq:ap}) as a function of the spin $a$, solution of
$\ell  \left(r_+(a_p)\right)=-\ell  \left(r_-(a)\right)$,
where the acceleration  $\ell(r_+)$ is the surface gravity. The outer and inner horizon are $r_+$ and $r_-$,  respectively. Upper right panel: The  frequencies $\omega_H^{\pm}$ of the outer and inner \textbf{BH} horizons. The quantities  $\omega_H^{\pm}$, $\ell_H^{\pm}$,  and the ratios are plotted as functions of  the \textbf{BH} dimensionless spin.
Bottom right panel:  The quantities  $r(s)$, $r_p(s)$ and the spin  $a(r)=a(r_p)$ from  Eq.\il(\ref{Eq:rrp}) and Eq.\il(\ref{Eq:alti-fo-scoper}) as functions of the parameter $s$.  There $\left(\omega(r_p),\ell(r_p)\right)=s \left(\omega(r),-\ell(r)\right)$.  Bottom right panel: The frequency $\omega$ and the acceleration $\ell$ evaluated on $r$ and $r_p$ are plotted as functions of $s$--see Eqs\il(\ref{Eq:oop}). }\label{Fig:nucleveropro}
\end{figure*}

We also investigate    the case   $\delta A_{area}^+=-\delta A_{area}^-$ occurring for  $\delta M=0$ (as  $\delta M^+=\delta M^-$).
In here we use notation $(\pm)$ to indicate quantities evaluated on  the horizons $r_{\pm}$. Moreover,  $\delta\Qa\equiv\Qa(1)-\Qa(0)$ denotes the change of the quantity $\Qa$ from  the initial $(0)$ to the final state $(1)$ of the transition. The point $a=M$ and  $r=r_p=M$ corresponds to the extreme Kerr \textbf{BH},  where   $\omega(r_p)\neq\omega(r)$ for   points $(r,r_p)$ of the   horizon curve.
For the surface gravity the situation is different. There can be  replicas on the curve of the horizons, which depend on $a$ and $r$ and  connect inner and  outer horizons; this is clearly illustrated in  Figs\il(\ref{Fig:nucleveropro}).

 For convenience, we report here some relations for  the frequencies:
 \bea&&\label{Eq:forma-uniq}
 \omega_H^{\pm}=\frac{\sqrt{2-r}}{2 \sqrt{r}},\quad
 \omega_H^{\mp}=\pm\frac{\sqrt{(r-1)^2}\pm1}{2 \sqrt{(2-r) r}}.
 \eea
  In the first equation of \il(\ref{Eq:forma-uniq}),  the frequency is defined as a function of a general point $r\in a_{\pm}$ of the inner horizon for $r\in[0,M]$ or the outer horizon for $r\in[M,2M]$. In the second expression, we explicit  the  inner and outer horizon frequencies.

\begin{figure}
 \centering
   \includegraphics[width=\columnwidth]{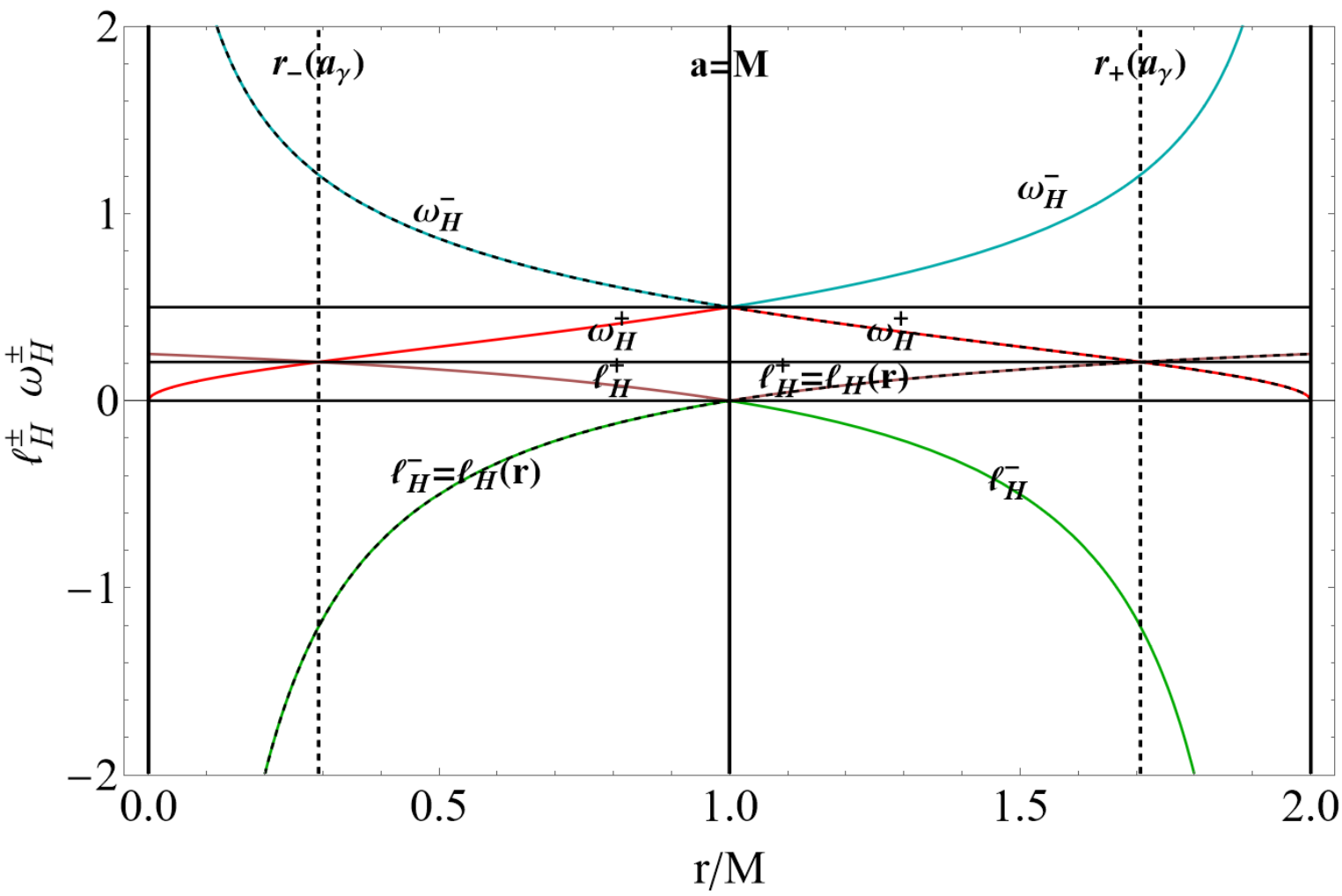}
   \caption{The accelerations $\ell_H^{\pm}$ and the frequencies $\omega_H^{\pm}$ on the horizon curve in the extended plane, as functions of a point $r\in[0,2M]$ on the horizon curve, defined  in Eqs\il(\ref{Eq:forma-uniq}) and (\ref{Eq:2009-eq}). Dashed lines are the  functions $\ell_H(r)$ and  $\omega_H(r)$ defined along the horizon curve $a_{\pm}(r)$. The radius $r\in [0,M]$ corresponds to the inner horizons and the radius $r\in [M,2M]$ to the outer horizon. Here $r_{\pm}=M$ is the horizon in the extreme Kerr \textbf{BH}. Note the symmetries of the functions $\ell_{H}^{\pm}$ in the entire range $r\in [0,2M]$. For the spin $a_{\gamma}\equiv \sqrt{1/2}$, we get  $\omega_{H}^{+}=\ell_H^+$ and $\omega_{H}^{-}=-\ell_H^-$.}\label{Fig:plotregul}
   \end{figure}

When calculated on the inner horizon curve in the extended plane, the surface gravity reads
 \bea&&\label{Eq:2009-eq}
 \ell_H^{\pm}=\frac{a^2\pm\sqrt{1-a^2}\mp 1}{2 a^2}=\frac{1}{2\left[1\pm\frac{1}{\sqrt{(r-1)^2}}\right]},\\&&\nonumber\mbox{and}\quad\ell_H^{\pm}(r)=\frac{r-1}{2 r}.
 \eea
  In first expression of  Eq.\il(\ref{Eq:2009-eq}), we denote with $\ell_H^{\pm}$ the  acceleration for the inner horizon $\ell_H^{-}$ and outer horizon $\ell_H^{+}$ as functions of the  spin $a$ or as functions of the radius $r$ on the horizon curve.  The second function, instead,  is the acceleration defined as function of a general point $r\in a_{\pm}$.
 We introduce the quantities   $\ell^-_+(r)\equiv \ell_H^-(r)/\ell_H^+(r)$ and
 $\omega^-_+(r)\equiv \omega_H^-(r)/\omega_H^+(r)$. Using the relation $\ell_+^-(r)=-\omega_+^-(r)$, we obtain
 \bea\nonumber&&
\{\ell_H^{-},\omega_H^{-}\}=\omega_+^-(r)\{-\ell_H^{+},\omega_H^{+}\},\quad
\{\ell_H^{+},\omega_H^{+}\}=\omega_-^+(r)\{-\ell_H^{-},\omega_H^{-}\},\\&&\label{Eq:eser-malabar}\mbox{where}\quad
 \omega_-^+(r)=\frac{1}{\omega_+^-(r)}.
 \eea
 -see Fig.\il(\ref{Fig:plotregul}).
There is $\ell_{H}^\pm=\pm\omega_H^\pm$ for $a=a_{\gamma}\equiv 1/\sqrt{2}$.

 An analogue relation holds for the  quantities evaluated as functions of the spin $a$.
 Considering  the dependence from the spin $a$ we find,
 \bea
 \ell_H^\mp(a)=\mp\frac{\sqrt{1-a^2}}{2r_\mp} \ ,
 \eea
 alternatively to  Eq.\il(\ref{Eq:2009-eq}).  Defining
 $\ell_-^+(a)\equiv\ell_H^+(a)/\ell_H^-(a)$ and $ \omega_-^+(a)\equiv\omega_H^+(a)/\omega_H^-(a)$, we obtain   $ \omega_-^+(a)=- \ell_-^+(a)$ and
   \bea&&
  \{\ell_H^+(a),\omega_H^+(a)\}=\omega_-^+(a) \{-\ell_H^-(a),\omega_H^-(a)\},\\&&
    \{\ell_H^-(a),\omega_H^-(a)\}=\omega_+^-(a) \{-\ell_H^+(a),\omega_H^+(a)\},
   \eea
   see also Eq.\il(\ref{Eq:eser-malabar})  and Figs\il(\ref{Fig:nucleveropro}).

 Here, we search for the relation  $\delta M(a_p)^{\star}=s  \delta M(a)^{\ast}$, where $a\in [0,M]$ and $a_p\in [0,M]$ are two \textbf{BH} geometries. For the particular case  $a=a_p$,  the couple $(\star,\ast)$ denotes that $\star$ or $\ast$ are $\pm$, relating to the inner-inner horizons of the  two geometries or outer-outer or  inner-outer horizons of the  two geometries $(a,a_p)$. The special case $a=a_p$ contains $(\star=\ast=\pm, s=1)$. The case considered for the quantities as functions of  $r$, where  $\star=-\ast=\pm$ and $s$, satisfies conditions (\ref{Eq:eser-malabar}).
 The apparent contradiction in this case is that for a fixed spacetime
  $a\in \textbf{BH}$, we have that $\delta M^+=\delta M^-$,  but the relation regulating this variation with the other characteristic  quantities depends on the point as  $\delta M^+=s\delta M^-$ (where we have used the notation $\pm$ to stress quantities in relations evaluated on the horizons $r_{\pm}$, respectively).

Then, we obtain
\bea&&\nonumber\delta M^+=\ell_H^+\delta A_{area}^++\omega_H^+\delta J^+=\delta M=\ell_H^+\delta A_{area}^++\omega_H^+\delta J =
\\
&&\label{Eq:prod-dpa-fraita}r\delta M^-\equiv
\ell_H^-\delta A_{area}^-+\omega_H^-\delta J^-.
%\\
%&&\nonumber\delta M=-\ell_-\delta A_{area}^++\omega_-\delta J =\kappa(\ell_+\delta A_{area}^++\omega_+\delta J)
\eea

We note also  the coincidence   $\omega_H^{\pm}(r)=\ell_H^{\pm}(r)=\Qa_{\ell\omega}^{+}\equiv{1}/{\sqrt{2}}-{1}/{2}$
 in the form of Eq.\il(\ref{Eq:forma-uniq}) and Eq.\il(\ref{Eq:2009-eq})  for    $r=r_+=  \left(\sqrt{2}+2\right)/2$. This  is  the outer horizon for the \textbf{BH} geometry with spin $a/M=a_\gamma=1/\sqrt{2}$, specifically, $\ell_H^\pm = \pm\omega_H^\pm$. Therefore, for this spacetime the mass variation has a special for when written in the extended plane $\delta M^\pm=\Qa_{\ell\omega}^{\pm}\delta\tilde{M}^{\pm}$
 (where $\delta\tilde{M}^{\pm}\equiv \pm\delta A^\pm_{area}+\delta J^\pm)$,   where $\Qa_{\ell\omega}^{-}=\omega_H^-={1}/{2}+{1}/{\sqrt{2}}=-\ell_H^-$.

We analyze the more general  problem  addressing firstly the special case $s=1$ and secondly the general case with $s\neq1$.
\begin{description}
\item[The case $s=1$]
This problem can be reduced to finding  replicas for $\ell  \left(r_+(a_p)\right)=-\ell  \left(r_-(a)\right)$, solved for
\bea\label{Eq:ap}
\frac{2 \sqrt{2}}{3}<a\leq 1,\quad a_p=a^{y}_p\equiv\frac{\sqrt{4 \left(1-a^2\right)^{3/2}+12 a^4-15 a^2+4}}{4 a^2-3}
\eea
-- Figs\il(\ref{Fig:nucleveropro}).
The cases $\ell  \left(r_\pm(a_p)\right)=-\ell  \left(r_\pm(a)\right)$ are solved for $a=a_p=M$, while $\ell_H^\pm(a)=\ell_H^{\pm}(a_p)$ for $a=a_p$. % (However we should note that an analogue relation can be found for  $\omega_H^{\pm}$)
The case  $a/M=\sqrt{{8}/{9}}$ implies $(\ell_H^-=-1/4,\ell_H^+=1/8)$ and $
\omega_H^-={1}/{\sqrt{2}},
\omega_H^+={1}/{2 \sqrt{2}}$. Therefore
$(\ell_H^-,\omega_H^-)-=2(-\ell_H^+,\omega_H^+)$.
\item[The case $s\neq1$.]
 As seen for the \textbf{BH} spin  $a/M=\sqrt{{8}/{9}}$, now we extend  the problem of finding replicas by searching for orbits that relate  the surface gravity and frequency as follows
\bea\label{Eq:kkrela}
 \omega_H(r_p)= s \omega_H(r)
 \quad{and}\quad
 \ell_H(r_p)=- s \ell_H(r)\eea
 for  $s\geq0$  (the situation for $s<0$, implying a change of the \textbf{BH} spin rotation, provides a  result analogue to the case $s>0$) that is a conformal transformation from an outer horizon $r_+$ to an inner horizon $r_-=r$. %, where the other relation   $ \omega_H(r_p)= s \omega_H(r)$ {and}  $
% \ell_H(r_p)=s \ell_H(r)$ is not satisfied a part in the trivial case \rtb{cambiare $k$ con $s$}.%or $k_-\leq0$.
 We obtain:
 \bea\label{Eq:rrp}
 r=\frac{2 s}{s+1}\quad\mbox{ and}\quad
 r_p\equiv \frac{2}{s+1},\eea
  therefore,  $r/r_p=s$, and the conformal relation can be also expressed as
\bea \omega_H(r_p)=(r/r_p) \omega_H(r)
 \; and \;
 -\ell_H(r_p)= (r/r_p)\ell_H(r).\eea

The frequencies and accelerations are
\bea\label{Eq:oop}&&
\omega(r)=\frac{1}{2 \sqrt{s}},\quad\omega(r_p)=\frac{\sqrt{s}}{2},\\
&&\nonumber  \ell(r)=\frac{s-1}{4 s}, \quad \ell(r_p)=\frac{1-s}{4}
\eea
then
\bea\label{Eq:alti-fo-scoper}&&
\left(\omega(r_p),\ell(r_p)\right)=s\left(\omega(r),-\ell(r)\right),\quad\mbox{where}\quad \frac{r}{r_p}=s\\&&\nonumber\mbox{with}\quad a(r)=a(r_p)=2 \sqrt{\frac{s}{(s+1)^2}}=\sqrt{r r_p}=\sqrt{s}r_p
\eea

The radii $r$ and $ r_p$  correspond to the  horizons of the \textbf{BH} spacetime  with  $a(r)=a(r_p)$.
It is clear that $(r,r_p)$ represent a horizon parametrization and $a(r_p)$ can be seen as analogue to the tangent curve of the horizon $a_g(\omega)$, function of the frequency $\omega$.
It is clear from  Eqs\il(\ref{Eq:rrp}) and (\ref{Eq:oop}) that there are two ranges,  $s<1$  and $s>1$, describing the same spacetime (horizontal line in the plot of $a(r_p)=a(r)$ as function of $s$). For $s<1$, we have that  $r=r_-$ and $r_p=r_+$  while for $s>1$, we obtain
 $r=r_+$ and $r_p=r_-$. The limiting case $s=1$ corresponds to the extreme Kerr \textbf{BH}.

Therefore, Eqs\il(\ref{Eq:kkrela}) hold  for two points $(r,r_p)$ only, the horizons of a \textbf{BH} spacetime with $a=a(r_p)$. There are two values of $s=\bar{s}>1$ and $s=1/\bar{s}<1$ for the same spacetime. The remarkable aspect of this relation is that the conformal factor is different according with the spin $\hat{s}_{\mp}\equiv[({2-a^2})\mp 2 \sqrt{({1-a^2})}]/{a^2}$, see Figs\il(\ref{Fig:Plotonethero3}).

\begin{figure}
\centering
  \includegraphics[width=\columnwidth]{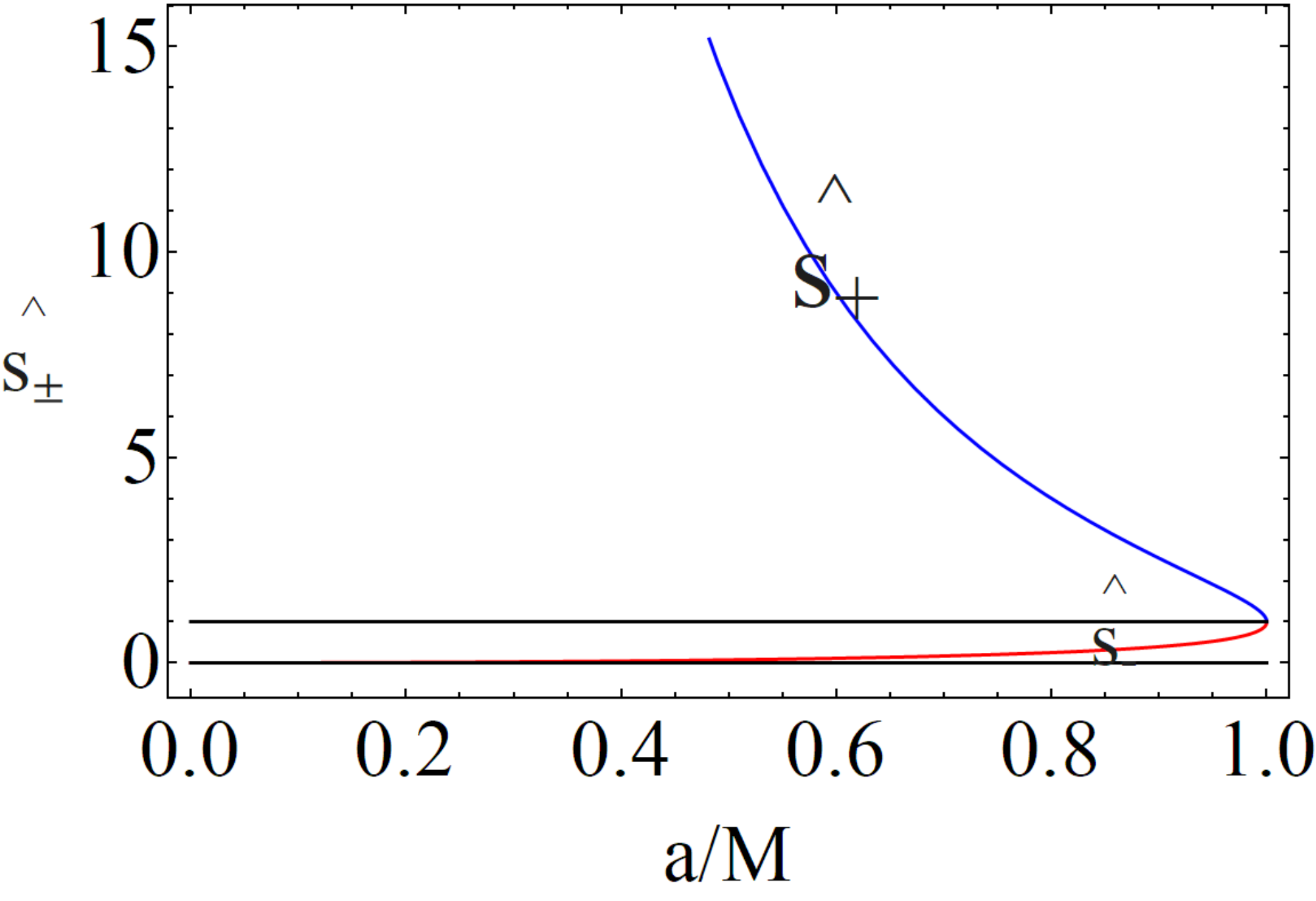}
  \caption{The quantities  $\hat{s}_{\mp}$ as functions of the \textbf{BH} dimensionless spin $a/M$, solutions of  Eqs\il(\ref{Eq:kkrela}): $ \omega_H(r_p)= s \omega_H(r)
 \quad{and}\quad
 \ell_H(r_p)=- s \ell_H(r)$, where $\ell_H$ and $\omega_H$ are the acceleration $\ell$  and the frequency $\omega$, respectively,  on the \textbf{BH} horizons $r_{\pm}$. The surface gravity is  $\ell$ evaluated on $r_+$.
 }\label{Fig:Plotonethero3}
\end{figure}
\end{description}
In  Sec.\il(\ref{Sec:comp-inner-outer}), we will study these relations in detail.
\begin{figure*}
\centering
  \includegraphics[width=6cm]{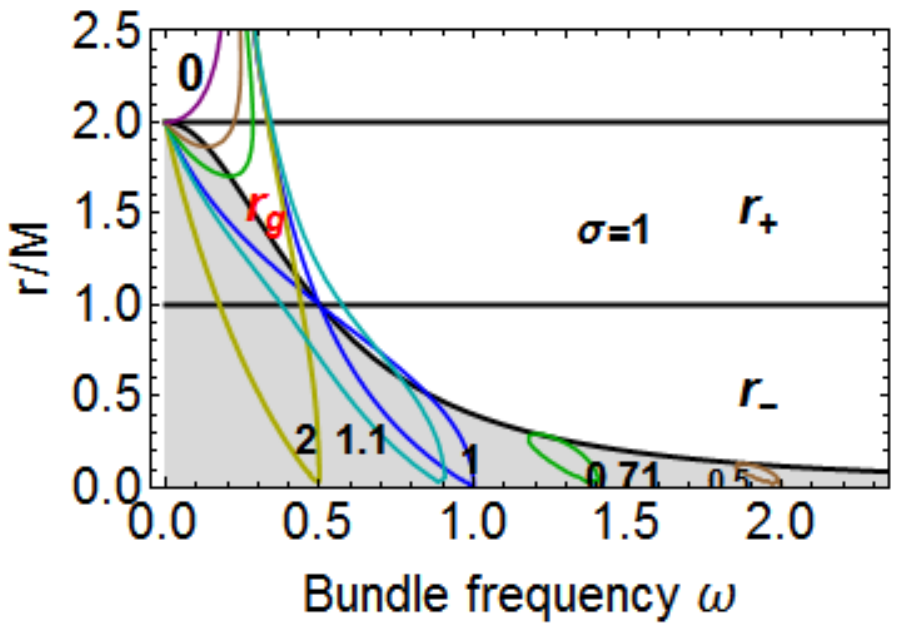}
   \includegraphics[width=6cm]{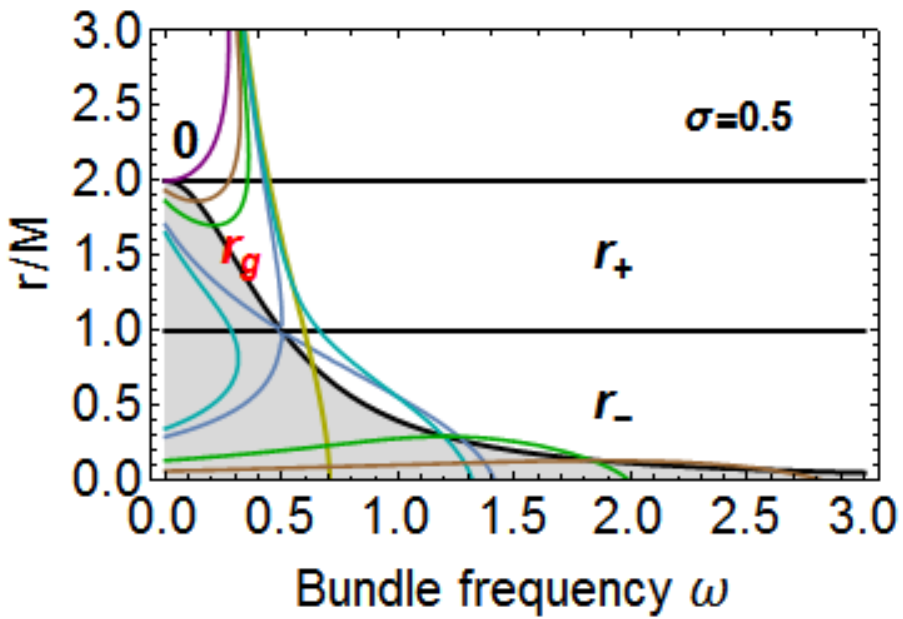}
           \includegraphics[width=5cm]{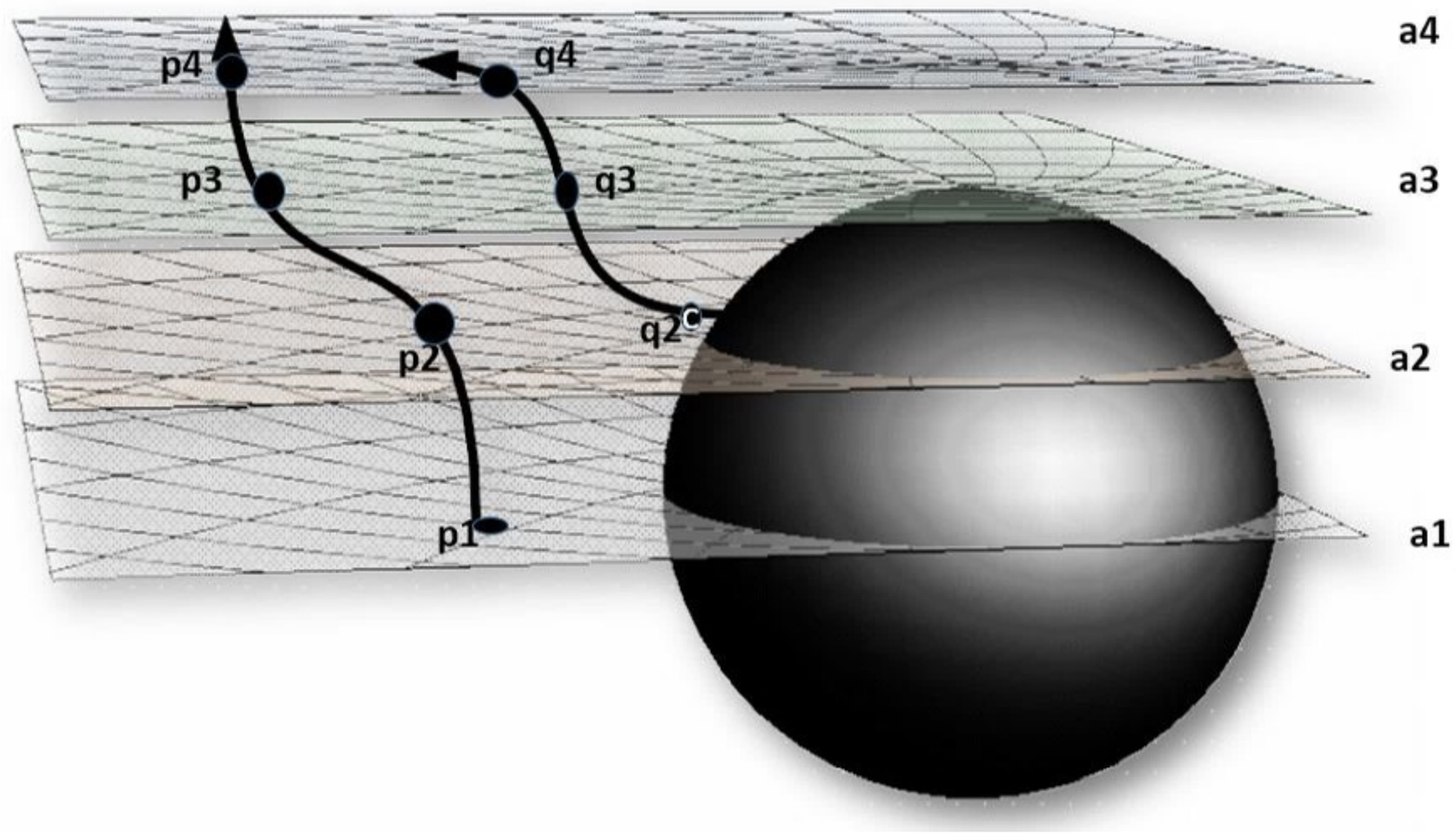}
  \caption{Left and center panels: Light surfaces $r_{s}^{\pm}$ as functions of the bundles  frequency $\omega$ for  different values (denoted on each curve) of the dimensionless spin  $a/M$, representing  \textbf{NS} and  \textbf{BH} geometries. The quantities
  	$r_s^{\pm}$ are solutions of $\laa\cdot \laa=0$, where $\sigma=\sin^2\theta$.
Left-panel: Equatorial plane $\sigma=1$. Right-panel: Plane  $\sigma=0.5$.
 $r_{\mp}$ are the inner and outer horizons in the extended plane, where
$r_+\in[M,2M]$, and $ r_-\in[0,M]$, while
$r_g$ represents the points of the metric bundles that are tangent to the horizon curve in the extended plane, i.e., it is the horizon  curve as function of the bundles characteristic (and horizons) frequency. Gray region is $r<r_g$. There $a_0$ is the bundle origin spin as function of the frequency  $a_0=1/\omega\sqrt{\sigma}$.
Note that in slowly rotating  \textbf{NSs}, an increase of the spin  $a_0$ leads to the appearance of the  Killing bottleneck (a restriction of the light surfaces).
Right panel: The central sphere represents \textbf{BHs} in the extended plane; each horizontal plane corresponds to a $a=$constant surface, i.e., a particular Kerr geometry, an the points on the black surface represent the inner and outer \textbf{BH} horizons. Planes not crossing the sphere are \textbf{NSs}. The sphere considers also the two rotation directions of the \textbf{BH} spin and the central equatorial plane is the Schwarzschild spacetime. The points $p_i$ and $q_i$ correspond to replicas, where $p_i$ denote geometries related by the same frequency $\omega$ and $q_i$ include a point of the horizon.
 }\label{Fig:consecentrA}
\end{figure*}
\section{Exploring  the laws of \textbf{BH} thermodynamics}\label{Sec:mass-termo-smarr}
{Israel's theorem} establishes that the event horizon must be spherically symmetric, implying  that  the sphericity of
an isolated and  static \textbf{BH} cannot be broken.
 Information  on the \textbf{BH} past,  in the sense of  \textbf{BH} "hair", is   to be considered as eliminated or
  inaccessible  to  observers.
(Counterexamples might be considered in some multi-dimensional spacetimes with   non-abelian fields).

In this section, we reformulate several aspects of  \textbf{BH} thermodynamics   in the extended plane and  on the bundles.

To explore the laws of \textbf{BH} thermodynamics  in the extended plane,
we consider the Smarr's formula  connecting $(M,J,A_{area})$, where $A_{area}$ is the horizon area.

In this section, we expand on the analysis of   Sec.\il(\ref{Sec:nil-base-egi}), where we considered masses and \textbf{BH} thermodynamics  in the extended  plane.
In Sec.\il(\ref{Sec:sud-mirr-egi}), we introduce the
 \textbf{BH} rotational energy and "rest"  mass in the extended plane, which is studied using
 the light surfaces in Sec.\il(\ref{Sec:cou.mass-irr}), where we investigate also
{horizon areas in the extended plane}
 and  \textbf{BH} thermodynamics.
Some of the concepts discussed here are the subject of  Sec.\il(\ref{Sec:constant-Mirr}). Moreover, we analyze the constant irreducible mass
and the quantities evaluated on the inner horizons in Sec.\il(\ref{Sec:alc-cases}).
\subsection{Rotational energy and \textbf{BH} "rest"  mass}\label{Sec:sud-mirr-egi}
%
%s
The horizon surface area (on the event horizon $r_+$)  is $A_{area}=16 \pi M_{irr}^2$ and  determines
the \textbf{BH} irreducible (or rest) mass $M_{irr}=\sqrt{a^2+r_+^2}/2$.

The total  \textbf{BH} mass $M$ of the Kerr \textbf{BH}
 can be decomposed into the  mass $M_{irr}$ and  the
 rotational energy.

From the first law of \textbf{BH}  thermodynamics, we obtain
  $M^2= M_{irr}^2+J^2/4M_{irr}^2$, where  $J$ is the  \textbf{BH} angular momentum in  units   of mass $M$, as measured in the asymptotical flat region.
  %where $(1)$ and $(0)$ indicates the final and starting point of stationary or quasi stationary process.

   The  maximum  rotational energy which
can be extracted from the black hole is $\xi\propto(M-M_{irr})$.  A result  of Christodolou and Ruffini sets an upper
limit on  the amount of  energy  that can be extracted   from a Kerr \textbf{BH} as  the  total rotational energy, assuming  a Schwarzschild \textbf{BH} as the final state, and the
 the bottom limit\footnote{The upper limit for  the energy extraction from a stationary process bringing the \textbf{BH} to the state $(1)$ is $M(0) - M_{irr}(M(0), J(0))$, where   $M(0)$ and  $ J(0)$ are the mass and  angular
momentum of the initial state and the \textbf{BH} angular momentum  after the process, the state $(1)$, is zero-- \cite{ella-correlation}.} of $M$  at the end of stationary  process of the energy extraction, has to be  $M_{irr}$.

Within  these assumptions, the maximum rotational energy which can be extracted  is the limit of $\xi=\xi_{\ell}\equiv \left(2-\sqrt{2}\right)/2$, hence  $\xi\in[0,\xi_{\ell}]$, where at the state $(0)$ (prior to the extraction) there is an extreme Kerr spacetime (with spin $a=M$).
% of the mass-energy of the black
Considering   $(0)$  as the state prior the extraction, from the law $M_{irr}^2=r_+(J,M)/2$, we obtain \bea
%\text{dJH}=\frac{a d}{M}=d L \text{M0}=\frac{\text{dMH} L}{\text{Ein}}
%\\
M(0)^2-M_{irr}(0)^2=\left(\frac{J(0) M(0)}{M(0) (2 M_{irr}(0))}\right)^2,\eea
{from the  variation  we find}
\bea\nonumber
\frac{\delta M_{irr}}{M_{irr}}=\frac{\delta M-\delta J \omega_H(0)}{\sqrt{M(0)^2-\frac{J(0)^2}{M(0)^2}}},\eea
where $
\delta M_{irr}\geq 0$, thus $(\delta M-\delta J \omega_H(0))\geq 0$  and $\omega_H\equiv\omega_H^+$ is the frequency of  the {outer} Killing horizon for the initial \textbf{BH}, imposing limits on the \textbf{BH}  spin-shift\footnote{Measuring $\xi$, therefore, will provide and indication of the \textbf{BH} spin. This method,  independent of the specific model of energy extraction,  was introduced   first in \cite{Daly0}, and applied in other analysis  in \cite{Daly0,Daly2,Daly3,
GarofaloEvans,ella-correlation}.
 relating the  dimensionless \textbf{BH} spin $a/M$ to   the dimensionless ratio $\xi$, where $\xi$ represents
the total released rotational  energy   versus \textbf{BH}  mass (measured by an observer at infinity),  assuming a  process ending with  the  total extraction of  the    rotational energy of the central Kerr \textbf{BH}.
}.
We obtain
$ M^2={J(0)^2}/({4 M_{irr}^2})+M_{irr}^2$,
and the  (extracted rotational) energy  is essentially
 \bea\nonumber\xi=M(0)-\sqrt{\frac{r_+(0)}{2}},\quad\mbox{or}\quad
\frac{\xi}{M(0)}=1-\frac{\sqrt{1+\sqrt{1-\frac{J(0)^2}{M(0)^4}}}}{\sqrt{2}}.
\eea
%simmetrie per esmepiob $\xi\to 2-\xi$
Note that here $\xi$ has units of mass,
the dimensionless spin  $a_g/M\equiv J(0)/M(0)^2$ refers to an initial state before the transition, defined by the mass  $M(0)$  and spin  $J(0)$, which is a function of the extracted rotational energy  $\xi/M(0)$, that is,   $1-M_{irr}(0)/M(0)$, and, therefore, of the ratio  $M_{irr}(0)/M(0)$.

On the other hand,  the radius  $r/M$ on the {extended plane} can be expressed in terms of  the total mass $M(0)$.
We can write the dimensionless spin as $a_g=a_{\xi}$ as
\bea\label{Eq:exi-the-esse.xit}
a_{\xi}(\xi)\equiv 2 \sqrt{(2-\xi) (\xi -1)^2 \xi}.
\eea
The function $a_{\xi}(\xi)$ links the former state spin $a_0$ to the rotational extraction in the subsequent phase,  where the \textbf{BH} is settled in a Schwarzschild spacetime.
The function $a_{\xi}$ is, therefore, an expression for the tangent spin in the extended plane as function of the rotational energy parameter $\xi$.
Here  and in the following, we shall  use dimensionless quantities.

 Equivalently, we can express the rotational energy parameter as
\bea\label{Eq:candd}&&
\xi^{\mp}_{\pm}=1\pm\sqrt{\frac{r_{\mp}}{2}}.\eea
{Considering} $a_{\xi}(\xi)\equiv a_s^{(\pm)}$, {solving for $\xi$ there is}\bea\label{Eq:xis-nilo}
&&\xi_s^{(\pm)}=1\flat\frac{\sqrt{1\natural\sqrt{1-(a_s^{(\pm)})^2}}}{\sqrt{2}},\quad\mbox{where}\quad \flat=\pm; \quad \natural=\pm,
\eea
relating directly the energy parameter $\xi$ to the horizons.
  As  $\xi\propto M-M_{irr}$  (here $\xi$ has units of mass $M$), only the solution $\xi_-^+$ has to be considered.
 The  general solution
$a_{\xi}(\xi)=a_{s}^{(\pm)}$, for a couple of spins $a_{s}^{(\pm)}$,  provides the eight functions
$\xi_s^{(\pm)}$.
For the point  $\xi=1$, it is  $a_{\xi}({\xi})=0$   with  $\xi=0$.
There is a maximum that depends on the spin, $\xi_{\ell}\equiv\left(2-\sqrt{2}\right)/2$ (and  $\xi_m\equiv \left(2+\sqrt{2}\right)/2$), where $a_{\xi}(\xi)=M$ and  $r_+=M$ (the extreme Kerr \textbf{BH}).
In general, it is  $a_{\xi}(\xi)\in[0,M]$ (dimensionless) and the  definition of the rotational energy as $1-M_{irr}/M$, with restricted range  $\xi\in[0,\xi_{\ell}]$, where $\xi_{\ell}\equiv\left(2-\sqrt{2}\right)/2$, limiting, therefore, the energy extracted to a maximum of  $\approx 29\%$ of the mass $M$.
 We can express  the extracted energy in terms of characteristic  frequency  of the bundle and through the tangency  condition of the bundle in the extended plane.
We introduce  the curves $\xi_{\tau\tau}^{\mp}$ and $\xi_{\tau}^{\mp}$ in the extension of the plane for extended values of $\xi$,
 considering  the  \textbf{BH} horizon frequencies $\omega_{H}^{\pm}$, the horizon curves   in the  plane $\xi-r/M$, and the functions $\xi_\mu^{\mp},\xi_{\nu}^{\mp}$ from the  other two solutions of $a=a_{\xi}(\xi)$:
\bea\label{Eq:xtautau}
&&
\xi_\mu^{\mp}=1\mp\sqrt{1-\frac{r}{2}},\quad \xi_\nu^{\mp}\equiv1\mp\sqrt{\frac{r}{2}}\quad\mbox{and}
\\\nonumber
&&
\xi_{\tau}^{\mp}\equiv 1\mp\frac{1}{\sqrt{4 \omega ^2+1}}= 1\mp\sqrt{\frac{r_g}{2}}=1\mp\sqrt{\frac{a_g}{4\omega}},\\&&\nonumber
\xi_{\tau\tau}^{\mp}\equiv 1\mp\frac{2 \omega }{\sqrt{4 \omega ^2+1}}= 1\mp2\omega\sqrt{\frac{r_g}{2}}=1\mp\sqrt{2\omega a_g},
\eea
where $r$  is a point of the horizon curve in the extended plane. The maximum rotational energy extractable is therefore expressed as function of  the light-surface frequencies (or equivalently the \MB origin $\la_0$).
For  the extreme case $a =
 M $, where the frequency is $\omega=  1/2$, we obtain   $ \xi=1\pm {1}/{\sqrt {2}}$.

The  horizon frequencies, which are also the  characteristic frequencies of the bundles, in terms of the dimensionless parameter $\xi$ are
\bea\label{Eq:show-g}
\omega_g^ {\pm} \equiv \frac {1  \pm \sqrt {4 (\xi -
            2)\xi (\xi - 1)^2 +
        1}} {4\sqrt {-(\xi - 2) (\xi - 1)^2\xi}};
\eea
$\omega_g^ {-} $ has a saddle point for energy $\xi=\left (1\pm \sqrt {2/3} \right)$, corresponding to the {spin  { $ a/M = {2\sqrt {2}}/{3}$} }and frequency\footnote{Notable spins are for $a=
1/\sqrt{2}$: $ r_{\gamma}^-=r_{\epsilon}^+$, and $a/M=
2(\sqrt{2}-1): r_{mbo}^-=r_{\epsilon}^+$,
$a/M=\sqrt{8/9}$ $r_{mso}^-=r_{\epsilon}^+$, and $
a/M=2\sqrt{2}$ for naked singularities,   where
$r_{mso}^-=r_{\epsilon}^+$, and $(r_{\gamma}^-,r_{mbo}^-,r_{mso}^-)$ are the corotating last circular orbit (photon orbit),  marginally bounded orbit, and marginally stable orbit, respectively. For  $a/M\geq \sqrt{8/9}$  co-rotating accreting disks close to the outer ergosurface,  centred on the \textbf{BH} equatorial plane must have their inner part (distance- torus  center--inner edge) in the ergoregion---\cite{dragged}.  This implies  the peculiarities highlighted here in  the extraction process should be evidenced  for these discs whose inner part  is subjected to the frame dragging of the ergoregion.} $\omega=({1}/{2\sqrt {2}},{1}/{\sqrt {2}}$).
Frequency $\omega_H^+=\pm{1}/{2 \sqrt{2}}$  for the outer horizon  $r_g/M= 4/3$   is also  the saddle point of $\xi_\tau^\pm$ as function of $\omega$, the maximum extractable  rotational energy decreases/increases faster with the  horizon (\MB) frequencies constrained  by  the limiting values $\omega_H^+=\pm{1}/{2 \sqrt{2}}$.

\subsection{BH irreducible mass}\label{Sec:cou.mass-irr}
We express the irreducible mass of a \textbf{BH} in terms of the horizon curve in the extended plane. Thus,    using also the results given  in  Sec.\il(\ref{Sec:nil-base-egi}), we rewrite the mass function in terms of the inner and outer horizon radii in the extended plane.
We then study  the areas and the relation  (\ref{Eq:prod-dpa-fraita}) in this representation, wiring  first law of \textbf{BH} thermodynamics  in this new frame.

We can express the \textbf{BH}  irreducible mass $M_{irr}$, a quantity defined on the outer \textbf{BH}  horizon, in terms of its area $M_{irr}=\sqrt{a^2+r_+^2}/2=\sqrt{{A_{area}}/{16 \pi}}$ as
\bea&&  \label{Eq:Mirr-all}M_{irr}(all)=\sqrt{\frac{\la_0^2}{\la_0^2+4}}=\sqrt{\frac{r_g}{2}}%=\sqrt{\frac{a_g\la_0}{4}}
=\sqrt{\frac{1}{4 \omega ^2+1}},
\\&&\nonumber M_{irr}^\mp=\sqrt{\frac{1}{2}\mp\frac{1}{2} \sqrt{\frac{\left(\la_0^2-4\right)^2}{\left(\la_0^2+4\right)^2}}}=
 M_{irr}(r_{\mp})=\sqrt{\frac{r_\mp(a_g)}{2}}.%\sqrt{\frac{\la_0^2+4+\sqrt{\left(\la_0^2-4\right)^2}}{2 (\la_0^2+4)}}
\eea
in terms of the radius $r_g$, the  frequency $\omega$, the bundle origin $\la_0=a_0\sqrt{\sigma}$ or the horizon curve $a_{g}$. (Note the dependence on $\theta$ through the bundle origin,  does not contradict  the rigidity of the black hole horizon, being a representation of the light-surfaces frequencies on different planes $\sigma$s). Therefore, the mass function is defined also in the  inner horizon.
The function $M_{irr}(all)$ denotes the irreducible mass evaluated  on the horizon curve in the extended plane, as function of the bundle origin $\la_0$, the tangent radius $r_g$ or the frequencies $\omega$. The notation $\pm$ refers to the radii $r_{\pm}$.

\begin{figure*}
\centering
  \includegraphics[width=7cm]{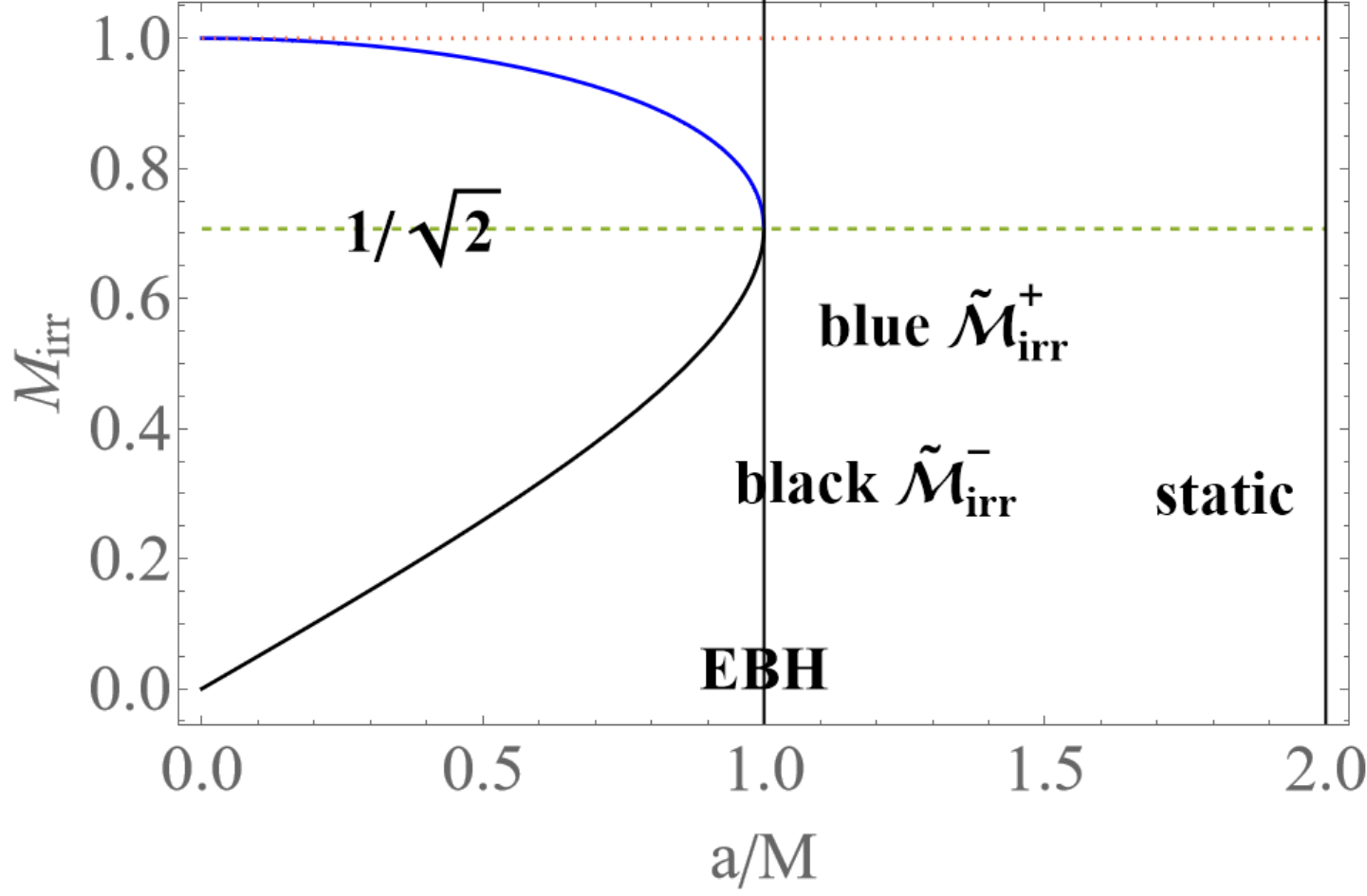}
   \includegraphics[width=7cm]{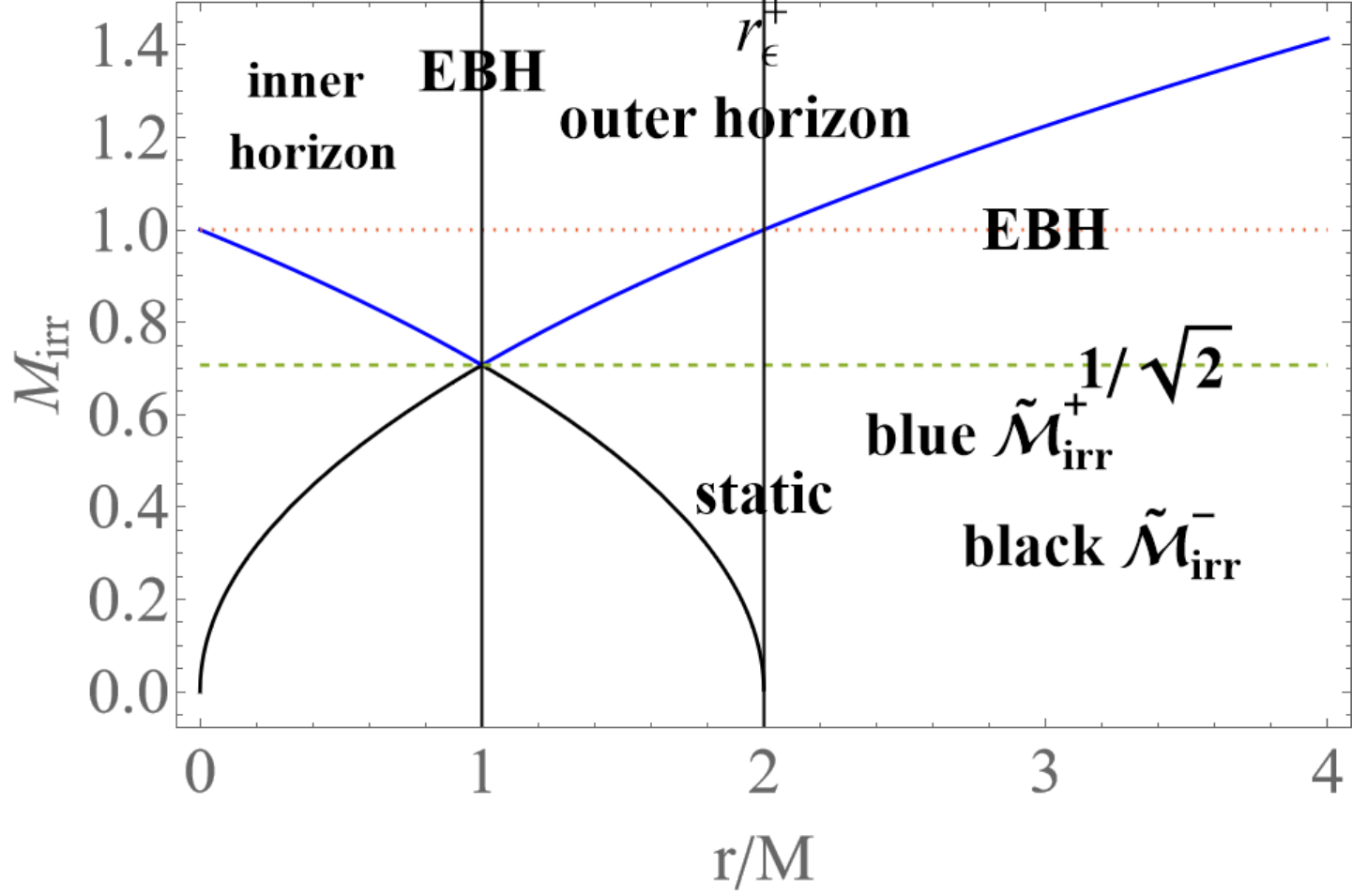}
    \includegraphics[width=7cm]{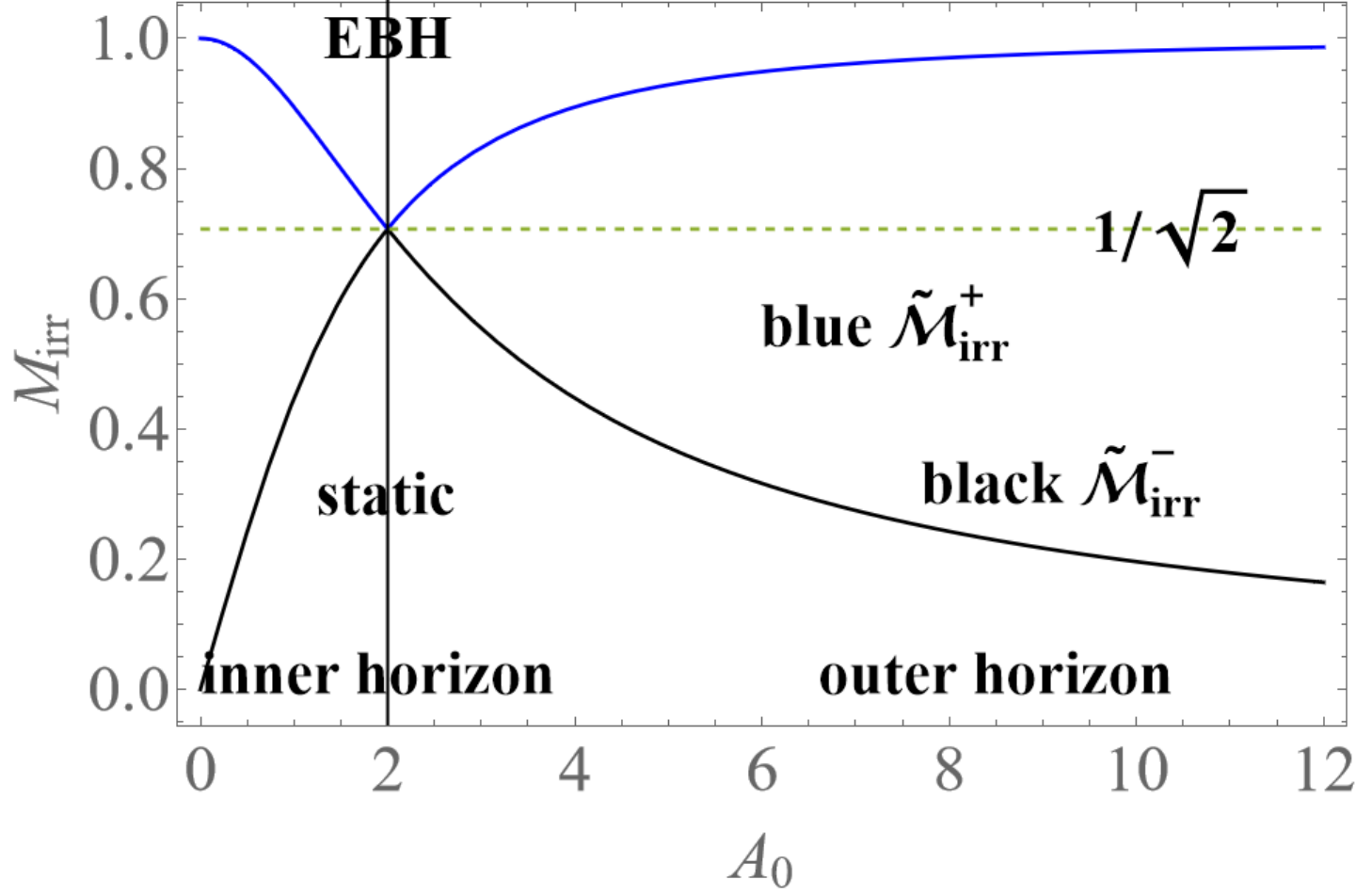}
      \includegraphics[width=7cm]{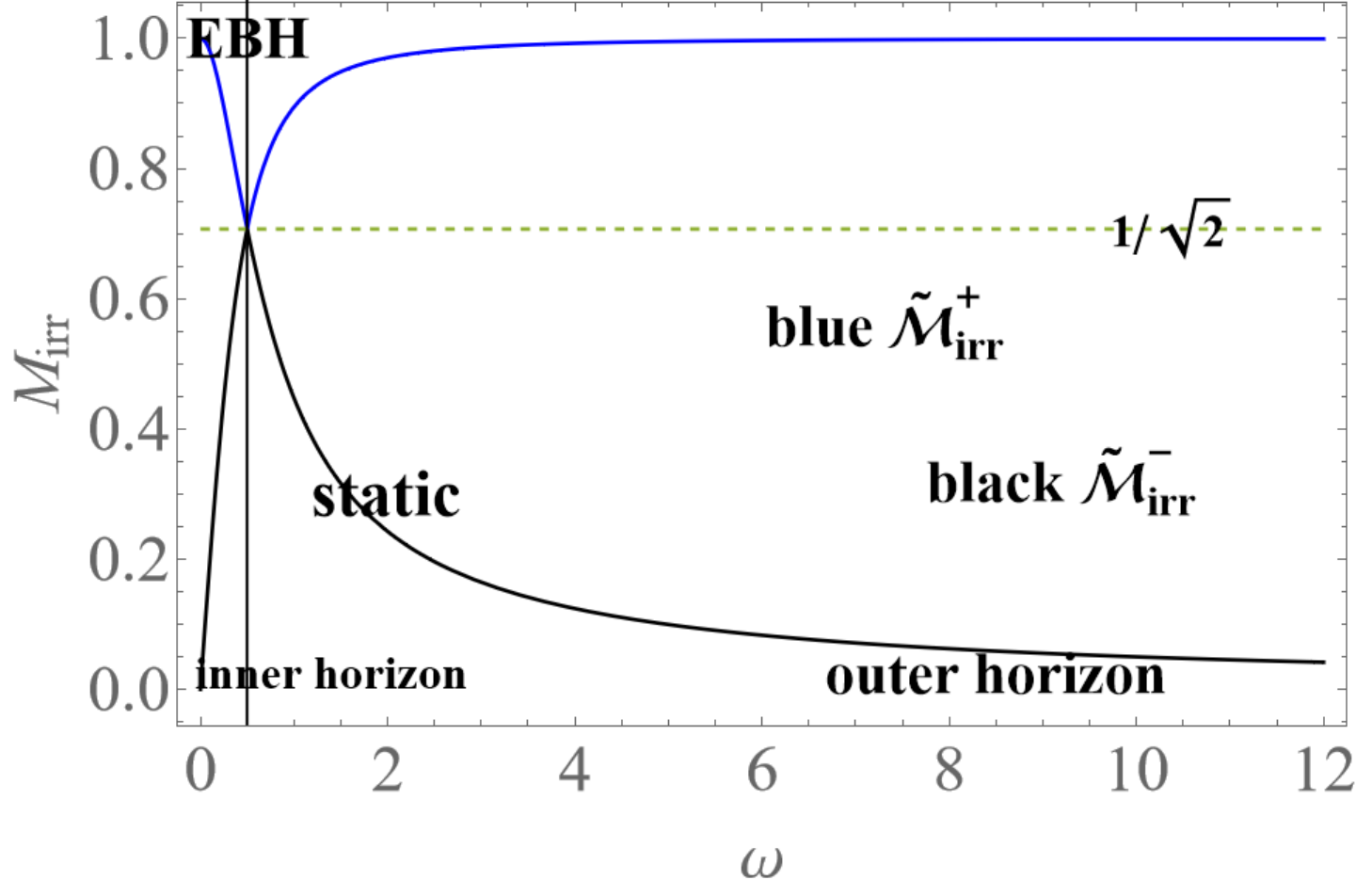}
  \caption{Irreducible mass $M_{irr}$ as function of
  the spin $a/M$, the radius on the extended plane $r/M$, the origin spin of the bundle $\la_0=a_0\sqrt{\sigma}$, where $\sigma\equiv \sin^2\theta$, and the bundle  characteristic frequency $\omega$.  Here,  \textbf{EBH} denotes the extreme \textbf{BH} spacetime. The regions of the inner and outer horizons in the extended plane are denoted, respectively, on the left and right of the \textbf{EBH} limit.}\label{Fig:zongiaaransboc}
\end{figure*}

The irreducible mass is,  therefore,
determined by the tangent radius in the extended plane.
Note that
\bea\label{Eq:narmer-reg}&&
\frac{M^+_{irr}}{M^-_{irr}}=\sqrt{\frac{r_+}{r_-}}=\sqrt{\frac{\omega_H^-}{\omega_H^+}}=\\
&&\nonumber
\sqrt{\frac{-\ell_H^-}{\ell_H^+}}=\frac{\sqrt{r_+}}{\sqrt{2-r_+}}=\sqrt{\frac{1}{4(\omega_H^+)^2}}
\eea
%.
relating the irreducible masses, the frequencies and the surfaces gravities.
Considering the expression for the irreducible mass in terms of the radius $r$, we obtain the expressions
 \bea
\widetilde{\mathcal{M}}_{irr}^{\pm }=\frac{\sqrt{r_{\pm }}}{\sqrt{2}},\quad
\widetilde{\mathcal{M}}_{irr}^{\pm}=\sqrt{\frac{1}{2}\mp\frac{1}{2} \sqrt{(r-1)^2}},
 \eea
 which are evaluated for $ a=a_{\pm}$,  respectively, see Figs\il(\ref{Fig:zongiaaransboc}).

We use  the concept of  maximum extractable energy $\xi$ introduced in Eqs\il(\ref{Eq:xtautau}), where  $M_{irr}=M-\xi$,   ($\xi$ has units of mass $M$) and
 $(M_{irr}/M)_{\min}=1-\xi_{\max}=1-\xi_{\ell}=1/\sqrt{2}$ for the extreme spacetime. We obtain
\bea&&\nonumber
\frac{M_{irr}}{M}=1-\xi^{\mp}_{\pm}=\mp\sqrt{\frac{r_{\mp}}{2}};\quad \frac{M_{irr(\nu)}}{M}\equiv1-\xi_\nu^{\mp}=\pm\sqrt{\frac{r}{2}}\quad\mbox{and}
\\&&\label{Eq:candd}
\frac{M_{irr}}{M}=1-\xi_{\tau}^{\mp}= \pm\frac{1}{\sqrt{4 \omega ^2+1}}= \pm\sqrt{\frac{r_g}{2}}=\pm\sqrt{\frac{a_g}{4\omega}}.
\eea

On the other hand, from the functions $\xi_\mu^{\mp}$ and $\xi_{\tau\tau}^{\mp}$ of Eqs.\il(\ref{Eq:xtautau}), which are  extensions of the rotational energy definition  in the extended plane, we obtain
\bea\label{Eq:xtautau}
&&
\frac{M_{irr(\mu)}}{M}=1-
\xi_\mu^{\mp}\equiv\pm\sqrt{1-\frac{r}{2}},\quad\mbox{and}\\&&\nonumber
\frac{M_{irr}}{M}=1-\xi_{\tau\tau}^{\mp}= \pm\frac{2 \omega }{\sqrt{4 \omega ^2+1}}= \pm2\omega\sqrt{\frac{r_g}{2}}=\pm\sqrt{2\omega a_g}.
\eea
Note that for $r=0$, according to Eqs\il(\ref{Eq:xtautau}), we find $M_{irr}/M=\{1,0\}$, with  $M_{irr}/M=1/\sqrt{2}$ for $r=M$, and   $M_{irr}/M=0$ for $r=2M$.
In Sec.\il(\ref{Sec:furth-mirr}), we analyze further aspects of the irreducible mass.

To enlighten the  properties of the mass  $M_{irr}$  in the extended plane, we can consider some  limiting values according to the \MB s structures.
From Eqs\il(\ref{Eq:candd}), we obtain
$M_{irr}/M=1/\sqrt{2}$ for the extreme Kerr spacetime. On the ergosurface and on the Schwarzschild horizon, we find   $M_{irr}/M=1$. For $\omega=+\infty$  (equivalently $r=0$), it is $M_{irr}/M=0$. From Eq.\il(\ref{Eq:Mirr-all}), we find
\bea
\label{Eq:o.r-childm}
&&\lim_{\la_0\to\mathbf{X}} M_{irr}(all)=\mathbf{Y}
\quad\mbox{and  }\quad
\mathbf{Y}=\left\{1,0,\frac{1}{\sqrt{5}},\frac{1}{\sqrt{2}},\frac{2}{\sqrt{5}}\right\},
\\\nonumber
&&\lim_{\la_0\to\mathbf{X}} M_{irr}(r_+)=\mathbf{Y}
\quad\mbox{and }\quad
\mathbf{Y}=\left\{1,1,\frac{2}{\sqrt{5}},\frac{1}{\sqrt{2}},\frac{2}{\sqrt{5}}\right\}, \\&&\nonumber\mbox{where respectively}\quad \mathbf{X}=\{+\infty,0,1,2,4\}.
\eea
Limits $M_{irr}={2}/{\sqrt{5}}$ or $M_{irr}={1}/{\sqrt{5}}$ refer to the definitions for inner and outer horizons for the tangent spin $a_g/M=4/5$.

\begin{figure*}
\centering
  \includegraphics[width=5.5cm]{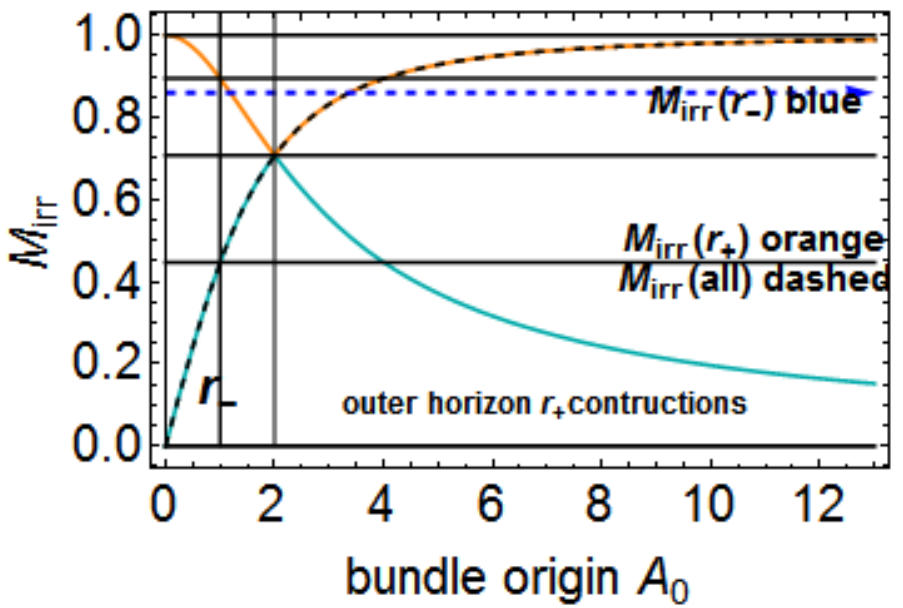}
  \includegraphics[width=5.5cm]{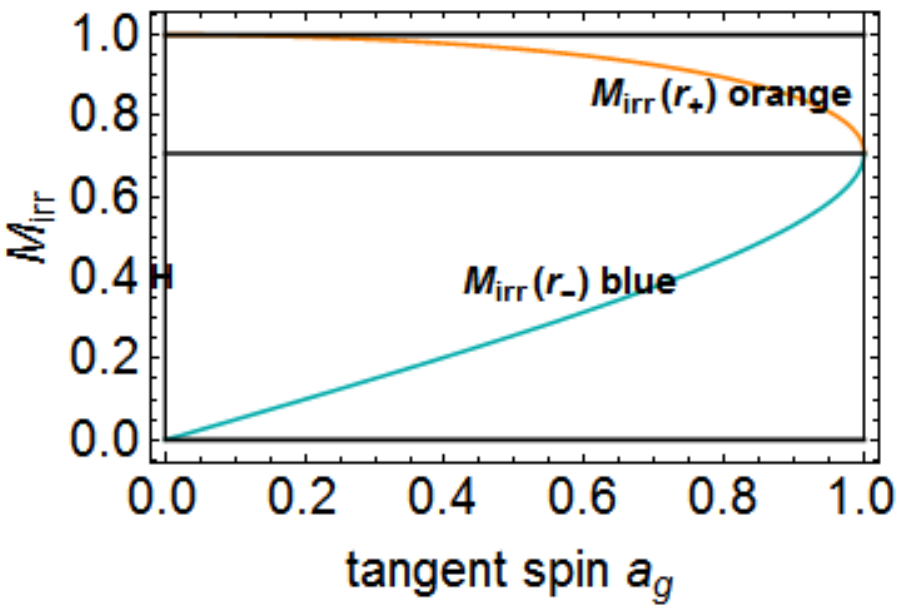}
  \includegraphics[width=5.5cm]{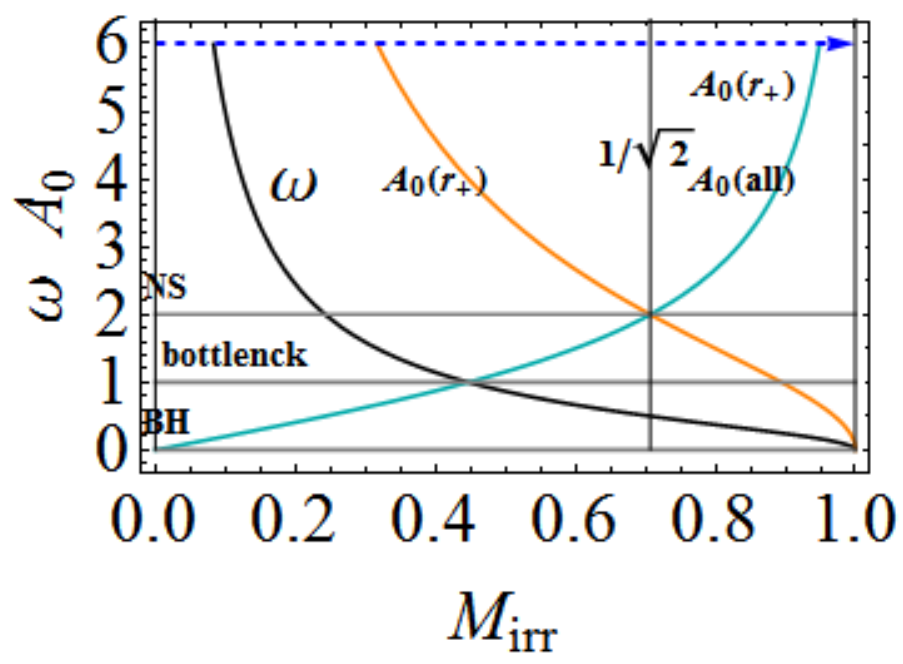}
  \caption{The irreducible mass $M_{irr}(a,r)$  in terms of  $\{r_{+}, r_-, r_g, a_g\}$, according to the analysis of   Eq.\il(\ref{Eq:o.r-childm}). $r_{\pm}$ are the \textbf{BH} horizons as functions of the \textbf{BH} spin,  $(r_g, a_g)$ are the tangent radius and tangent spin of the metric bundles, i.e., Killing horizons  as functions of the origin $\la_0=a_0\sqrt{\sigma}$ (left panel) or tangent spin $a_g$ (center panel). Right panel: Analysis of Eqs\il(\ref{Eq:bullionss}). The bundle origin spins $\la_0$ and frequency $\omega$ as functions of the irreducible mass. The \textbf{BH}, \textbf{NS} and bottleneck regions as well as the special value $1/\sqrt{2}$ are highlighted. }
  \label{Fig:bullionss}
\end{figure*}

To clarify  the relation between  bundle  \textbf{NS} origin and the \textbf{BH} irreducible mass  through the  tangency properties of the bundles, we invert some of the previous relations as  functions of    $M_{irr}$, obtaining the  bundle origin and the characteristic  frequency as the expressions
\bea&&\label{Eq:bullionss}
\la_0(all)= \frac{2 M_{irr}}{\sqrt{1-M_{irr}^2}},%\quad \la_0(r_+)= \frac{2 \sqrt{1-M_{irr}^2}}{M_{irr}},\quad\omega = \frac{\sqrt{1-M_{irr}^2}}{2 M_{irr}},&&
\\&&\mbox{therefore}\quad\omega=\frac{1}{\la_0(all)}={4\la_0(r_+)},
\eea
which are illustrated in Figs\il(\ref{Fig:bullionss}).
These  expressions relate the origin spin $\la_0$ to  the irreducible mass $M_{irr}$.

   Below we discuss this limiting value in relation of the \textbf{BH} transitions, revisiting the considerations of Sec.\il(\ref{Sec:sud-mirr-egi}).
 From the relation
   $M_{irr}^2={r_+}/{2} $, it follows that
  % \bea
%
%
\bea\label{Eq:mass-irr-rampi}&&
\frac{M^+_{irr}}{M}=%1-\frac{\xi}{M(0)}=
+\frac{\sqrt{1+\sqrt{1-\frac{J(0)^2}{M(0)^4}}}}{\sqrt{2}},%
\eea
and using Eqs\il(\ref{Eq:xis-nilo})
\bea
%&&
\frac{M^\flat_{irr}}{M}=%1-\xi_s^{(\pm)}=
\flat\frac{\sqrt{1\natural\sqrt{1-(a_s^{(\pm)})^2}}}{\sqrt{2}},\quad\mbox{where}\quad
\flat=\mp; \; \natural=\pm\ ,
\eea
where we use dimensionless units and $(\delta M-\delta J(0) \omega_H^+)\geq 0$. In particular,  we find $M_{irr}/M=\{0,1\}$ for
for $a=0$ and
 $M_{irr}/M=\pm1/\sqrt{2}$ for $a=M$ --see discussion in Sec.\il(\ref{Sec:furth-mirr}).

\medskip

\textbf{BH horizon areas in the extended plane}

The \textbf{BH} inertial mass  can be found from the  \textbf{BH} area:
\bea&&
A_{area}=\int\limits_{0}^{\pi}\int\limits_{0}^{2\pi} \sqrt{g_{\phi\phi}g_{\theta\theta}} d \theta d\phi=4 \pi  \left(a^2+r^2\right).
\eea
Here we use the extended definition $A_{area}(a_{\pm})=8\pi r$. Note that this expression depends explicitly on $r=r_g$.
Considering also the relations analyzed in Sec.\il(\ref{Sec:nil-base-egi}), we obtain
\bea&&\label{Eq:prop-time}
\delta M^{\mp}=\frac{\ell^{\mp}}{8\pi} \delta A_{area}^{\mp}+ \omega_H^{\mp} \delta J^{\mp},
\quad
da_{\pm}=\mp\frac{\ell(a_{\pm})}{\omega_\pm(a_{\pm})} dr,\\&&\nonumber(or \quad a_{\pm} \delta a_{\pm} =(M-r)\delta r ) \ .
\eea
The last relation can be written also as \\
\({\delta a=
[{\delta M r+\delta r (M-r)}]/{\sqrt{r (2 M-r)}}
}\).
In  Eq.\il(\ref{Eq:prop-time}),  we considered  the variation of the horizon area (and the  irreducible masses)  for the ADM mass $M$, the area $A_{area}$ and the momentum $J$. The horizons frequencies  $\omega_H^{\pm}$ are the  characteristic bundle frequencies.
 We consider other regions of the extended plane.
In Sec.\il(\ref{Sec:furth-mirr}), there are also further considerations on the transitions with $\delta M_{irr}=0$ and $\delta M_{irr}>0$.

From the expressions for $A_{area}(\la_0)$ and $A_{area}(\omega)$, we obtain the saddle points
 \bea\label{Eq:areassaddle-points}
 &&
 \mbox{for}\quad A_{area}(\la_0)={8 \pi r_g(\la_0)},\quad \mbox{it is }\\&&\nonumber
 \la_0=\frac{2}{\sqrt{3}} \quad\mbox{where}\quad A_{area}(\la_0)=4\pi,
 \\
 &&
  \mbox{for}\quad A_{area}(\omega)={8 \pi r_g(\omega)}\quad\mbox{it is }\\&&\nonumber
  \omega=\frac{1}{2\sqrt{3}}\quad\mbox{where}\quad  A_{area}=12\pi
 \eea
 Frequency $\omega=\omega_H^+={1}/{2 \sqrt{3}}$ is the outer horizon frequency for the \textbf{BH} with spin $a/M={\sqrt{3}}/{2}$, where $\omega_H^-={\sqrt{3}}/{2}=1/\mathcal{A}_0=2/\sqrt{3}$.
 For $a/M= {\sqrt{3}}/{2}$ there is $
r_+/M={3}/{2},r_-/M={1}/{2}$, in this geometry
there is $\partial_r \ell=0$ on $r_+$
 and $\partial_a \ell=0$ on  $r_-$ where $\ell=({r^2-a^2})/{\left(a^2+r^2\right)^2}$ is the acceleration, defined for $r>0$ equal the the \textbf{BHs} surfaces gravities when evaluated on the horizons.

{The saddle point $\la_0=2/\sqrt{3}$}   (\textbf{NS}) for the area     is also  the  saddle point for the angular momentum,   corresponding to the frequency  $\omega=1/(L_f)$  of the inner horizon  $r_-=1/2$  for the spacetime $a=\omega$, whose frequency of the outer horizon is a saddle point for the frequency function).
 It is easy to see that
 \bea&&\nonumber
 A_{area}^{\pm}\left(r_{\pm }\right)=\pm 8\pi r_{\pm } \quad\mbox{and}\quad \delta A_{area}^- =-\delta A_{area}^+ \\&&\mbox{i.e.}\quad
 \partial_a A_{area}^{+}\delta a =-\partial_a A_{area}^{-}\delta a\eea
which follows from relation $\delta r_ += -\delta r_ -$, where
 $r_ + r_ -= a^2$ and $ r_++ r_ -= 2M$ and, therefore, it holds if $\delta M=0$.

 Analogously, we obtain
\bea&&\nonumber
A_ {area} (a_ {\pm}) = 8\pi r,\quad\mbox{and}\quad
\ell(r_\flat,a_{\pm})=\frac{1}{2+\flat\frac{2}{\sqrt{(r-1)^2}}}
,\\&&
(\flat=\pm),\quad
\omega_H^{\mp}(a_{\pm})=\frac{\sqrt{r(2-r) }}{2\mp 2 \sqrt{(r-1)^2}}\ .\eea
The total  frequencies in terms of the radius $r/M$   is
 \bea&&\nonumber
 \omega=\frac{\sqrt{2-r}}{2 \sqrt{r}},\quad
 \ell(r,a_\pm)=\frac{r-1}{2 r},\quad\mbox{and there is }\\&&\ell=\omega_{\mathbf{EBH}}^2-\omega ^2,\quad
 %\ell(a,r_{\mp})=\frac{a^2-r_{\pm }}{2 a^2}
 \eea
 or, equivalently, $\omega= \frac{1}{2} \sqrt{1-4\ell}$, that is, $\omega_{\mathbf{EBH}}$ is the frequency of the extreme \textbf{BH} --
 see Eq.\il(\ref{Eq:inse-elem-nov-contro}).

\medskip

\textbf{Notes on  \textbf{BH} thermodynamics}

\medskip

We can now  analyze  \textbf{BH} thermodynamics in the extended plane,  considering  the mass variations and the interpretation of Eq.\il(\ref{Eq:prop-time}) for the inner horizon curve $(-)$.
However, in  the extended plane,  quantities relevant to  \textbf{BH} transitions may be different, when referring to  a point   $r$ on the horizon curve for a \textbf{BH} state.
(For simplicity, in some of the expressions  we do not consider the factor  $8\pi$. Moreover,  we will consider in some expressions
  spin, radius, characteristic frequency and origin spin as dimensionless parameters).
  In these relations, we express  the area (or the irreducible mass) in terms of the radius in the extended plane,  considering  dimensionless parameters.

Then,
\bea&&\label{Eq:minacc-faraoana}
\delta M=\frac{\delta A_{area} (r-1)}{2 r}+\frac{\delta J \sqrt{2-r}}{2 \sqrt{r}},\quad\mbox{and}\\&&\nonumber \delta M=\delta A_{area} \left(\frac{1}{4}-\omega ^2\right)+\delta J \omega,\quad  \delta M=\left(\frac{1}{4}-\frac{1}{\la_0^2}\right) \delta A_{area}+\frac{\delta J}{\la_0}.
\eea
This the  the mass variation is constrained by the  \MB  tangent point  $(r=r_g)$ or characteristic null frequency $\omega$ (origin $\la_0$) regulating therefore the \textbf{BHs} transitions.
Where $ r=r_g\in [M,2M]$,
when evaluated on the outer curve. In the extended plane the relations are satisfied  also for the inner horizons terms  and we use again this property in  Sec.\il(\ref{Sec:non-iner-extr-surf})  where we will study the doubled metric for the inner and outer  horizon by writing two different metrics.

In Sec.(\ref{Sec:nil-base-egi}), we discussed the relation between quantities defined on inner and  outer horizon points in the extended plane. Here, we consider this  relation in terms of surface gravity and frequencies.
 From   Eq.\il(\ref{Eq:narmer-reg}),  it follows that    $(\omega_+^-)^{-1}=\omega_-^+= \omega_H^+/\omega_H^ -= -\ell_H^+/\ell_H^ -= s (a)$.
Also,
   $A_{area}^-= -A_{area}^++ 4 M^2$; thus,
we can write   $\delta A_{area}^- = -\delta A_{area}^+ + 8 M \delta M$,
  and   $\delta A_{area}^+=-\delta A_{area}^-$ for   $\delta M = 0$, where
i.f  $\delta J=0 $ then
$	\delta M={\delta A_{area}^+  \ell_H^+}/{2 M}$%4 \delta M  \ell_H^--\frac{\delta A_{area}^+  \ell_H^-}{2 M}.

\begin{figure}
\centering
  \includegraphics[width=\columnwidth]{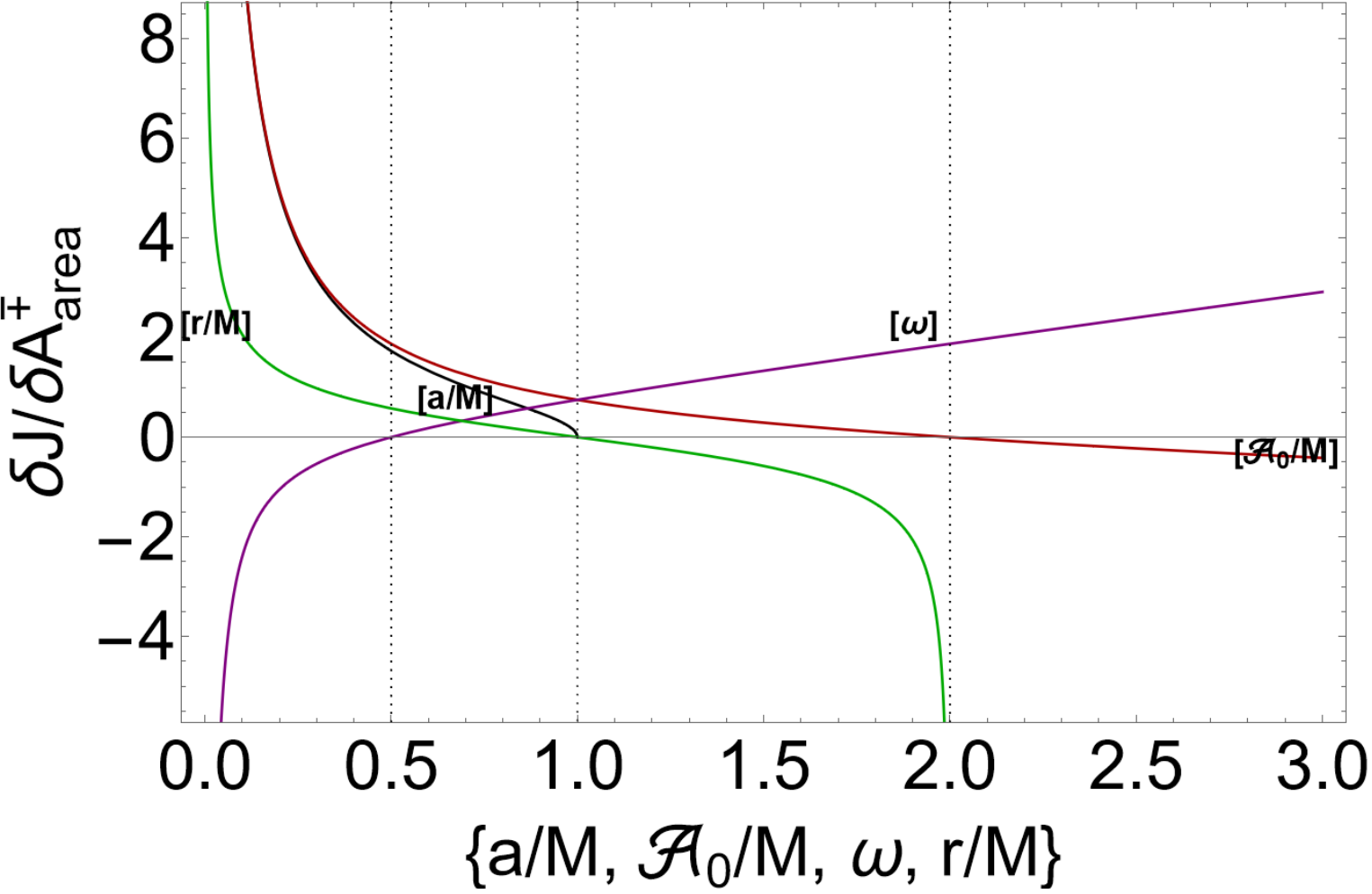}
  \caption{Variation
${\delta J}/{\delta A_{area}^{\mp}}$ of Eq.\il(\ref{Eq:variation-null}), as functions of the tangent spin $a=a_g\in[0,M]$, the origin spin $\mathcal{A}_0\in [0,+\infty]$, the characteristic \textbf{BH} horizons frequencies $\omega$, the tangent radius $r=r_g\in [0,2M]$.  The \textbf{BH} area is $A_{area}$ ($A^\pm_{area}$ for $r=r_\pm$ \textbf{BH} horizons), $J$ is the \textbf{BH} spin.}\label{Fig:Plotexplimit}
\end{figure}

A more precise analysis is provided in Sec.\il(\ref{Sec:alc-cases}). Here we consider  the  condition $\delta A^+_{area}=-\delta A_{area}^-$  (i.e. $\delta M=0$, invariant masses). Then,
\bea&&
0=\delta M=\frac{1}{2} \delta A_{area}^+ (\ell^+-\ell^-)+\frac{1}{2} \delta J (\omega_H^-+\omega_H^+),\\&&\nonumber
\mbox{and}\quad\frac{\delta J}{\delta A_{area}^+} = \frac{\ell^-+\ell^+}{\omega_H^--\omega_H^+}
\eea
From the definition of horizon in the extended plane, we find
$
\delta M=	(\omega
\delta J+\ell \delta r)/{(3 M-r)},
$
with the distinguished value  $r=3M$, which is related to the Schwarzschild  case \cite{bundle-EPJC-complete}.
Considering  $\delta M=0$, we obtain
\bea\label{Eq:variation-null}&&
\frac{\delta J}{\delta A_{area}^{\mp}}=(\pm)\frac{\sqrt{1-a^2} }{a}=
\frac{1}{\la_0}-\frac{\la_0}{4}\\
&&\nonumber=\omega-\frac{1}{4 \omega}=\frac{1-r}{\sqrt{r(2-r)}}
\eea
--see Figs\il(\ref{Fig:Plotexplimit})
In this way, we can   consider the variation of masses, areas and angular momentum of the \textbf{BHs}   in terms of the  quantities defined on the  horizon curves, for instance,
the first law of \textbf{BH} thermodynamics in terms of
the tangent point $r$ on the horizons, the origin $\la_0$ and the frequency $\omega$, describing the outer  horizon  $r\in [M,2M]$, $\la_0\in[2M,+\infty]$, $\omega\in[0,1/2]$ or the inner horizon for $r\in[0,M]$, $\la\in[0,2M]$, $\omega\in[1/2,\infty]$, respectively.
Such relations express the  variations  in terms of the angular momentum of the horizons,  corresponding   in some cases  to    naked singularities.
We can write Eq\il(\ref{Eq:minacc-faraoana}) in terms of quantities relative to the extreme $(ext)$ Kerr  \textbf{BH},  considered as reference
with $r_{ext}=M$, $\omega=1/2$ and $\la_0=2M$, as
\bea\nonumber
&&\delta M=\frac{1}{2 r}\delta \tilde{M}\quad\mbox{where}\quad\delta \tilde{M}\equiv\delta A_{area} (r-r_{ext})+\delta J a_{\pm}(r),
\\
&& \delta M=\frac{1}{(L_f)^2_{extr}}\delta\tilde{M}=\left[\frac{1}{(L_f)^2_{extr}}-\frac{1}{(L_f)^2}\right] \delta A_{area}+\frac{\delta J}{(L_f)},
%&
\eea
where $(L_f)=1/\omega$,  BH angular momentum, coincident with the bundle origin.
The case of extreme Kerr \textbf{BHs} in this sense is a limit where
$\delta M=\delta J/2$.
\begin{figure*}
\centering
  \includegraphics[width=5.6cm]{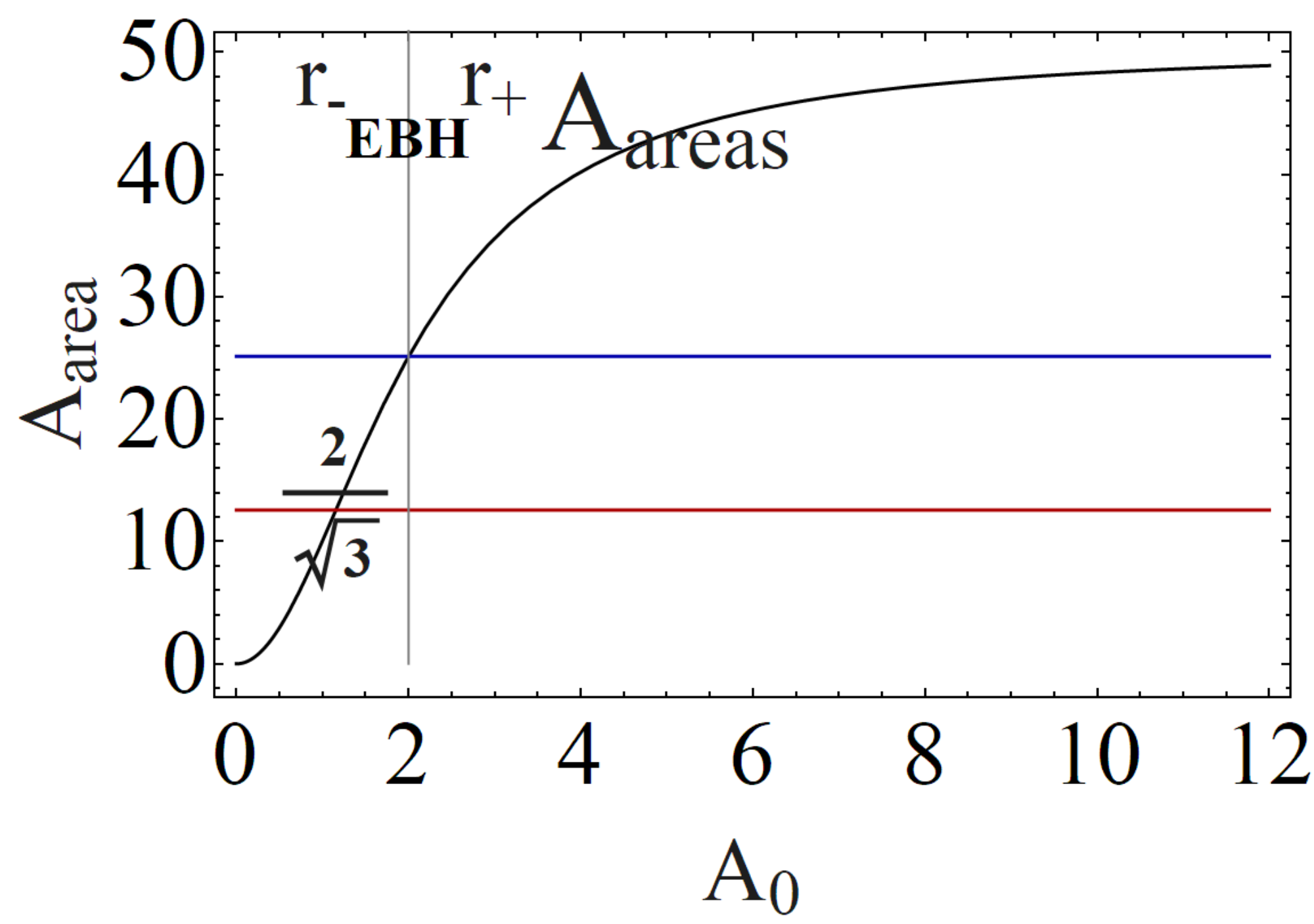}
   \includegraphics[width=5.6cm]{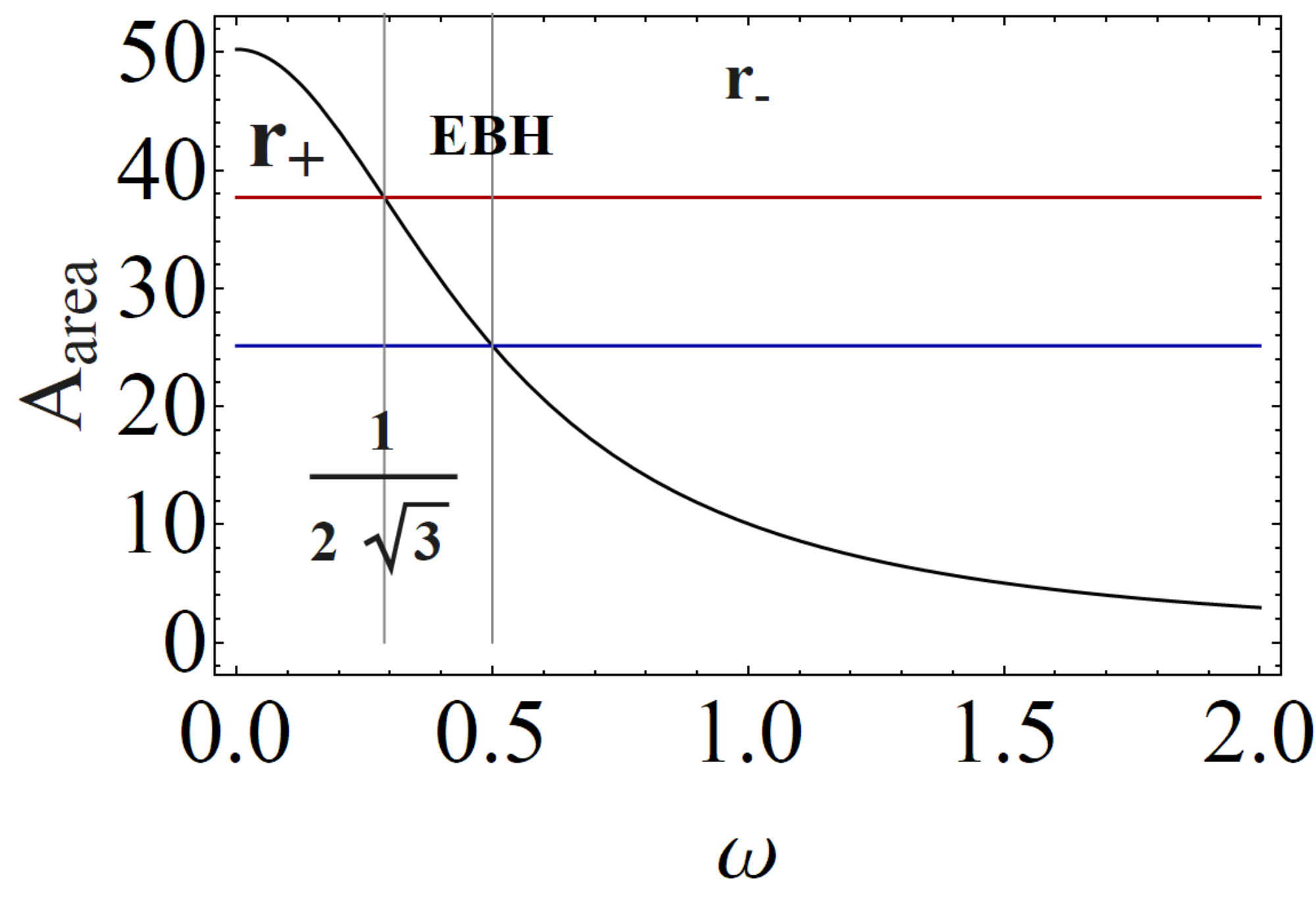}
    \includegraphics[width=5.6cm]{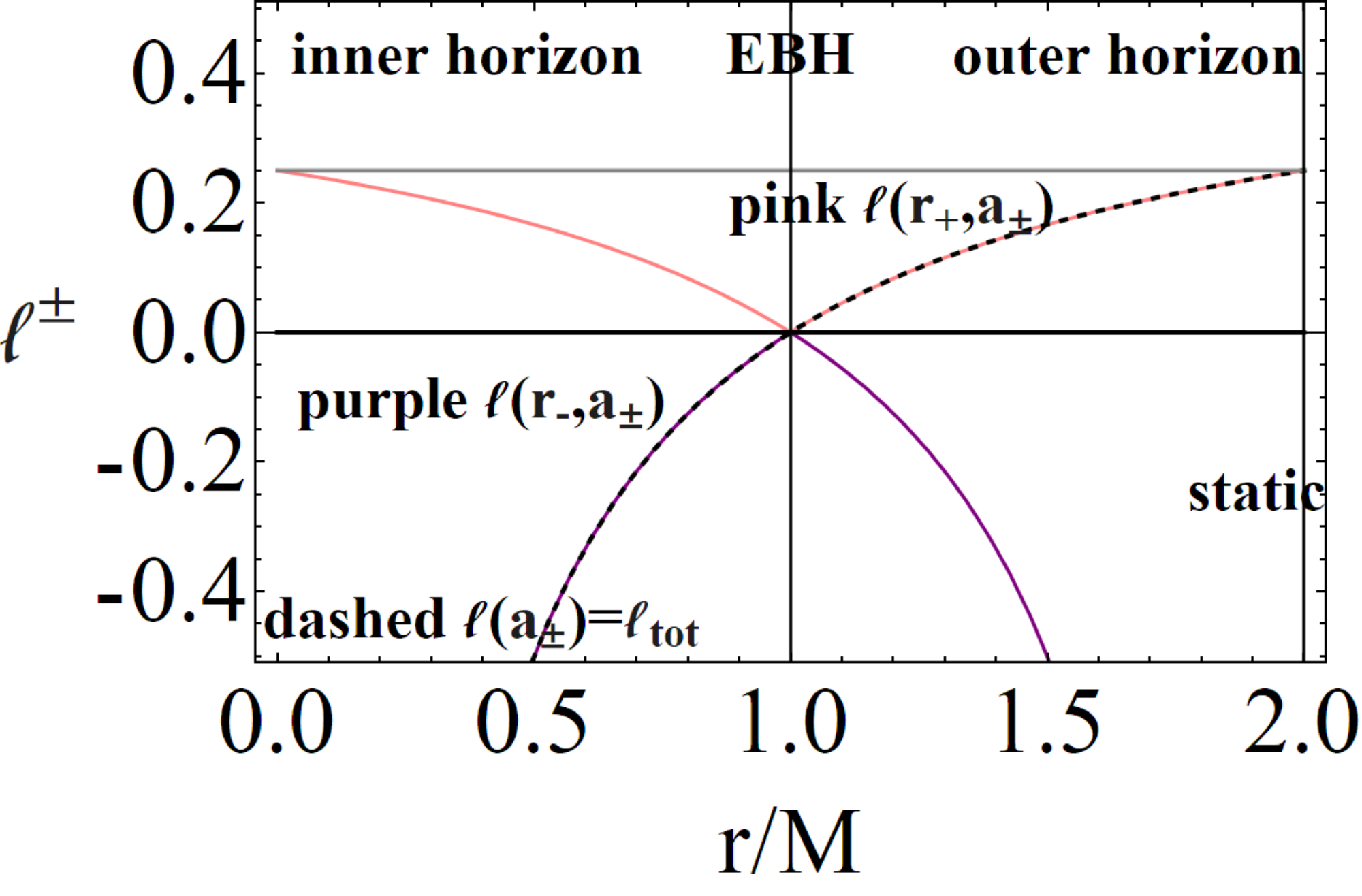}
       \includegraphics[width=5.6cm]{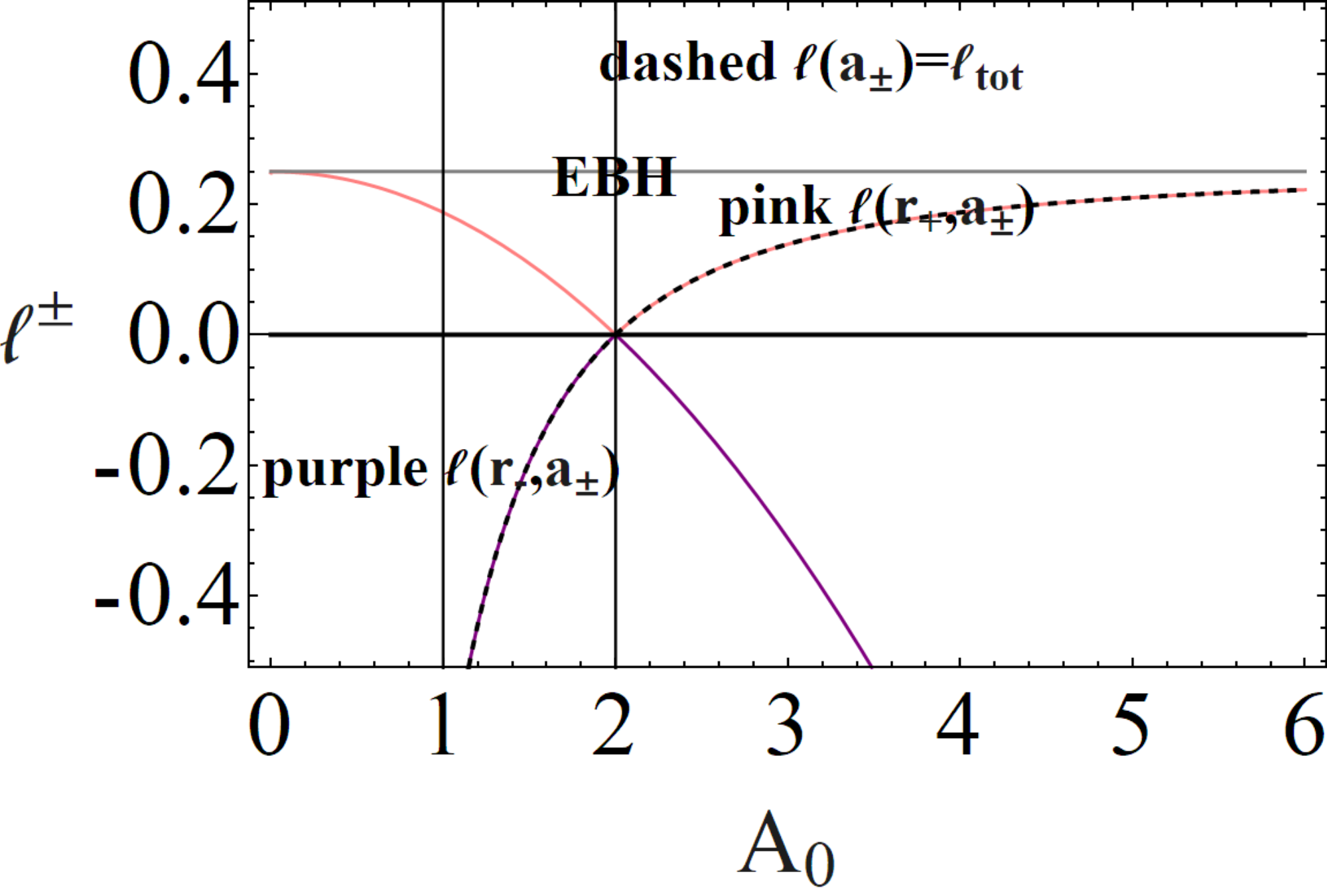}
      \includegraphics[width=5.6cm]{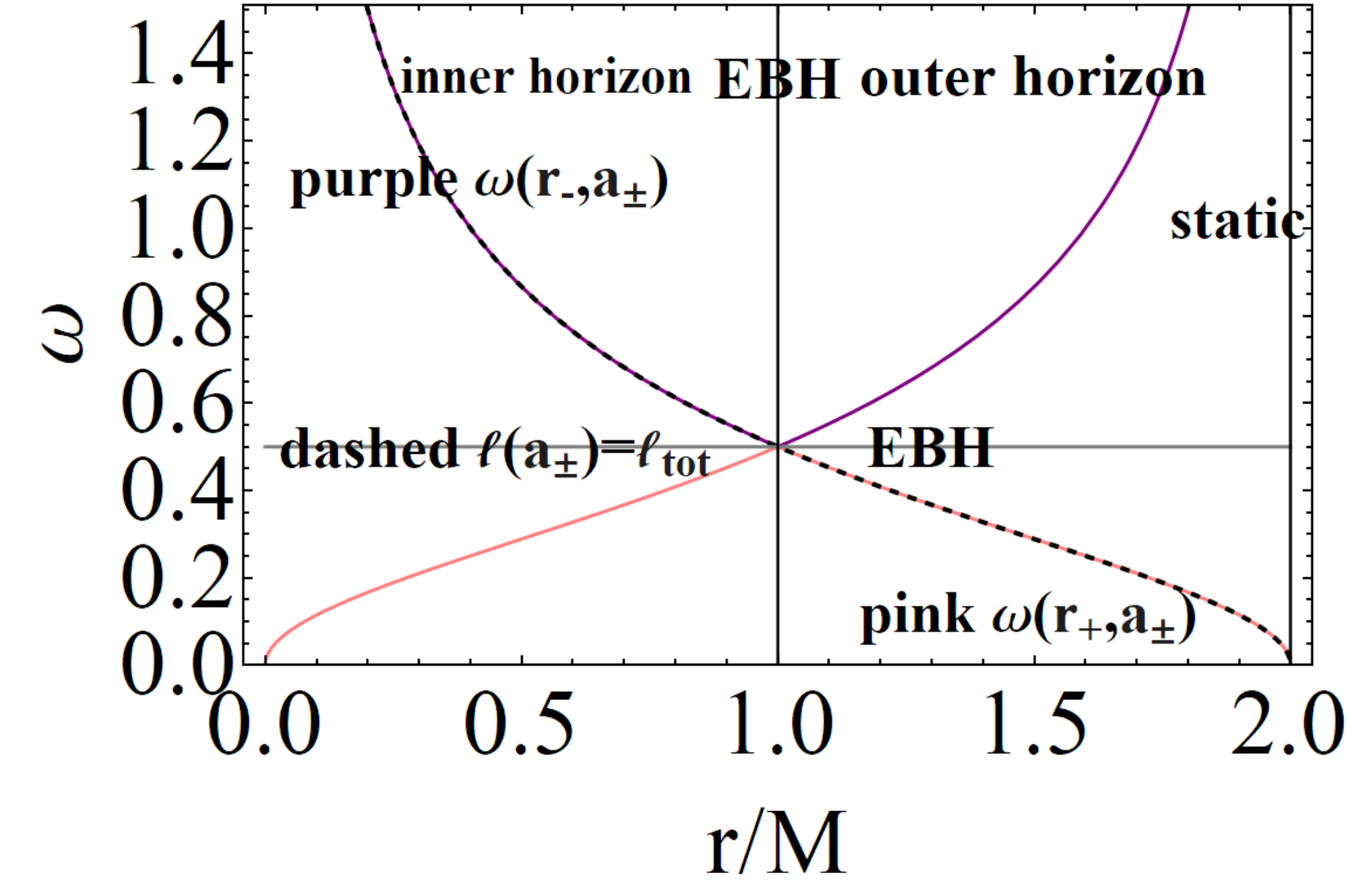}
    \includegraphics[width=5.6cm]{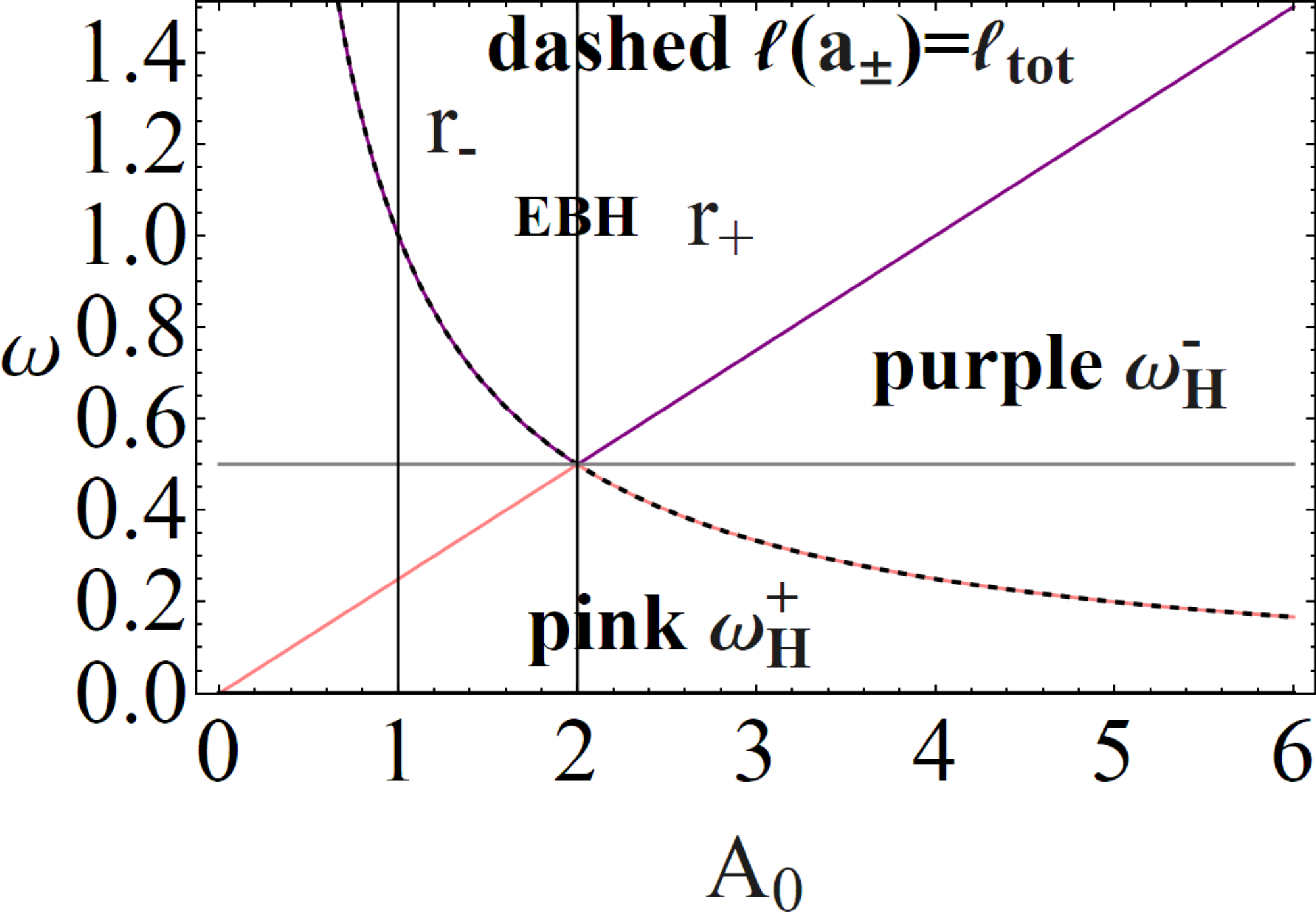}
  \caption{The \textbf{BH} area $A_{area}$ ($A^+_{area}$ for $r=r_+$) as function of the bundle origin spin $\la_0\equiv a_0\sqrt{\sigma}$, where $\sigma\equiv \sin^2\theta$, and $\omega$ is the bundle characteristic frequency and the horizon frequency. \textbf{EBH} denotes the extreme Kerr \textbf{BH}. The inner $r_-$ and outer horizons  $r_+$ in the extended plane are also shown.
  The points $\la_0=2/\sqrt{3}$ and $\omega=1/(2\sqrt{3})$ are saddle points of the curves corresponding to inner and outer horizons, respectively. The accelerations $\ell^{\pm}$ are evaluated on the horizon curves $r_{\pm}$, where $\ell^+$ is the \textbf{BH} surface gravity. They are represented  as functions of $r/M$ on the extended plane and as functions of the bundle origin. The functions $\ell(a_{\pm})$ represent the accelerations on the horizon curve $a_{\pm}$ in the extended plane.
  Center and right panel of bottom line:  The characteristic bundle frequency and horizon frequency $\omega$ in the extended plane as function of $r/M\in[0,2]$. Here, $r_{\epsilon}^+\equiv 2M$ represents the static limit for $\sigma=1$ and the static Schwarzschild spacetime.  $\omega_H^{\pm}$ are the horizon frequencies.}\label{Fig:zongiaarano}
\end{figure*}
The figures(\ref{Fig:zongiaarano}) and
(\ref{Fig:zongiaaransboc}) outline the four zones identified  through the inertial mass as functions of the spin,  $\la_0$, of the characteristic frequency of the bundles.

\section{On BH transitions}\label{Sec:furth-mirr}

\subsection{Relating inner and outer horizons in  BH transitions}
\label{Sec:comp-inner-outer}

\begin{figure*}
\centering
     \includegraphics[width=4.14cm]{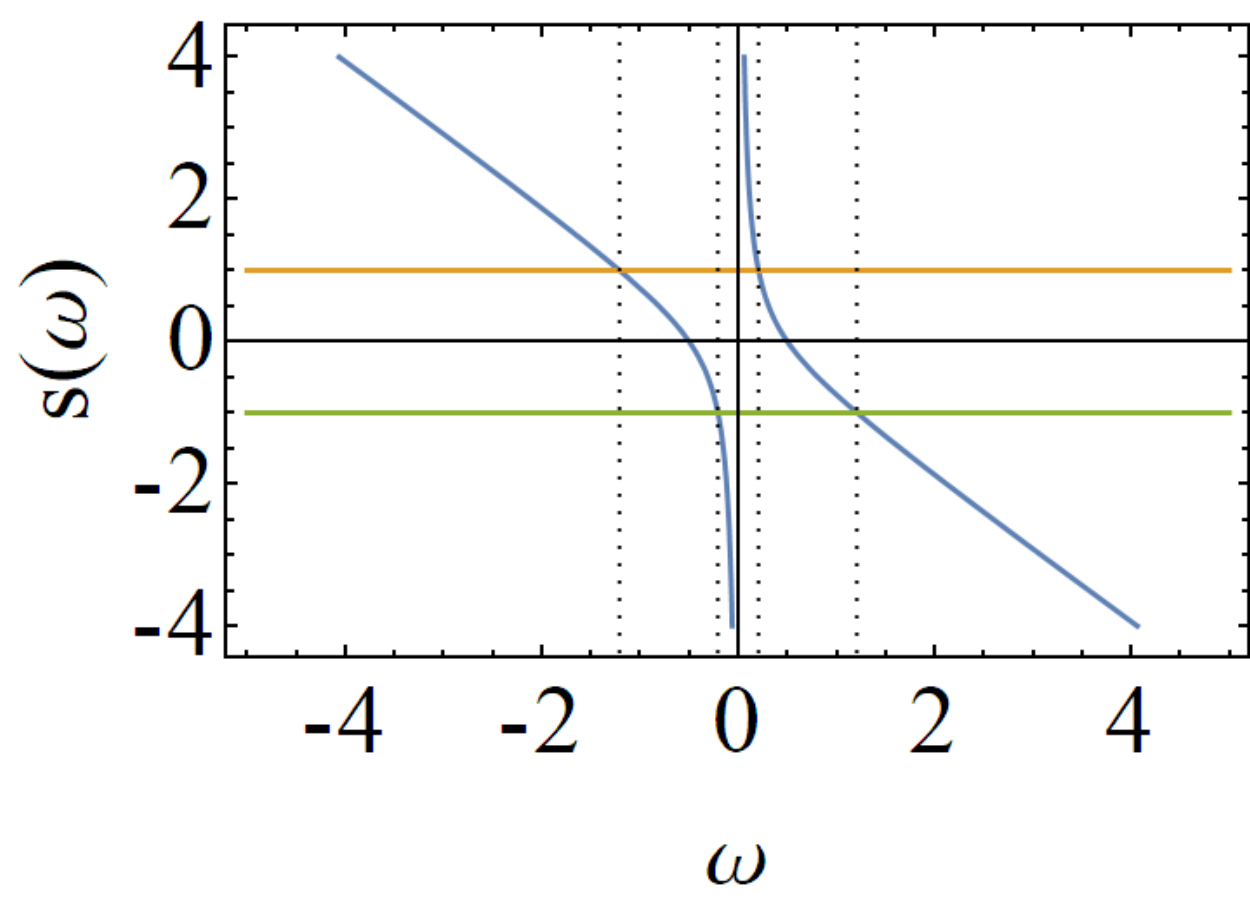}
           \includegraphics[width=4.14cm]{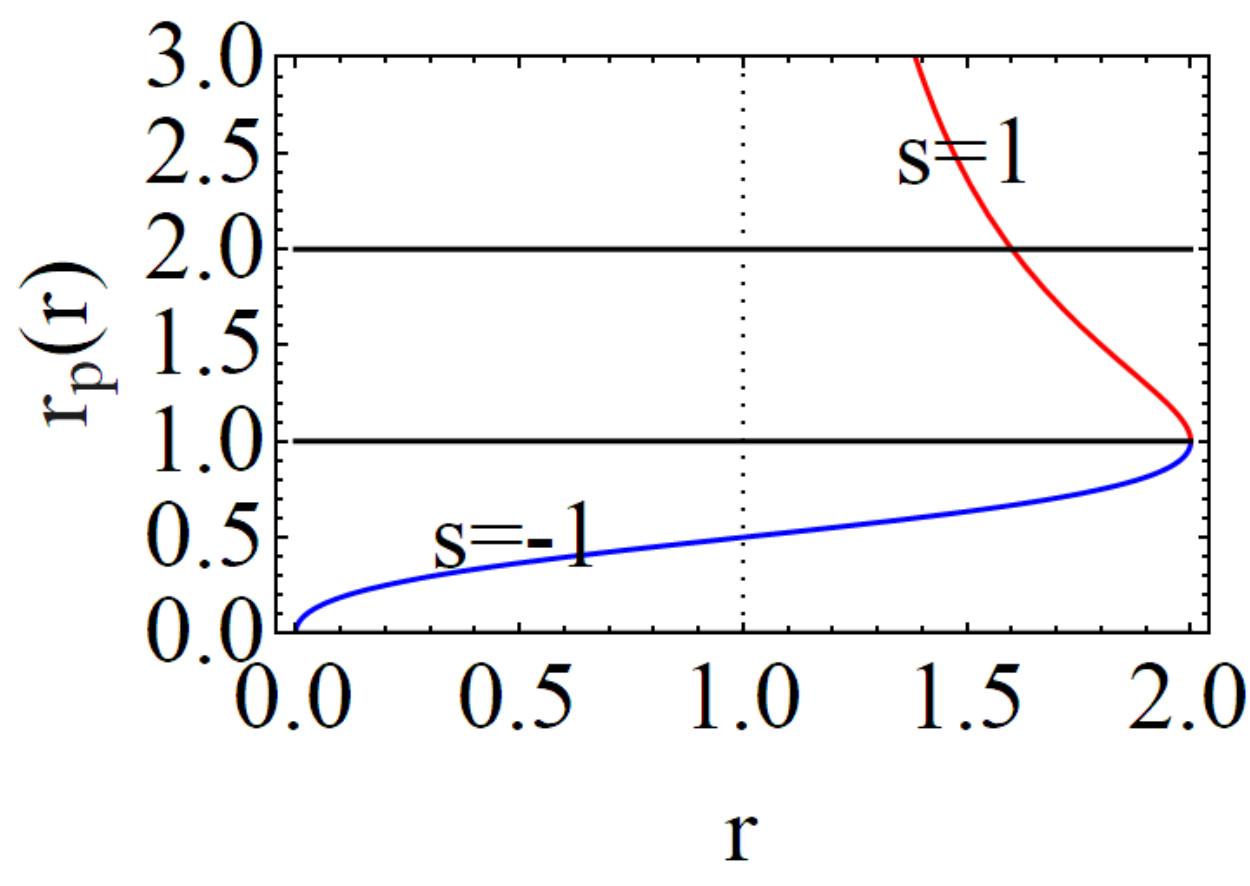}
            \includegraphics[width=4.14cm]{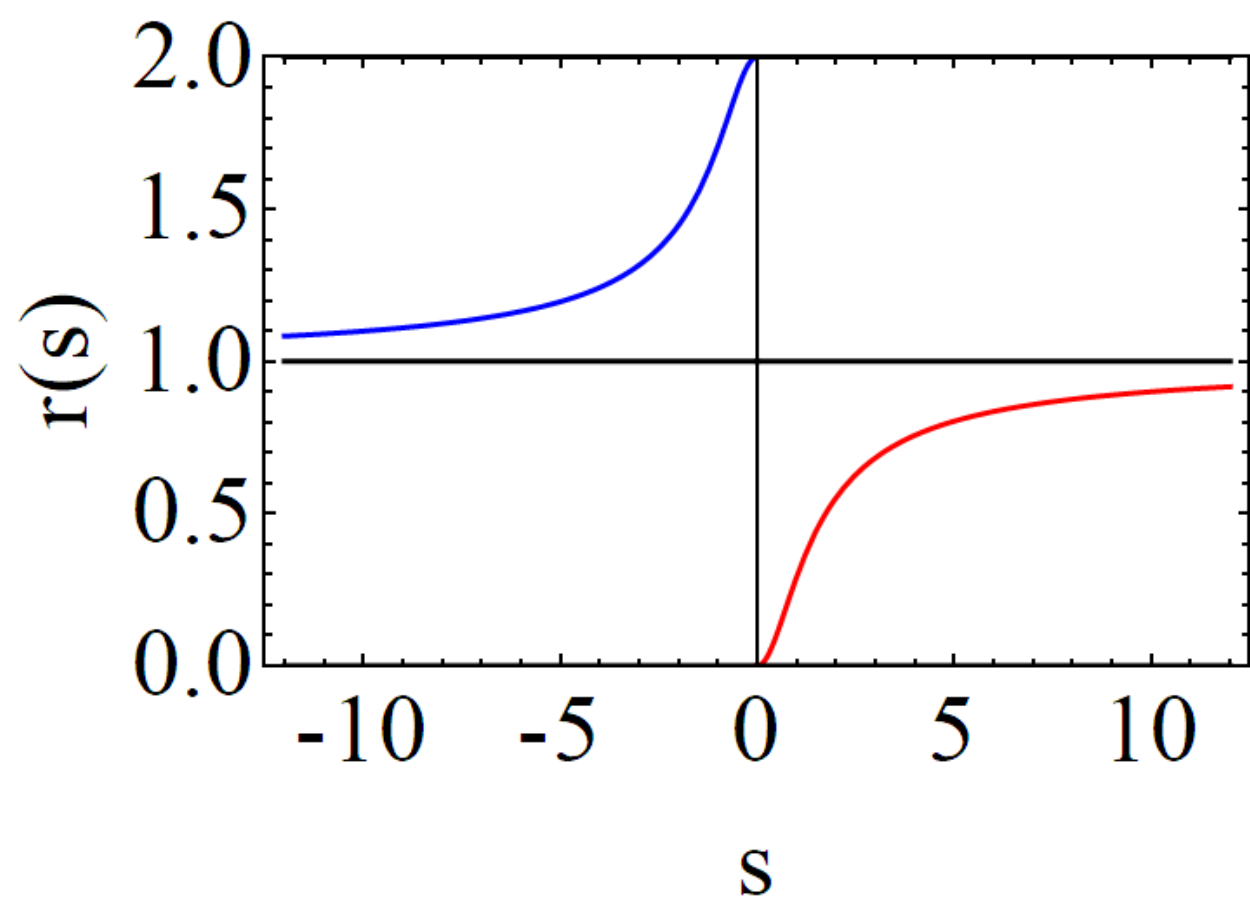}
   \includegraphics[width=4.14cm]{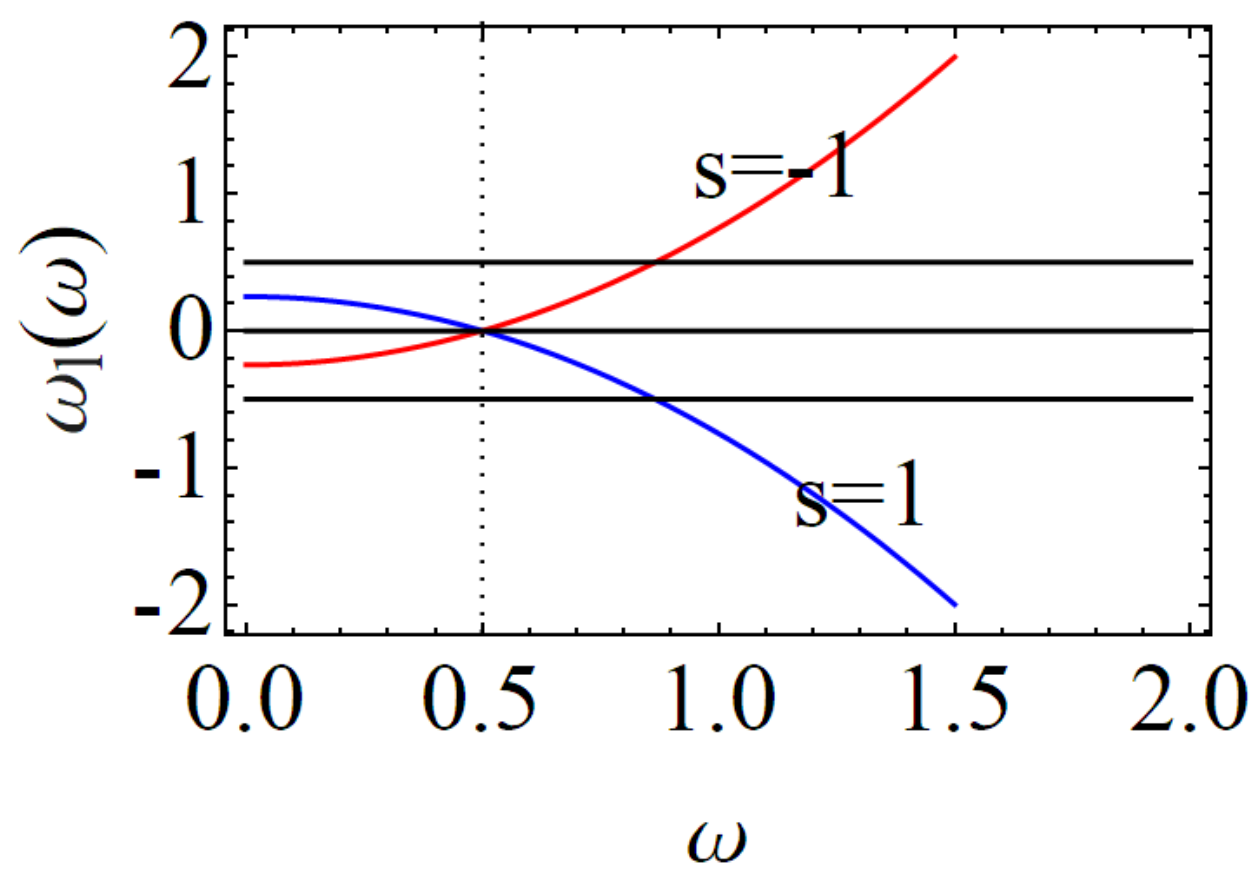}
  \caption{Left panel: The function $s(\omega)$ introduced in Sec.\il(\ref{Sec:comp-inner-outer}), solution of  $\ell_H^{\pm} (\omega) =
 s\omega$, is shown. Horizontal lines are for $s=\pm1$. The function  $\ell$ is the acceleration on the \textbf{BH} horizons, where $\ell_H^+$ is the \textbf{BH} surface gravity, $\omega$ is the \textbf{BH} angular frequency and bundle characteristic frequency. Second  panel:
 the functions $r_p(r)$ for $s=\pm1$: $\omega (r) = s \ell (r)$ are defined in Eqs\il(\ref{Eq:spe-o-t-su1}) and  Eqs\il(\ref{Eq:spe-o-t-su2}). Third panel: The function $r=r_p(s)$ defined Sec.\il(\ref{Sec:comp-inner-outer}), solution of $\omega(r)=s \ell(r)$, is shown. The value $r=M$ corresponds to the extreme Kerr \textbf{BH}. Right panel: The function $\omega_l(\omega)$ defined in Sec.\il(\ref{Sec:comp-inner-outer}), solution of $\ell(\omega)=s\omega_l$ for $s=\pm1$, is shown. $\omega=\pm1/2$ corresponds to an extreme Kerr \textbf{BH}. }\label{Fig:pltkepsimpl}
\end{figure*}

Considering the discussion of   Sec.\il(\ref{Sec:nil-base-egi}),
we use the expression  $\ell_H^{\pm} (\omega)$  in the extended plane
and solve the equation  $\ell_H^{\pm} (\omega) =
 s\omega$ for a given frequency $\omega$.
 Analyzing  this special \textbf{BH},  transition  we look for a solution
$\ell$  in a fixed spacetime and   obtain
   $ s = ({1-4\omega^2})/{4\omega} $, which is represented in Figs\il(\ref{Fig:pltkepsimpl})  for any $\omega$.
  The condition $s < 0 $ implies that  $\ell$ is evaluated on the inner horizon (we assume $\omega>0$) and,  therefore,   $\omega^2 >
 1/4 $, bounded by the limiting value of the extreme Kerr spacetime. Then,
$\ell =1/4 - \omega^2 $. However, we are interested in the cases
$s=$constant: $(\ell, \omega) =
   \omega\left[({1-4\omega^2})/{4\omega},1\right]$.
For $s = \pm1$ it is  $\omega =
\pm\left (1\pm\sqrt {2} \right)/2$  corresponding to $a_{\gamma}/M =
   1/\sqrt {2}$.

Here we solve the more general problem where the frequencies are not equal, that is, where  quantities $(\ell,\omega)$ and $\omega_l$ are related by $\ell(\omega)=s \omega_l$, with $\omega\neq \omega_l$, and obtain
$\omega_l\neq0$ and  $\omega=\pm 1/2$ (extreme Kerr \textbf{BH}) or  $\omega_l\neq0$ and   $s=({1-4 \omega^2})/{4 \omega_l}$;
for  $s=\pm1$ we obtain  $\omega_l=\pm \left(1-4 \omega^2\right)/4$, which is represented in  Figs\il(\ref{Fig:pltkepsimpl}).
In this way  we connect \textbf{BHs}, as initial and final state of these particular transformations.

We now consider the relation  between frequencies and surface gravity as a function of a radius $ r$  of the horizon curve in the extended plane (that is, in this way  we express the conditions for the inner and outer horizons relations), where  $\omega (r) = s \ell (r)$ at the same point $r$ ($r=r_g$, tangent point on the horizon curve). We obtain $r=1\mp\sqrt{{1}/({s^2+1})}$ for $s<0$  and $s\geq0$, respectively. This  relates  a \textbf{BH} inner and outer horizon (at fixed $a$).
For  $s=\pm1$ we obtain   $r=r_p=r_\pm=1\pm{1}/{\sqrt{2}}$, outer and inner horizons, respectively, for  the \textbf{BH} with spin $a_{\gamma}/M=1/\sqrt{2}$.

Consider the more general case where $r\neq r_p$.

{There is: $ \omega(r)=s \ell(r_p)$  for
\bea&&\label{Eq:offtamp}
(r=2M, s=0, r_p>0),\quad\mbox{and}\\&&\nonumber \left(r=\frac{2 r_p^2}{r_p^2+(r_p-M)^2 s^2}, [(r\in]0,M], s<0), (s>0, r_p\geq M)]\right)
\eea
alternatively, defining $\bar{s}_\beta\equiv \sqrt{\frac{2M-r}{r}} (r_p/({r_p-M})$ we can write the results as follows
\bea&&
s>0:\quad (r=2M, r_p=M, s>0), \\&&\nonumber \mbox{and}\quad\left(r_p>M, r\in]0,2M[, s=\bar{s}_\beta\right)
\\&&
s \leq 0: \quad
(r=2M, r_p=M, s\leq 0), \\&&\nonumber\left[((r_p\in]0,M[, r\in]0,2M]), (r_p>M, r=2M)), s=\bar{s}_\beta\right].
\eea
In particular,
for the static case there is   $(r=2M, s=0, r_p>0)$.}
For the special cases $s=\pm1$ we obtain
\bea&&\label{Eq:spe-o-t-su1}
\mbox{for}  \;  s=1 \quad  \mbox{it is } \quad  r\in ]M,2M]\, \mbox{and} \; r_p=\frac{M}{M-\frac{\sqrt{2M-r}}{\sqrt{r}}},
\\
&&\label{Eq:spe-o-t-su2}
\mbox{for}  \;   s=-1 \quad \mbox{it is}\quad  r\in ]0,2M] \; \mbox{and}\;  r_p=\frac{M}{M+\frac{\sqrt{2M-r}}{\sqrt{r}}}.
\eea
 represented in  Figs\il(\ref{Fig:pltkepsimpl}), where the relation between the \textbf{BH} horizons
(where the range $r=r_g$ has been extended)  connect  couple of geometries as final and initial states of these special transitions.
\subsection{Constant irreducible mass}\label{Sec:constant-Mirr}
A more complicated transition occurs when  the final state is not a  Schwarzschild static \textbf{BH},  implying  $J(1)=0$, where $(1)$ is for final state of transition and $(0)$ for the initial state.
We start by noting that the condition $M_{irr}=$constant and  $M(1)=M(0)=M$  implies  $J(1)=J(0)=J$, which can also be interpreted as   $J(1)^2=J(0)^2=J^2$ and  consider the case of  change in rotation orientation, implying a state to be considered  a static \textbf{BH}, that is, with a state  $J=0$  with a consequent increase of mass,
(for  $\delta M_{irr}\geq0$).

Introducing the ratios $(k_m,k_j)$, with
$J_1= k_j J_0 $ and $M_1= k_m M_0$ (we use notation $\Qa(0)=\Qa_0$ or $\Qa(1)=\Qa_1$), the condition
 $\delta M_{irr}=0$ is verified for
\bea&&\nonumber
k^{\pm}_j=(\pm)\sqrt{\frac{2 \left(k_m^2-1\right) M_0^4}{J_0^2}\pm2 \sqrt{\frac{\left(k_m^2-1\right)^2 M_0^4 \left(M_0^4-J_0^2\right)}{J_0^4}}+1},
\eea
(we considere with the sign $(\pm)$ the possibility of $J_1 J_0<0$). Eventually, it can also be  written in compact form as
\bea k
=\pm
\frac{\sqrt{2 \left(k_m^2-1\right) M_0 r_+(0)+a(0)^2}}{a(0)}\ ,
\eea
where  $r_+(0)$  is the outer horizon radius at state $(0)$. Therefore, $k_j=\pm1$ for $k_m=1$ (unaltered  masses).
Assuming now  $k_j \equiv  k_m k_{jm}$ (with  $k_m>0$), the condition $\delta M_{irr}=0$ implies
\bea&&\nonumber
k_{jm}=\pm\frac{\sqrt{\frac{2 \left(k_m^2-1\right) M_0 r_+(0)+a(0)^2}{k_m^2}}}{a(0)} \quad\mbox{where}\\&& k_{jm}\in k_m \left[-\sqrt{\frac{M(0)^2}{a(0)^2}},\sqrt{\frac{M(0)^2}{a(0)^2}}\right].
\eea
%
% then if  $k_m=1$  there is $k_{jm}=\pm
Then, there are
%\bea&&
%there is
\bea&&
\left(k_m=\frac{1}{\sqrt{2}},\quad
 k_{jm}=0\right),\\&&\nonumber \left(k_m\in\left]\frac{1}{\sqrt{2}},1\right[,\quad k_{jm}\pm \sqrt{2-\frac{1}{k_m^2}}\right),
\eea
where it is described a total or partial energy extraction respectively.

Let us introduce the quantities
\bea&&\nonumber
\tilde{k}_\pm\equiv\sqrt{-\frac{2 M_0^2}{a(0)^2 k_m^2}\pm 2 \sqrt{\frac{\left(k_m^2-1\right)^2 M_0^2 \left(M_0^2-a(0)^2\right)}{a(0)^4 k_m^4}}+\frac{2 M_0^2}{a(0)^2}+\frac{1}{k_m^2}},
\\&&\mbox{and}\quad
c_a\equiv \sqrt{\frac{r_+(0)}{M_0}}
\eea
Note that $c_a/\sqrt{2}$ is related to the definition of irreducible mass--Sec.\il(\ref{Sec:cou.mass-irr}). For the  Schwarzschild case we obtain     $(a(0)=0, k_m=1)$. For $ a(0)\in ]0,M_0[$ we find
\bea&&\nonumber
(k_m=c_a/\sqrt{2}, k_{jm}=0),\quad \mbox{or}\quad  \left(k_m\in c_a]1/\sqrt{2},1],k_{jm}=\pm \tilde{k}_\pm\right),
\\\nonumber
&&\mbox{for}\quad a(0)=M_0,\quad \mbox{it is }\quad (k_m=c_a, k_{jm}=0),\\&&\nonumber\mbox{and}\quad\left(k_m\in]c_a,1],\quad k_{jm}=-\tilde{k}_\pm\right).
\eea
In particular,   the condition
$k_m = 1$  implies  $a(0) = 0$ or   $a(0)\in ]0,M]$   and   $k_ {jm} = \pm1
$.
For  $k_ {jm} =
 0$, that is, the final state of a Schwarzschild \textbf{BH}, or for     $a(0)\in [0,
    M_0]$ it is  $k_m =c_a/\sqrt{2}$, which for an extreme \textbf{BH}, $a(0)=M_0$,
     is $k_m={1}/{\sqrt {2}}$.
For $k_ {jm} =
 1$, it  is   $k_m = 1$, implying  $k_j=1$; in other words, the condition $k_j =
   k_m$ implies that the \textbf{BH} state is immutable.
   Analogously, $k_m = 1$
   for  $k_ {jm} = -1$.

Let us now introduce the  spins
{\small
\bea\label{Eq:j1jajpm}&&
J_a\equiv M_1\sqrt{2 M_0^2 -M_1^2},\quad
J_i\equiv2 M_1\sqrt{M_0^2-M_1^2}, \\&& \nonumber\tilde{J}_{\pm}\equiv \sqrt{J_0^2\pm 2 \left[\sqrt{\left(M_0^4-J_0^2\right) \left(M_0^2-M_1^2\right)^2}+M_0^2 (M_1^2-M_0^2)\right]}
\eea}
We obtain the following conditions for a transition $\delta M_{irr}=0$:
\bea&&
\mbox{for}\quad
M_0=\frac{M_1}{\sqrt{2}}: \quad  (J_0=0,  J_1=\pm M_1^2);
\\\nonumber&&M_0\in\left]\frac{M_1}{\sqrt{2}}, M_1\right[: \quad ( J_0\in ]0,J_a[, J_1=\pm \tilde{J}_{+}); \\&&\nonumber \left(J_0=J_a, J_1=\pm M_1^2\right)
\\
&&\nonumber\mbox{for}\quad M_0=M_1:\quad (J_0=0, J_1=0); \\&&\nonumber (J_0\in ]0,J_a[, J_1=\pm \tilde{J}_{-}); \\&&\nonumber \left(J_0=J_a, J_1=\pm M_1^2\right);
\\&&\nonumber
\mbox{for}\quad\frac{M_0}{M_1}\in \left]1,\sqrt{2}\right[:\quad  (J_0=J_i, J_1=0); \\&&\nonumber \left(J_0\in ]J_i, M_0^2], J_1=\pm  \tilde{J}_{-}\right);
\\
&&\mbox{for}\quad M_0=\sqrt{2} M_1:\quad  (J_0=M_0^2,  J_1=0).
\eea
Finally, we see also the special case:
\bea\label{Eq:camb-vibe}
&& \mbox{for}\quad \frac{M_0}{M_1}=\sqrt{\frac{3}{2}}: \quad (J_0=J_a, J_1=0); \\&&\nonumber \left(J_0\in]J_a, M_0^2], \quad J_1=\pm \tilde{J}_{-}\right),
\eea
(note the symmetries in the state $(0)\leftrightarrow (1)$).
In this case $J_a=J_i=\sqrt{8/9} M_0^2$ while  $\tilde{J}_{\pm}=\sqrt{J_0^2\pm \frac{2}{3} \left(\sqrt{M_0^8-J_0^2 M_0^4}-M_0^4\right)}$, and in this transition, $a=\sqrt{8/9} M_0$ is the \emph{initial} \textbf{BH} state.

If the transition ends  with a static \textbf{BH}, $J_1 =
   0$, then $J_0 =
    0$ and  $M_1 =
     M_0$, that is, there is no transition   with  $M_1=M_0/\sqrt {2}$ or  with  $J_0=M_0^2$, i.e., an extreme Kerr \textbf{BH} as initial state. Viceversa,  in the case $J_0\in ]0,M_0^2[$ we have
   $M_1 = \sqrt{r_+(0)}/\sqrt{2}$.
 If  $J_0 =
   0$, that is, the starting state  is a Schwarzschild \textbf{BH}, then  $J_ 1 =
    0$ with  $M_0 = M_ 1$. Therefore,  there is no transition for    $M_1\in] M_0, \sqrt {2} M_0]$
  with  $J_1 = 2 M_0\sqrt{M_1^2-M_0^2}$ (note the symmetry with solution  $J_i$ in Eq.\il(\ref{Eq:j1jajpm})). In the case,  $M_1=\sqrt{2}M_0$, we get $J_1=2 M_0$ implying an extreme Kerr spacetime ($J_1=M_1^2$).

In the extreme case, where  $M_1^2 =J_1$, that is, a transition ending in a Kerr extreme \textbf{BH}, for
$J_0\in [0,M_0[ $,   we obtain  $M_1 = \sqrt{r_+(0)}$. Then, for  $J_0 =
     M_0^2$, that is, starting from a Kerr extreme \textbf{BH}  and ending in Kerr extreme  \textbf{BH}, we get  $M_0=M_1$.
Note that the condition  $J_1/M_1^2 =
  J_0/M_0^2$, that is,   the \textbf{BH} dimensionless spin is  invariant,  implies that   $J_0=J_1$  and $ M_ 1 =
 M_ 0$, that is, there is no transition from $(J_0,M_0)$ with $M_{irr}=$constant and $a/M=$constant. This is particularly relevant for the case of Kerr extreme \textbf{BH}.
If the initial state is  an extreme Kerr \textbf{BH}, i.e.  $J_0=M_0^2$, the transformation can lead to a  Schwarzschild \textbf{BH}, i.e. $J_1=0$,  if
$M_1 = {M_0}/{\sqrt{2}} $; or can lead to a Kerr \textbf{BH} with
\bea\label{Eq:bat} \frac{M_1}{M_0}\in \left] \frac{1}{\sqrt {2}}, 1\right] \quad\mbox{and}\quad
J_1 =\pm M_0^2 \sqrt {\frac{2M_1^2}{M_0^2} - 1},
\eea
(note the symmetry with respect to the case $J_0=\{0,1\}$  or $J_1=\{0,1\}$considered above).
In the  particular case {$M_1/M_0=\sqrt{{2}/{3}}$}, we obtain
  \bea\label{Eq:brum-field}&&
  \frac{J_0}{M_0^2}=\frac{2\sqrt{2}}{3},\quad\mbox{and}\quad J_1=0,\quad\mbox{or}\quad
\frac{J_0}{M_0^2}\in\left] \frac{2
\sqrt{2}}{3}, 1\right],\\&&\nonumber\mbox{and}\quad J_1=\pm\sqrt{J_0^2-\frac{2M_0^2r_+(0)}{3}}.
\eea
\subsection{On the inner horizon relations in \textbf{BH} thermodynamics}\label{Sec:alc-cases}
Here and in the following analysis, we use the notation $M=M_\pm$ and $J=J_{\pm}$ to stress the quantities defined on a point of the outer or inner horizon, respectively.
From the definition  $M^{\mp}_{irr}\equiv A^{\mp}_{area}/2$ per inner and
outer horizon,    we obtain the two relations
\bea&&\nonumber
\xi_{\pm}\equiv M_{\pm}-\frac{1}{2} \sqrt{\left(\frac{J_\pm}{M_\pm}\right)^2+r_\pm^2},\quad\mbox{equivalently}\\&& \xi_ {\mp}(a) =
 1 - \frac {\sqrt {1\mp\sqrt {1 - a_\mp^2}}} {\sqrt {2}}
\eea
constituting the  two branches of $a(\xi)$ functions and $M_{irr}$
 in Figs\il(\ref{Fig:Plinng}) and Figs\il(\ref{Fig:zongiaaransboc}). The second relation is in terms of the dimensionless spin of the \textbf{BH} and dimensionless energy $\xi$.
These are solutions of  $a(\xi)=
  2\sqrt {-(\xi - 2) (\xi - 1)^2\xi}$ for  $ \xi_-\in \left[ \frac {1} {2}\left (2 - \sqrt {2} \right),
   1\right]$ and  for  $ \xi_+\in [0,\frac {1} {2}\left (2 - \sqrt {2}\right)] $, respectively.
Solving $a(\xi)=a$, we obtain the solutions:
 \bea&&
\xi^+_{\mp}\equiv M_{\pm}+\frac{1}{2} \sqrt{\left(\frac{J_\pm}{M_\pm}\right)^2+r_\pm^2},\\&&\nonumber \mbox{equivalently}\quad\xi_ {\pm}^+(a) =
 1 + \frac {\sqrt {1\mp\sqrt {1 - a_\mp^2}}} {\sqrt {2}}
\eea
defining the   new functions
$\bar{M}_{irr}^{\pm}=-\sqrt{{r_{\pm}}/{2}}$ and constituting the other two branches of Figs\il(\ref{Fig:Plinng} and Figs\il(\ref{Fig:zongiaaransboc})) (at $\xi>1$).

Then,   $(M_{irr}^+)^2=M^2-(M_{irr}^-)^2  $  where
$\delta A_{area}^+=2 M \delta M -\delta A_{area}^- $.
Condition  $( < ) _M : \delta M_ {irr}^+ M_ {irr}^+ <
          M \delta M$ occurs in the following cases:
        By using the ratio  $ {\delta J}/{\delta M}_+$, we find that condition $ ( < ) _M$ holds for  $J_ + \in ] 0, M^2] $ and $\delta M <
   0$ (mass decreasing) for ${\delta J}/{\delta M}_ + <  {2M_+^3 r_ -}/{J_+}$, and $ \delta M >
   0$ (mass increasing) for  ${\delta J}/{\delta M}_ + >  {2M_+^2 r_ -}/{J_+} = 4\omega_H^+ =L_f^-$,
or explicitly
	 \bea&&\nonumber
J_+\in]0,M_+^2[,\quad \left[\delta M_+\leq 0, \delta J_+>\delta J_+^+\right];
\\&&\nonumber \mbox{and}\quad \left[\delta M_+>0, \delta J_+>\delta J_+^-\right],
\\
&&
\mbox{where}\quad \delta J_+^{\pm}\equiv \frac{2 M_+ \left(\delta M_+ M_+^2\pm J_+ \sqrt{\delta M_+^2 \left(\frac{M_+^4}{J_+^2}-1\right)}\right)}{J_+}.
\eea
In these relations we assumed  $\delta M_{irr}$, for the variations $ (\delta M, \delta J)$, to be null. However, during the \textbf{BH} evolution  the irreducible mass can also increase; in this case, we get the relation
$ \delta M_ {irr}^+\geq0$ for $  \delta M^+- \omega^+_ 0 \delta J^+\geq 0$  and then
$\delta J^+/\delta M^+\leq 1/\omega_ 0^+ = \la_ 0 (0) = (L_f)$, that is, the origin of its bundle (where $\la(0)=a_0\sqrt{\sigma}$) constrains  the transition. This relation also holds for the angular momentum of the horizon.

   However,
   \bea\label{Eq:sosp}&&(M_{irr}^\pm)^2 = M^2 - (M_{irr}^\mp)^2,\quad\mbox{then}\\&&\nonumber
 \xi=M-\sqrt{M^2 - (M_{irr}^-)^2}=M-M_{irr}^+.
\eea
The condition  $\delta M_{irr}^->0$ with $\delta M_{irr}^+ > 0$ holds for
 \bea
  M > M_{irr}^+,\quad \delta M_{irr}^+\in \left[0, \frac{\delta M M}{M_{irr}^+}\right]
  \eea
  viceversa,
 the condition  $\delta M_{irr}^-<0$ with $\delta M_{irr}^+ > 0$ holds for
 \bea\nonumber
M>M_{irr}^+,\quad  \left[(\delta M\leq 0, \delta M_{irr}^+\geq 0),\quad \left(\delta M>0,  \delta M_{irr}^+\geq \frac{\delta M M}{M_{irr}^+}\right)\right].
 \eea
	While the condition  $\delta M_ {irr}^+\geq 0 $, from Eq.\il(\ref{Eq:sosp}), holds for
\bea&&(M = M_{irr}^+,\quad\delta M\geq 0);
\\
          &&\nonumber M >M_{irr}^+,\quad\mbox{and}\quad \left (\delta M >
          0;\quad \delta M_{irr}^-\leq -\sqrt{\frac{\delta M^2 M^2}{M^2 - (M_{irr}^+)^2}}\right);\\&&\nonumber\left(\delta M\leq 0;\quad\delta M_{irr}^-\leq\sqrt {\frac{\delta M^2 M^2}{M^2 - (M_{irr}^+)^2}}\right).
		\eea
	Interestingly,    $\delta M_ {irr}^+ > 0$ for
		$
 J_+\in]0, M_+^2[$, $\delta M_+=0$, $\delta J_+<0$.
In other words,
\bea\nonumber
&&		\delta M_{irr}^+>0,\quad \mbox{for}
\\\nonumber
&&J_+=0,\quad\mbox{and}\quad \delta M_+>0;\\\nonumber
&&  J_+\in ]0,M_+^2[,\quad\mbox{and}\quad  \left(\delta M_+<0,\quad \frac{\delta J_+}{\delta M_+}>\frac{1}{\omega_H^+}\right);\\&&\nonumber \left(\delta M_+>0,\quad \frac{\delta J_+}{\delta M_+}<\frac{1}{\omega_H^+}\right)
		\eea
		and
\bea\nonumber&&		\delta M_{irr}^->0,\quad \mbox{for} \\
		&& J_-\in ]0,M_-^2[,\quad\mbox{and}\quad  \left(\delta M_-<0,\quad \frac{\delta J_-}{\delta M_-}<\frac{1}{\omega_H^-}\right);\\&&\nonumber \left(\delta M_->0,\quad \frac{\delta J_-}{\delta M_-}>\frac{1}{\omega_H^-}\right)
	\\	\nonumber
&&
 J_-\in ]0,M_-^2[,\quad\mbox{and}\quad  \delta M_-=0,\quad \delta J_->0.
\eea
		where $\omega_H^\pm= {J_\pm}/{2 M_\pm^2 r_\pm}$ and we consider the cases $\delta M_{irr}^\pm\gtrless 0$. In these relation we note how the bundle origin   $\la_0$  constrains the transition, distinguishing  \textbf{NS} origin with  $\la_0>2$, for the relations in terns of the  \textbf{BH} inner horizon (in the extended plane inner and   outer horizons relations have to be considered).

\medskip

Below we summarize some properties of the  \textbf{BH} spacetimes highlighted in this analysis relation to the \MB s characteristic.
 In the \textbf{BH} geometry with spin $a_{\gamma}\equiv \sqrt{1/2}$, it is  $\ell_{H}^\pm=\pm\omega_H^\pm$. In this spacetime the transitions are essentially regulated by the characteristic bundle frequency. In this case,  $r_{\gamma}^-=r_{\epsilon}^+$, that is, the outer ergosurface on the equatorial plane of the \textbf{BH}  is a  geodesic orbit of the (corotating) photon. In this geometry,   $\omega (r) = s \ell (r)$ at the same point $r$ \MB tangent point with the horizon curve. We obtain $r=1\mp\sqrt{{1}/({s^2+1})}$ for $s<0$  and $s\geq0$, respectively.
For  $s=\pm1$, we obtain   $r=r_\pm=1\pm{1}/{\sqrt{2}}$, outer and inner horizons, respectively, for  \textbf{BHs} with spin $a/M=1/\sqrt{2}$--Eq.\il(\ref{Eq:offtamp}).
 In other words, if we consider   $\ell_H^{\pm} (\omega) =
 s\omega$,
 then for $s = \pm1$,  we obtain $\omega =
\pm\left (1\pm\sqrt {2} \right)/2$,  corresponding to $a/M =
   1/\sqrt {2}$.
Furthermore,
$\partial_a^{(2)}\ln s=0$ for $ a= a_{\gamma}\equiv {M}/{\sqrt{2}}$, where $s\equiv \omega^+_-\equiv \omega_H^+/\omega_H^-$.
The analysis of the extremes of the accelerations $\ell$,  solutions $\partial_r \ell=0$, enlightens the role of the \textbf{BH}  spin $a/M=\sqrt{3}/2$.
 The radius    solution of $\partial_r\ell=0$, is on the horizon curve in the extended  plane, where    $r= 3/2M $,
 which is  the outer horizon of  the \textbf{BH}  spacetime  with spin {$a/M = {\sqrt{3}}/{2}$}.
   Restricting to the horizon curve, we find the solution for  $r=
 M/2 $, which is the inner horizon of the \textbf{BH} spacetime with  $a/M = \sqrt {3}/2$. Therefore, the extremes  of the acceleration $\ell$  for $a$ and $ r$ constitute, respectively, the inner
and outer  horizon  for the \textbf{BH} with spin $a/M = \sqrt {3}/2$. (That is $\partial_a \ell =
 0$ for $ r_a = a/\sqrt {3}$, while $\partial_r \ell =
   0$ for $r_r = \sqrt {3} a$; however,
   $r_a = r_ - $ and $r_r = r_ +$ for $ a/M = \sqrt {3}/2$).
The saddle point of  $(L_f)(\omega^{\pm})$ (\textbf{BH} angular momentum or bundle origin) as function of $r$ is $r_-/M=1/2$, which is  the  inner horizon of the \textbf{BH} spacetime, where  the momentum and frequency are
$(L_f)(\omega^{\pm})={2}/{\sqrt{3}}=1/\omega$. (Note that in this case $a=\omega_-$). The saddle point  of the function  $\omega(r)$ is for the same \textbf{BH} spacetime  on the outer horizon  $r_+/M={3}/{2}$ where  $(L_f)=2 \sqrt{3}=1/\omega$.
The horizon (metric bundles and light surfaces) frequency
	$\omega_g^ {-} $  of  Eq.\il(\ref{Eq:show-g}) has a saddle point for $\xi=\left (1\pm \sqrt {2/3} \right)$ corresponding to the {spin  { $ a = {2\sqrt {2}}/{3}$} } and frequency $\omega=({1}/{2\sqrt {2}},{1}/{\sqrt {2}}$).
 For
$a/M=2\sqrt{2}/3$,  we obtain also  $r_{mso}^-=r_{\epsilon}^+$, that is, the marginally stable orbit of the corotating particle is the outer ergosurface  on the equatorial plane.
Spin $a/M= {2 \sqrt{2}}/{3}$ solves the problem of replicas for $\ell  \left(r_+(a_p)\right)=-\ell  \left(r_-(a)\right)$--see Eq.\il(\ref{Eq:ap}).
On the other hand, Eq.\il(\ref{Eq:brum-field}) shows the relevance of this spin
 $ {J_0}/{M_0^2}=2\sqrt{2}/3$ in relation to
 \textbf{BH} transition with masses ratio  {$M_1/M_0=\sqrt{{2}/{3}}$}.
 The origin spin $\la_0=2/\sqrt{3}$  is a saddle point  for the area,  Eq.\il(\ref{Eq:areassaddle-points}),
   and it is also  the  saddle point for the angular momentum   corresponding to the frequency  $\omega=1/(L_H)$  of the inner horizon  $r_-/M=1/2$  for the spacetime $a=\omega$, whose frequency of the outer horizon is a saddle point for the frequency.
\section{Inertial mass, extractable rotational energy, and surface gravity}\label{Sec:non-iner-extr-surf}
\begin{figure*}
\centering
  \includegraphics[width=7cm]{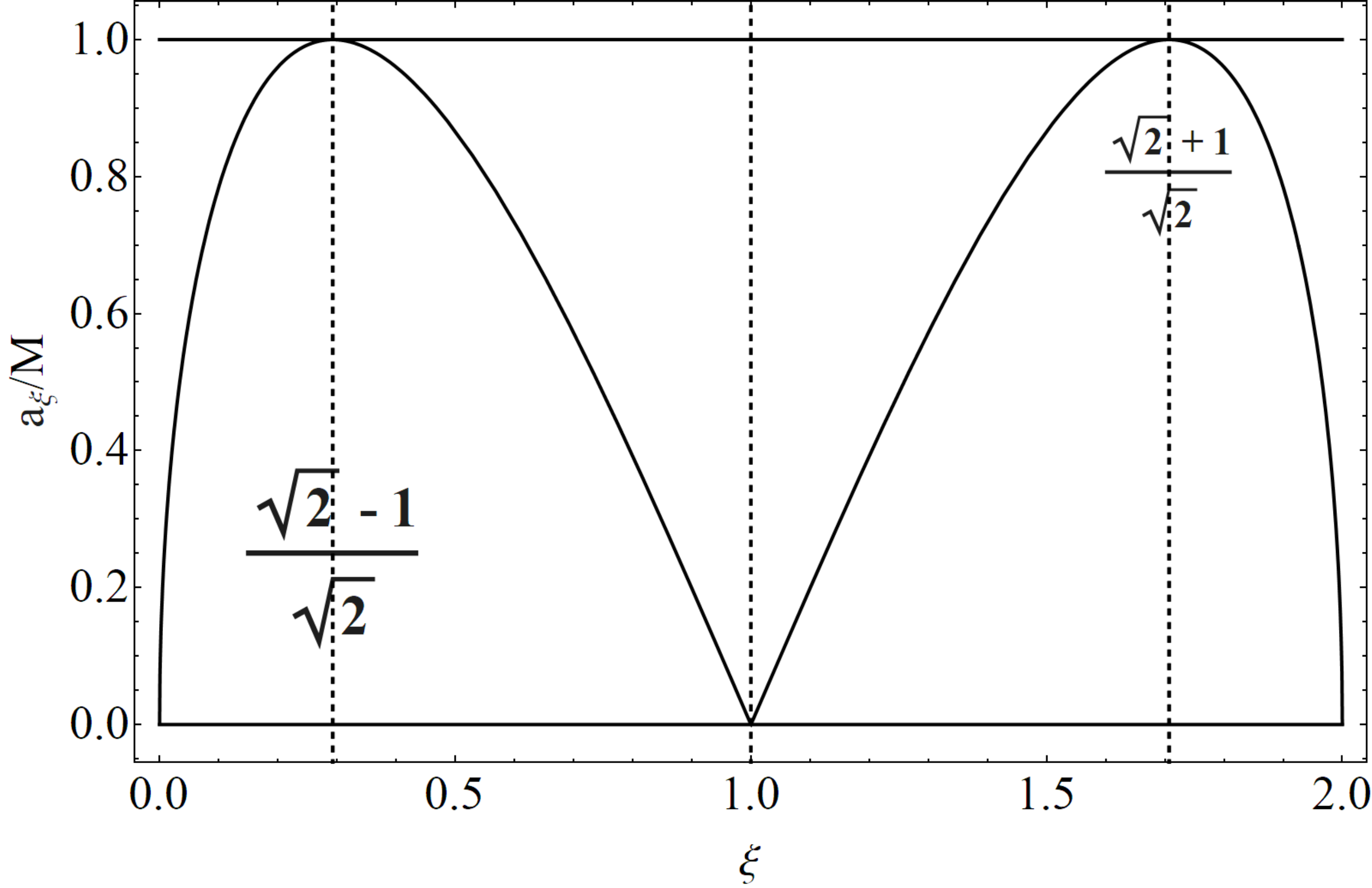}
  \includegraphics[width=7cm]{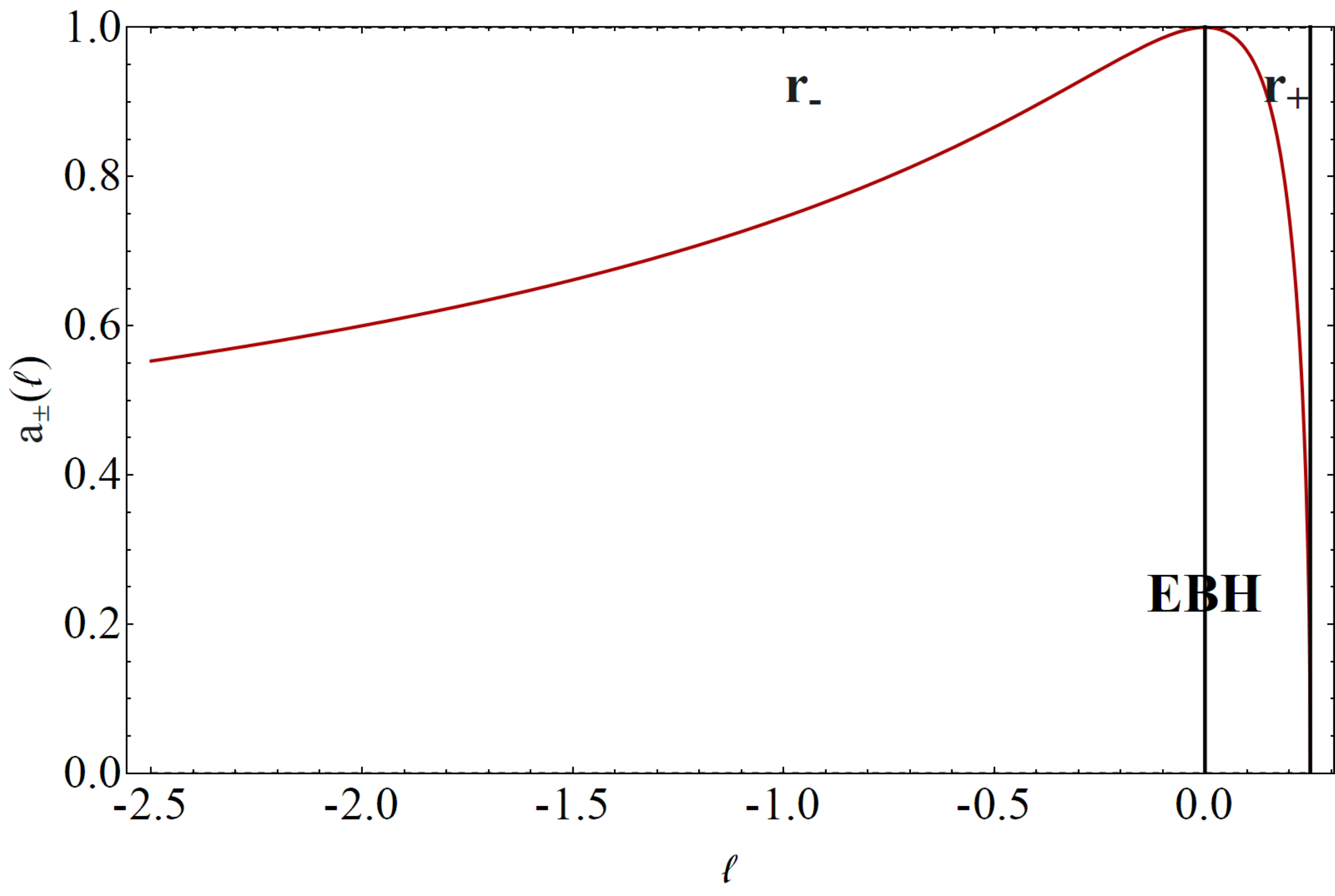}
    \includegraphics[width=7cm]{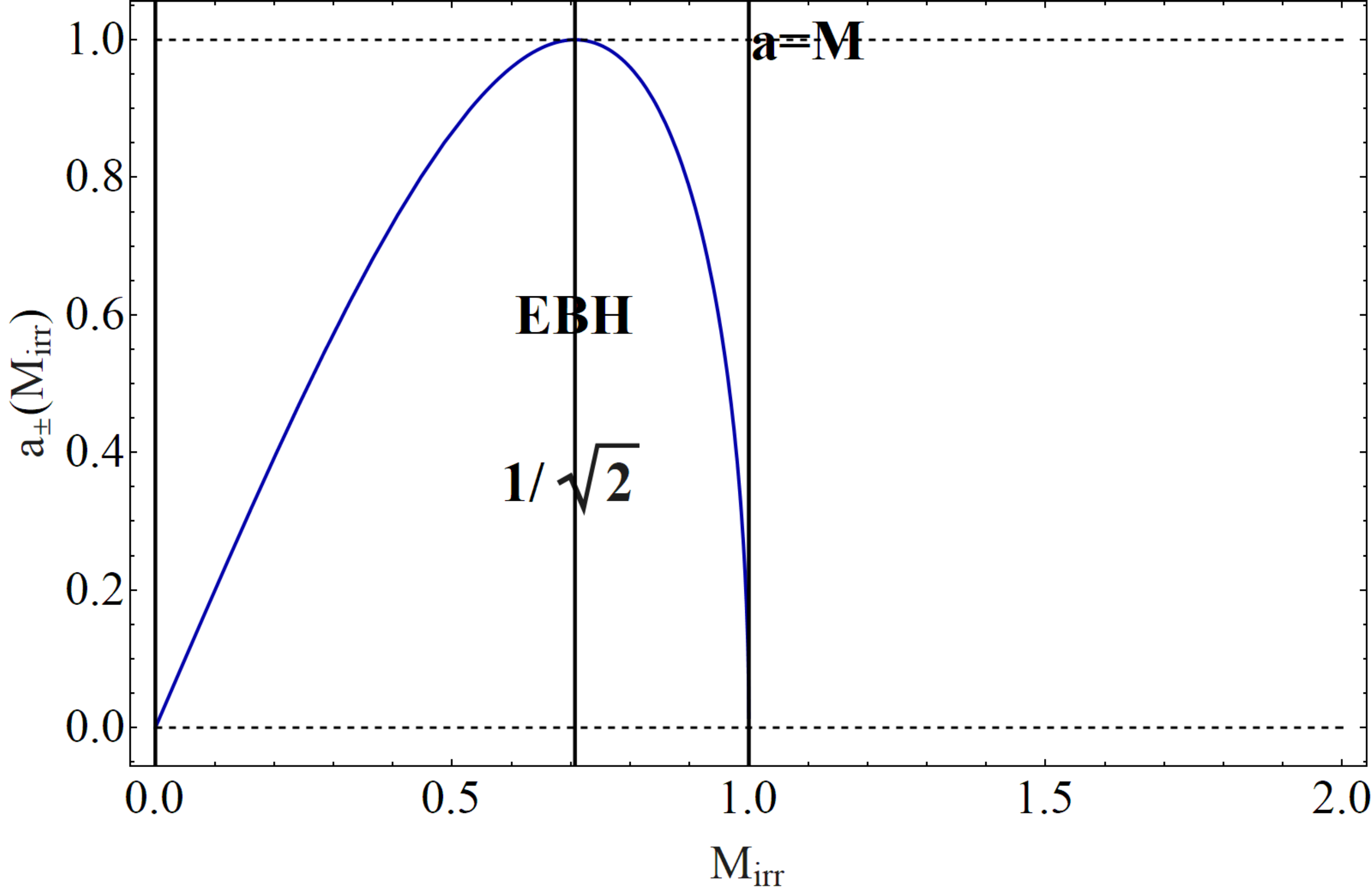}
 \includegraphics[width=7cm]{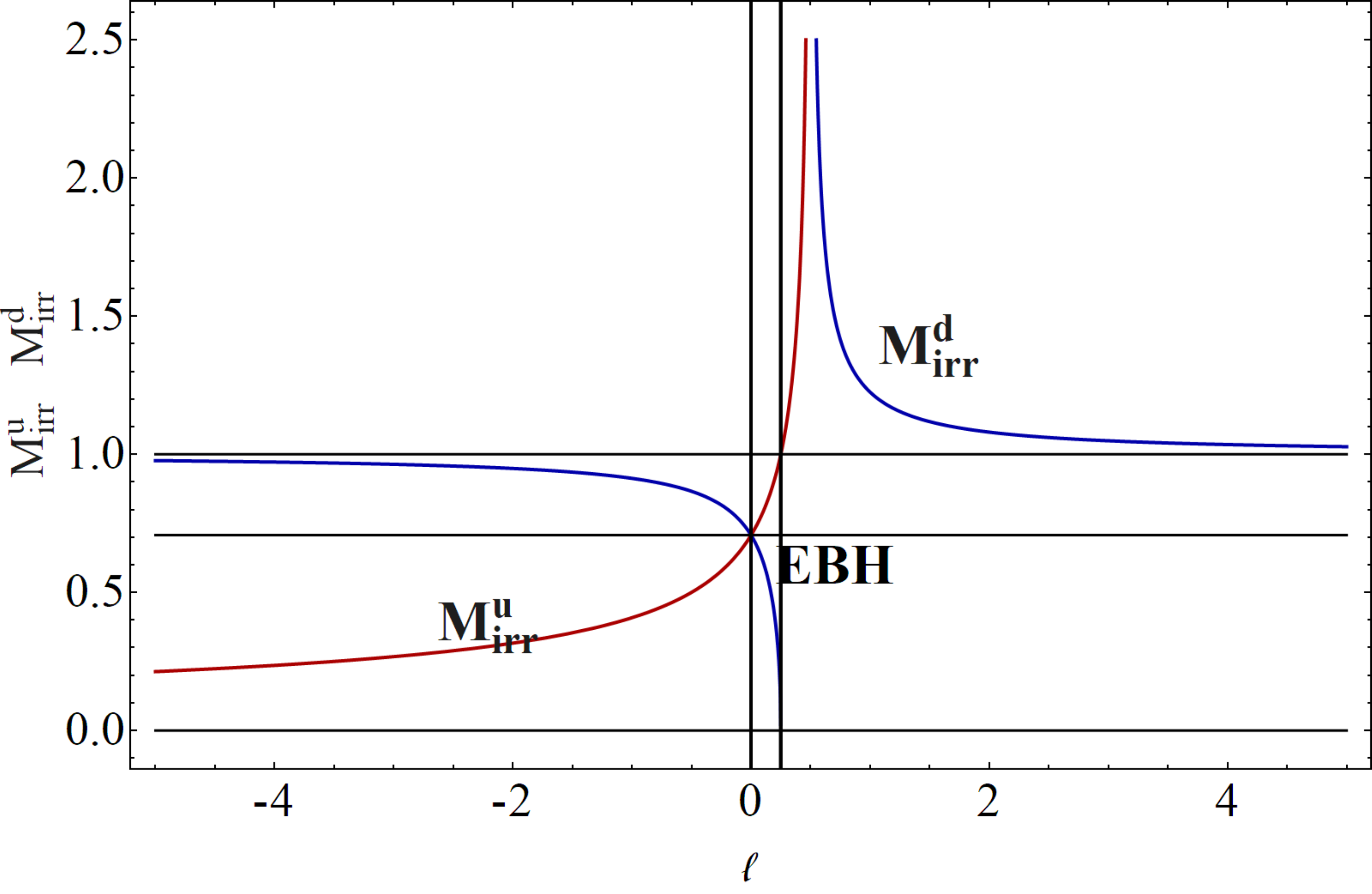}
  \caption{
Upper-left panel: The \textbf{BH} spin $a_{\xi}$  of  Eq.\il(\ref{Eq:exi-the-esse.xit})  versus extractable rotational energy $\xi$.  The Schwarzschild geometry ($a=0$) for $\xi=\{0,1,2\}$ and the extreme Kerr spacetime $a/M$ for  $\xi=\xi_\ell$ and $\xi_m$ are shown. In Sec.\il(\ref{Sec:non-iner-extr-surf}), we used the extended range $\xi\in [0,2]$ for the re-parametrization of the metric in the extended plane.
Upper-right panel:  The \textbf{BH} spin function $a_{\pm}(\ell)$ as function of the surface gravity $\ell=\ell^{\pm}$ for the outer and inner horizons. Bottom-left panel: The spin  $a_{\pm}(M_{irr})$ as function of  irreducible mass $M_{irr}$.  Bottom-right panel:  The irreducible masses, $(M_{irr}^u,M_{irr}^d)$ as functions of the surface gravity $\ell$, see Eqs\il(\ref{Eq:cour-sev}).The static  spacetime and the limit of extreme Kerr \textbf{BH}, \textbf{EBH}, are represented.
}\label{Fig:Plinng}
\end{figure*}

We can connect two \textbf{BH} states, before and after a transition,  representing them in the same extended plane.
We analyze  the  tensor  $\textbf{g}(\cdot,\cdot)$ in the extended plane by evaluating  the metric  along the horizon curve in terms of thermodynamic quantities.
In the first case,   we  use
the function   $a=a_{\xi}(\xi)$ of  Eq.\il(\ref{Eq:exi-the-esse.xit}) for the \textbf{BH} spin  in terms of the maximum extractable  rotational   energy $\xi$. The tensor describes a set of \textbf{BH} geometries. Considering Figs\il(\ref{Fig:Plinng}), where the function $a_\xi(\xi)$ is represented,  there are multiple copies of the geometry, up to four,  for the parameter $\xi$ in the range $\xi\in]0,2[$ and $a_{\xi}\in]0,1[$. The maximum extractable energy is for an  initial extreme Kerr \textbf{BH}  with parameter $\xi_\ell$, we consider however  the range  $\xi\in]0,2]$.
 In the second  case, we use the concept of irreducible  mass  with the  relations Eq.\il(\ref{Eq:Mirr-all}),
Eq.\il(\ref{Eq:candd}),
and Eq.\il(\ref{Eq:mass-irr-rampi}). In the third case, we  use the acceleration $\ell$ and the surface gravity.
This re-parametrization provide a way to connect different states ofter with multiple solutions, as in the extended plane, for different values of the parameters, creating a sort of \textbf{BH} cluster metric.   Notably   there are no  \textbf{NS}  solutions.

\begin{description}
\item[Rotational energy $\xi$]

Considering  the spin-function  $a=a_{\xi}(\xi)=2 \sqrt{-\xi  (\xi -2) (\xi -1)^2}\in[0,1]$ of  Eq.\il(\ref{Eq:exi-the-esse.xit}), we obtain up to four copies of  the geometry for each $\xi$ value.
In this way we use the symmetries of the function to create copies.
The metric components are
{
\bea\label{Eq:polit-soc}
&&g_{tt}=
 -\frac{4 (\xi -2) \xi  (\xi -1)^2 (\sigma -1)+(r-2) r}{4 (\xi -2) (\xi -1)^2 \xi  (\sigma -1)+r^2},\\&&\nonumber g_{t\phi}= -\frac{4 \sqrt{-(\xi -2) (\xi -1)^2 \xi } r \sigma }{4 (\xi -2) (\xi -1)^2 \xi  (\sigma -1)+r^2} \\\nonumber
&&g_{rr}=\frac{4 (\xi -2) (\xi -1)^2 \xi  (\sigma -1)+r^2}{(r-2) r-4 (\xi -2) (\xi -1)^2 \xi },\\&&\nonumber g_{\theta\theta}=4 (\xi -2) (\xi -1)^2 \xi  (\sigma -1)+r^2\eea}
\begin{strip}
{\small
\bea\nonumber
&&
 g_{\phi\phi}= \frac{\sigma  \left[\left(r^2-4 (\xi -2) (\xi -1)^2 \xi \right)^2+4 (\xi -2) \xi  (\xi -1)^2 \sigma  \left[(r-2) r-4 (\xi -2) (\xi -1)^2 \xi \right]\right]}{4 (\xi -2) (\xi -1)^2 \xi  (\sigma -1)+r^2}.
\eea}
\end{strip}
The metric describes a \textbf{BH} geometry for $\xi\in [0,\xi_\ell]$, where $\xi_\ell$ is the extreme \textbf{BH} spacetime. In the extended range $\xi\in ]0,2M]$, the metric  tensor  (\ref{Eq:polit-soc}) describes multiple copies of the \textbf{BHs} with $a\in[0,M]$. The extreme Kerr \textbf{BH} corresponds to  $\xi_{\ell}$ and     $\xi_m\equiv\left(2+\sqrt{2}\right)/2>\xi_{\ell}$.

Then, the Kerr  extreme \textbf{BH} and the   Schwarzschild static metric are the limits
\bea\nonumber
\lim\limits_{\xi\rightarrow\xi_\ell}g=\lim\limits_{\xi\rightarrow\xi_m}g=\mathbf{g_{Kerr}}; %\equiv
%\left(
%
%\right)
\quad
\lim\limits_{\xi\rightarrow0}g=\lim\limits_{\xi\rightarrow1}g=\lim\limits_{\xi\rightarrow2}g=\mathbf{g_{Schw}}%\equiv
%\left(
%
\eea
%
%\eea
\item[Irreducible mass $M_{irr}$]

We use the function  \\ $a=a_{irr}\equiv2 \sqrt{M_{irr}^2 \left(1-M_{irr}^2\right)}$  for the spin; then,
 the metric tensor describes \textbf{BH} geometries\footnote{The irreducible mass is defined in Eqs\il(\ref{Eq:Mirr-all}), where $M_{irr}=\sqrt{r^2+a^2}/2$. Therefore, $M_{irr}^{\pm}=\sqrt{r_{\pm}/2}$. All the quantities are dimensionless.}. Note that the  condition  {$a_{irr}=0$} is a limiting value in the extended plane, corresponding  (as for $M_{irr}$) to the static limit.
 In fact, the function  $a_{irr}$ has been evaluated from the horizon curve in the extended plane, writing the irreducible mass for any point of the horizon curve. Therefore, it describes \textbf{BH} spacetimes ($a_{irr}\in [0,1]$). From Figs\il(\ref{Fig:Plinng})  it is also clear that there are two copies of one  \textbf{BH} geometry $a$ for $M_{irr}=M_{irr}^{\pm}(a)$.
 Explicitly,
 {\small
 \bea&&\label{Eq:thisp-dang}
 g_{tt}=\frac{2 r}{4 \left(M_{irr}^2-1\right) M_{irr}^2 (\sigma -1)+r^2}-1,\quad g_{t\phi}= -\frac{4 \sqrt{M_{irr}^2(1-M_{irr}^2)} r \sigma }{4 \left(M_{irr}^2-1\right) M_{irr}^2 (\sigma -1)+r^2} \\&&\nonumber
 g_{rr}= \frac{4 \left(M_{irr}^2-1\right) M_{irr}^2 (\sigma -1)+r^2}{4 M_{irr}^2(1- M_{irr}^2)+(r-2) r},\quad  g_{\theta\theta}=4 \left(M_{irr}^2-1\right) M_{irr}^2 (\sigma -1)+r^2  \\&&\nonumber
g_{\phi\phi}= \frac{\sigma  \left[\left[4 M_{irr}^2(1- M_{irr}^2)+r^2\right]^2-4 M_{irr}^2 \left(M_{irr}^2-1\right) \sigma  \left(2 M_{irr}^2-r\right) \left(2 M_{irr}^2+r-2\right)\right]}{4 \left(M_{irr}^2-1\right) M_{irr}^2 (\sigma -1)+r^2}.
 \eea}

The limits for the Schwarzschild spacetime and for the   the extreme Kerr spacetime are
\bea&&
\lim\limits_{M_{irr}\rightarrow0}g=\lim\limits_{M_{irr}\rightarrow1}g=\mathbf{g_{Schw}},\quad
 \lim\limits_{M_{irr}\rightarrow{1}/{\sqrt{2}}}g=\mathbf{g_{Kerr}}
\eea

The singular points of the line elements are
\bea
&&\nonumber
g_{tt}=0:\quad
\left[r=2: \left((M_{irr}=0, \sigma \in [0,1]); (\sigma =1, M_{irr}\in ]0,1[)\right)\right];
\\\nonumber
&& \left[ M_{irr}=\frac{1}{\sqrt{2}}: \left[(r=1, \sigma =0);  \left(\sigma\in]0,1[, \left(r=1\pm\sqrt{\sigma }\right)\right)\right]\right];
\\\nonumber
&&\sigma \geq 0:\quad \left[r=r_{Mirr}^+:\left[\sigma <1,\left(M_{irr}\in ]0,1[,M_{irr}\neq \frac{1}{\sqrt{2}}\right)\right]\right.,\\&&\nonumber\left.\left(M_{irr}\geq 1, \sigma \leq 1\right)\right],
\\\nonumber
&&\left[r=r_{Mirr}^-:\sigma <1,\left(M_{irr}\in ]0,1[,M_{irr}\neq \frac{1}{\sqrt{2}}\right)\right],
\\\nonumber
&&g_{rr}=0:
\quad[r=0, \sigma =1, (M_{irr}\in ]0,1[; M_{irr}>1)];\\&&\nonumber [r=r_{Mirr}^s, \sigma\in [0,1[, M_{irr}>1],\\\nonumber
&&g_{tt}^{-1}=0:
\quad\sigma\in [0,1[, M_{irr}>1, r=r_{Mirr}^s,\\\nonumber
&&g_{rr}^{-1}=0:
\quad\sigma\in[0,1],\left[\left(M_{irr}=0, r=2\right);\right.\\&&
\nonumber \left(M_{irr}=\frac{1}{\sqrt{2}}, r=1\right);\quad\left. \left(r=2 M_{irr}^2,\left(M_{irr}\neq \frac{1}{\sqrt{2}}\right)\right)\right);\\ &&\nonumber\left. \left(r=2(1- M_{irr}^2),\left(M_{irr}\in ]0,1[,M_{irr}\neq \frac{1}{\sqrt{2}}\right)\right)\right],
\eea
(note the possibility of $M_{irr}>1$) where
\bea&&\nonumber
r_{Mirr}^s\equiv2 M_{irr} \sqrt{\left(M_{irr}^2-1\right) (1-\sigma )},
\\&& r_{Mirr}^{\mp}\equiv 1\mp\sqrt{1+(r_{Mirr}^s)^2}.
\eea
\item[Surface gravity $\ell$]

We use  $a=a_{\pm}(\ell)={\sqrt{1-4 \ell}}/{\sqrt{(2 \ell-1)^2}}$,  from the definition $\ell_H^{\pm}(a)=\ell$. Then,  $a=M$ for $\ell=0$ and $a=0$ for $\ell=1/4$.
Note that  the metric tensor describes \textbf{BH} geometries and $a_{\pm}(\ell)$ is not well defined for the asymptotic value $\ell=1/2$. In terms of the irreducible mass using Eq.\il(\ref{Eq:thisp-dang}) with the two solutions $M_{irr}^u\equiv{1}/{\sqrt{2(1-2 \ell)}}$  and $M_{irr}^d\equiv1/\sqrt{(1-4 \ell)^{-1}+1}$ (Figs\il(\ref{Fig:Plinng})), the tensor is
{\small
\bea&&\nonumber\mathbf{g}=
\left(
\begin{array}{cccc}
 \frac{2 (1-2 \ell)^2 r}{\widetilde{\mathcal{P}}}-1 & 0 & 0 & -\frac{2 \sqrt{1-4 \ell} \sqrt{(1-2 \ell)^2} r \sigma }{\widetilde{\mathcal{P}}} \\
 0 & \frac{\widetilde{\mathcal{T}}}{\widetilde{\mathcal{X}}-2 r} & 0 & 0 \\
 0 & 0 &\widetilde{\mathcal{T}} & 0 \\
 -\frac{2 \sqrt{1-4 \ell} \sqrt{(1-2 \ell)^2} r \sigma }{\widetilde{\mathcal{P}}} & 0 & 0 & \frac{\sigma  \left[\widetilde{\mathcal{X}}^2+\frac{(4 \ell-1) \sigma  (2 \ell (r-2)-r+1) ((2 \ell-1) r+1)}{(1-2 \ell)^4}\right]}{\widetilde{\mathcal{T}}} \\
\end{array}
\right),
\\\nonumber
&&\widetilde{\mathcal{P}}\equiv4 \ell \left[(\ell-1) r^2+\sigma -1\right]+r^2-\sigma +1,\quad  \widetilde{\mathcal{T}}\equiv\frac{(4 \ell-1) (\sigma -1)}{(1-2 \ell)^2}+r^2,\\&& \widetilde{\mathcal{X}}\equiv \frac{1-4 \ell}{(1-2 \ell)^2}+r^2.\label{Eq:cour-sev}
\eea}
The extreme Kerr and the Schwarzschild   \textbf{BH} limit are
\bea&&
\lim\limits_{\ell\rightarrow 0}\mathbf{g}=\mathbf{g_{Kerr}},\quad
 \lim\limits_{\ell\rightarrow1/4}\mathbf{g}=\lim\limits_{\ell\rightarrow \pm\infty}\mathbf{g} =
\mathbf{g_{Schw}}.
\eea
There are the following singular points for the metric

\bea&&
g_{tt}=0:\quad\sigma =0; \quad\left(\left(\ell=0, r=1\right);\left(r=r_{\alpha}^+,\ell\neq\left\{ \frac{1}{2},0\right\}\right);\right.
\\
&&\nonumber\left. \left(r=r_{\alpha}^{-},\ell<\frac{1}{4};\ell\neq 0\right)\right)
\\\nonumber
&&\ell\neq \frac{1}{2}: (\sigma =1, r=2), \left(\sigma \in]0,1[, \left[r=r_{\beta}^+, \left(r=r_{\beta}^-,  \ell<\frac{1}{4}\right)\right]\right),\\\nonumber
&&g_{rr}=0:\quad \left(\sigma =1,\left[\left(\ell=\frac{1}{2}, r\geq 0\right),\left(r=0,\ell\neq\left\{\frac{1}{2}, \frac{1}{4}\right\}\right)\right]\right);
\\
&&\nonumber\left(\sigma \in[0,1[, r=r_{z},   \ell>\frac{1}{4},\ell \neq \frac{1}{2}\right),
\\
&&g_{tt}^{-1}=0:\quad \left(\sigma\in[0,1[, \ell> \frac{1}{4}, \ell\neq\frac{1}{2},\quad  r=r_{z}\right).
\\\nonumber
&& g_{rr}^{-1}=0:\quad \left(\sigma\in [0,1], (\ell=0, r=1),\quad  \left[r=r_{\pm}(\ell), \left(\ell<\frac{1}{2}, \ell\neq 0\right)\right]\right.,\\\nonumber
&& \left.\left. \left[r=2-r_{\pm}(\ell), \left(\ell<\frac{1}{4},\; \ell\neq 0,\; \ell>\frac{1}{2}\right)\right)\right]\right)
\eea
with
\bea\nonumber&&
r_{\alpha}^{\pm}\equiv 1\pm 2 \sqrt{\frac{\ell^2}{(1-2 \ell)^2}},\quad r_{\beta}^{\pm}\equiv 1\pm\sqrt{\frac{4 \ell (\ell-\sigma )+\sigma }{(1-2 \ell)^2}},\\&&\label{Eq:ralphabetal} r_{z}\equiv\sqrt{\frac{(4 \ell-1) (1-\sigma)}{(2 \ell-1)^2}}, \quad r_{\pm}(\ell)\equiv\frac{1}{1-2 \ell},
\eea
where $r_{\pm}(\ell)$ are the horizon curves as functions of the acceleration $\ell_{H}^\pm$ --see Figs\il(\ref{Fig:Plotonethero12}).
\end{description}
\begin{figure}
\centering
  \includegraphics[width=6cm]{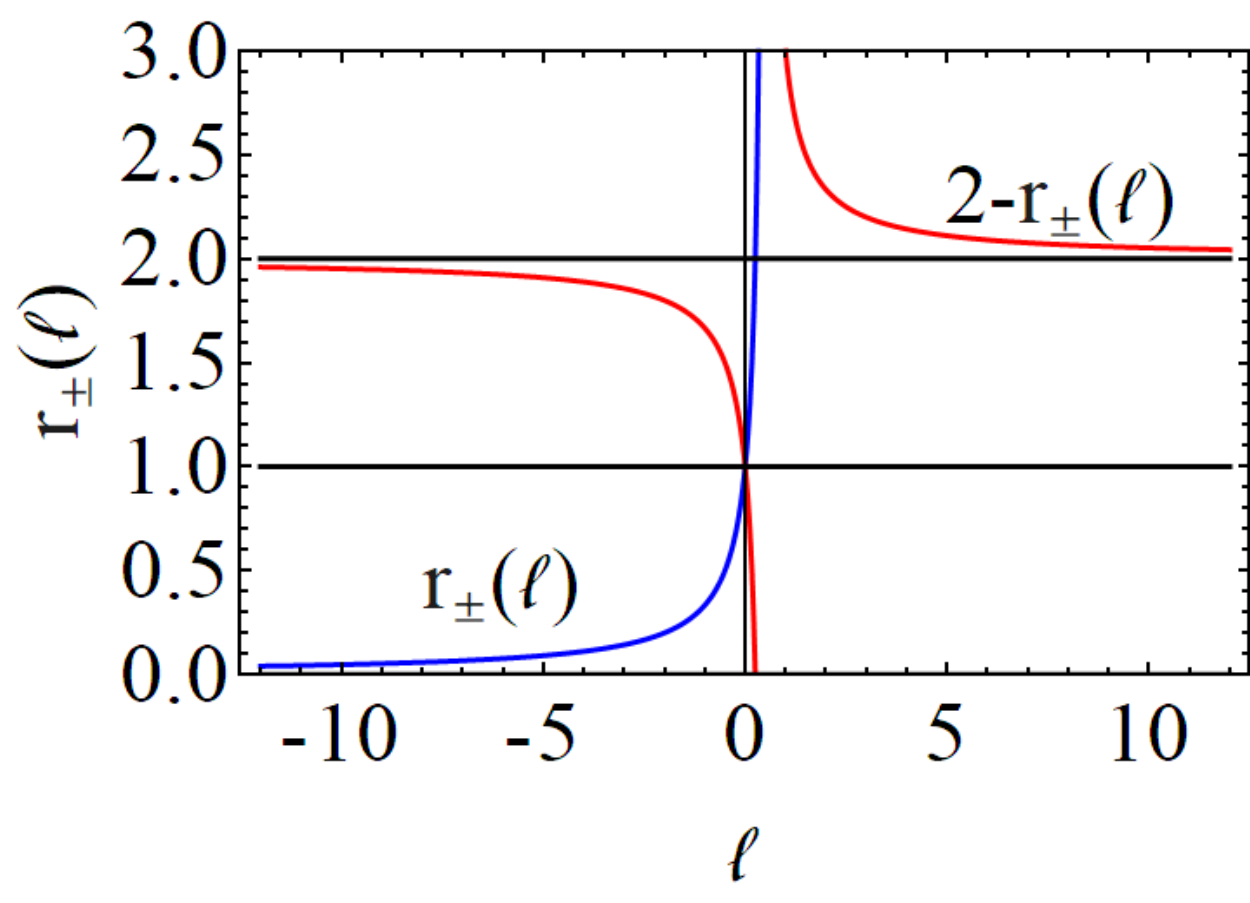}
  \includegraphics[width=6cm]{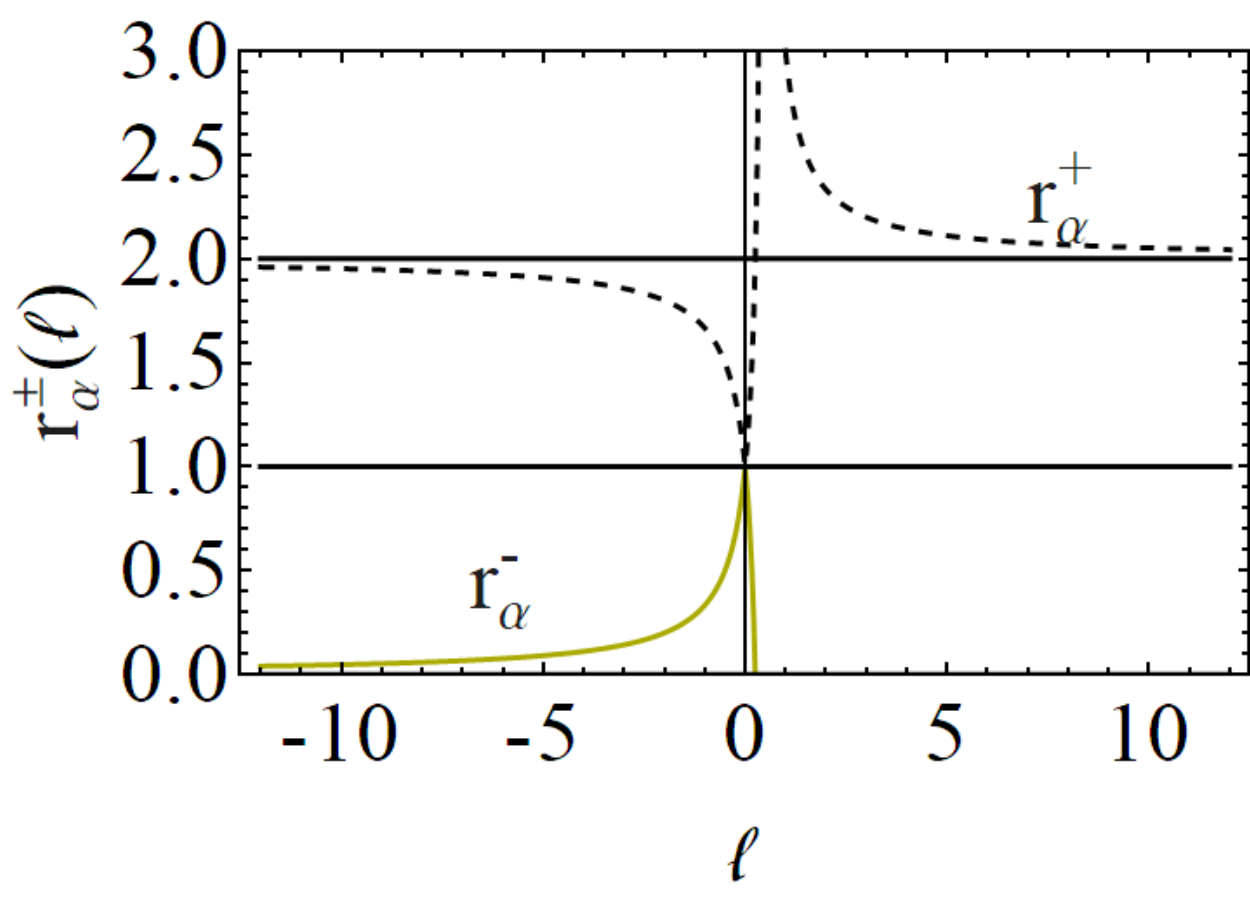}
   \caption{The radii $r_{\alpha}^{\pm}$  and \textbf{BH}  horizons $r_{\pm}(\ell)$ of Eq.\il(\ref{Eq:ralphabetal}) as functions of the acceleration $\ell$. These are the singular points of the metric line (\ref{Eq:cour-sev}).
   Note the horizons replicas.
}\label{Fig:Plotonethero12}
\end{figure}
\section{Discussion and final remarks}
\label{Sec:conclu}
The laws of \textbf{BH} classical thermodynamics have  a geometric nature determined by  the \textbf{BH} horizon, which has a prominent role  in  \textbf{BH} physics.
In this work, we studied  \textbf{BH}  thermodynamics by using certain light surfaces, grouped in  structures defined  in metric bundles  (\MB s),  tightly connected to the definitions of Killing horizons.
The introduction of the extended plane allows an overview of all the possible \textbf{BH}   (and possibly \textbf{NS}) states and their properties.
It can be used to  describe  \textbf{BH} transitions    from the \MB s view-point, for instance, a transition from  the tangent spin $a_{g}(0)$ for the \textbf{BH}(0) to $a_{g}(1)$ for the \textbf{BH}(1), corresponding to a transition between the respective tangent bundles.
Consequently, we can also describe  transitions  in terms of the origin frequency and the tangent radius of the bundle.

In this new frame we reformulated,  \textbf{BH} thermodynamics on the replicas in terms of the light surfaces, exploring the thermodynamic properties of the \textbf{BHs} geometries. Fundamentals of classical \textbf{BH}  thermodynamics are discussed    interpreting  \textbf{BH} transformations  according to the characteristic frequency which is also the \textbf{BH} horizon frequency.
 In particular, we considered the transformations with constant irreducible mass.

    In the analysis of the thermodynamical properties,  we apply the concepts of rotational energy  parameter $\xi$,   surface gravity,  frequency $\omega$,  irreducible mass $M_{irr}$, which we use to re-express the first law as well as the metric line element.
A special metric line parametrization, describing only \textbf{BH} spacetimes, is given in terms of the maximum extractable rotational energy  $\xi$ or,   correspondingly, the inertial mass parameter $M_{irr}$,  replacing the metric spin parameter.
In this work,  we change the  perspective in the analysis of some aspects of the Kerr solutions as seen in the extended plane, that is, we study  properties unfolding along all the solutions; therefore, we re-consider the line element adapting it to this new frame.
In Sec.\il(\ref{Sec:non-iner-extr-surf}),  we  rewrite the  metric tensor on parts of the extended plane,  providing  alternative    representations of the \textbf{BH}  geometries, creating clusters of \textbf{BH} copies with respect to a new metric parameter, recovered from the bundles representations as curves in the extended plane.
We  used the  irreducible   mass  $M_{irr}$ of the \textbf{BH}--Figs\il(\ref{Fig:zongiaaransboc}),
the \textbf{BH} rotational energy $\xi $  (Figs\il(\ref{Fig:Plinng})),  the \textbf{BH}
surface gravity  (Figs\il(\ref{Fig:zongiaarano})).
The extended plane and  \MB s concepts  include  \textbf{NSs};   however, no transformation  allows
a \textbf{BH-NS} transition. \textbf{NSs}, as \MB s components, can be used to represent some properties of the \textbf{BH} geometry.
  With these re-parameterizations with functions defined in the extended plane, we obtain  structures constituted by  replicas of the metric tensor describing the same geometries
or portions of the same geometry in the extended plane.
We plan in a future work to study the cluster metric with extended range of parameters.

  Replicas of light-surfaces  with equal photon frequencies, are useful to connect different points of the spacetime, and different geometries following the \textbf{BH} transition with energy release.
 Here, we use the function  $\xi(r)$, maximum  extractable  rotational energy--see \cite{ella-correlation,Daly0,Daly2,Daly3,Daly:2008zk,GarofaloEvans}.
In this context,  we discuss how  $ \omega$, the bundle characteristic frequency, is tied to the extracted  rotational energy  $\xi$.
The frequency, in fact,  also appears as a component of the "work term" in the laws of \textbf{BH} thermodynamics. We use extensively this correspondence to formulate \textbf{BH} thermodynamics in terms of  metric bundles.
  More precisely, the  energy extraction process can take place regulated by the light surfaces which  are, in general,  two   surfaces, one surface  located outside the  horizon  $r_+$ and one  within the ergoregion $]r_+,r_{\epsilon}^+[$.
 (In the case of the inner horizon, that is, of points characterized by the frequency $\omega_H^-\equiv \omega_{\pm}(r_-)$, where $r_-$ is the inner Killing horizon,  the situation is more complex and it is not always possible to find a replica of the horizon \cite{remnants}. Interestingly, the dependence of the orbit from  the poloidal angle $ \theta $, allows a more accurate study of the regions close to the \textbf{BH} rotational axis,  revealing   interesting observational implications\cite{remnants}.

In Sec.\il(\ref{Sec:mass-termo-smarr}),  we reformulated aspects of \textbf{BH} thermodynamics     on the bundles, exploring  the properties of the   rotational energy and  "rest"  mass in the extended plane.  The transition from a state $(0)$ to a new state $(1)$  is essentially regulated by the characteristics of the initial state, such  as the \textbf{BH} surface gravity $\ell_H^+(0)$ evaluated on the outer horizon, entering the entropy/temperature terms, and by the  work term, which is expressed through the bundle  frequency $\omega_H^+$. We have expressed the relation between these quantities in terms of the tangency conditions of the bundles with the horizon curve and in terms of the bundle curves in the extended plane, leading to the alternative view-point in terms  of replicas, i.e.,   orbits of the bundles located at $r>r_+$ (and some being in the  $r<r_-$ region). In this sense, we describe the state transitions  in terms of metric bundles transitions.
 In this regard, as   bundles connect  different points and geometries in the extended plane, we   evaluated  the line element along the bundles and in other portions of the extended plane, thus, connecting metrics particularly from the state with $(\ell(0),\omega(0))$  to the state with parameters  $(\ell(1),\omega(1))$.

We  highlighted   classes of \textbf{BHs} related to the fact that a transition between initial and final state is subject to particular laws.
The results point out
 \textbf{BH} spacetimes with spins
$a/M= \sqrt {8/9}$ and
$a/M=1/\sqrt{2}$ showing  anomalies in the rotational energy extraction, and surface gravity variation after \textbf{BH} state transitions and \textbf{BH} spin $a/M=\sqrt{3}/2$ relevant in the \textbf{BH} area variations.
We also enlightened    \textbf{BHs} masses ratios  $M_1 = {M_0}/{\sqrt{2}}, $  and $M_1/M_0=\sqrt{{2}/{3}}$ (distinguishing spin $a/M=\sqrt{8/9}$), regulating the \textbf{BH} transitions studied here,  and  representing limiting values in the \textbf{BH} transformations--see for example Eq.\il(\ref{Eq:bat}) and Eqs\il(\ref{Eq:camb-vibe}).

%%%%%%%%%%%%%%%%%%%%%%%%%%%

\section*{Acknowledgements}
This work was partially supported  by UNAM-DGAPA-PAPIIT, Grant No. 114520,
Conacyt-Mexico, Grant No. A1-S-31269.

\appendix
\section{Black holes in the extended plane}\label{Sec:BHsextended}
\begin{figure}
\centering
  \includegraphics[width=\columnwidth]{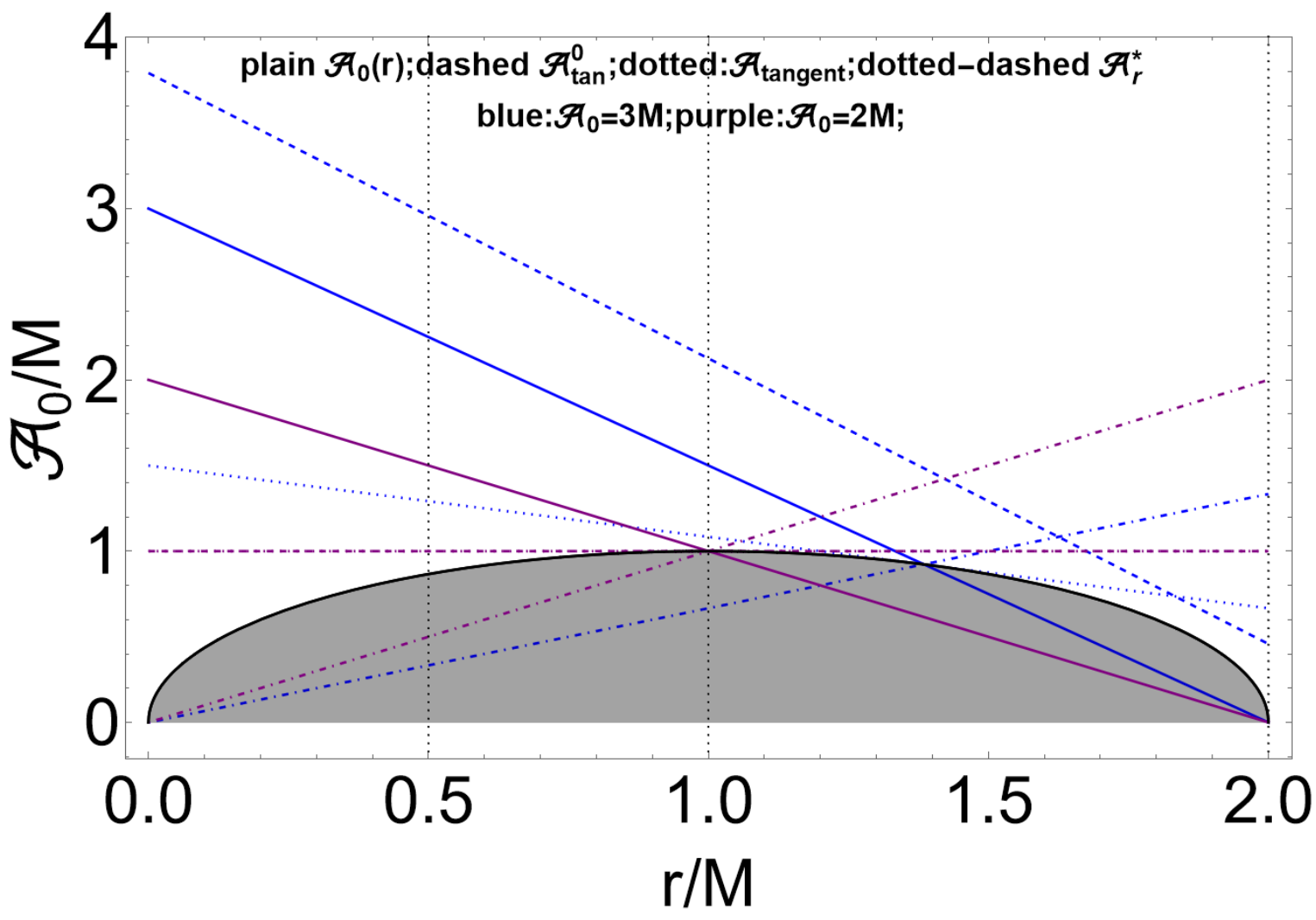}
   \caption{Functions $(\la_0,\la_{r}^*,\la_{tan}^0,\la_{tangent})$ of Eqs\il(\ref{Eq:fac-ko}) in the extended plane, for different bundle origin $\la_0$. Horizons curve is also plotted (black curve) $a_\pm$, the region $a<a_\pm$ is shaded.
}\label{Fig:Plotexplimita}
\end{figure}

In this section, we discuss some aspects of the   extended plane   \cite{remnants,bundle-EPJC-complete}. Here, we present the set of functions
{\small
\bea\label{Eq:fac-ko}
&&
\mathbf{\pp}_1:\quad\la_0(r)\equiv \la_0\left(1-\frac{r}{2M} \right),\\
 &&\nonumber \mathbf{\pp}_2:\quad\la_{r}^*(r)\equiv \frac{2M}{\la_0}r,\quad\mathbf{\pp}_3:\quad\la_{tan}^0\equiv \frac{\la_0^2-2}{2a_g(\la_0)}+\left(\frac{4-\la_0^2}{\la_0}\right) r,\quad\mbox{and}\\\nonumber
 &&\mathbf{\pp}_4:\quad\la_{tangent}(r)\equiv\frac{{\left(\la_0^4-16\right)a_g(\la_0)}+32 \la_0}{8 \left(\la_0^2+4\right)} -\frac{a_g(\la_0) \left(\la_0^4-16\right)}{16 \la_0^2} r,
%\\
\eea}
see  Fig.\il(\ref{Fig:Plotexplimita}).
%
%
%\end{figure}
The four regions of  Figs\il(\ref{Fig:zongiaaransboc}),   outlined in this analysis  also through the inertial mass,  the surface gravity $\ell$ and the  \textbf{BH} area  in the extended plane,  can be obtained as functions of the origin spin $\la_0$,  the characteristic frequency of $\omega$ of the bundles, and  the tangent radius (the first law of \textbf{BH} thermodynamics in the extended plane is defined along the boundary of these regions).
Considering the curves of  Eq\il(\ref{Eq:fac-ko}) and
 using  $r_g=r_g(\la_0)$ and
 $a_g=a_g(\la_0)$, with   the origin spin
$\la_0={2r_{\pm}}/{a_g}$ and $\la_0={2\sqrt{r_g}}/{\sqrt{2-r_g}}$, we can express the curves as
 \bea\nonumber\{\pp_{\sharp}\}\equiv\{\mathbf{\pp}_1:\la_0(r),
 \mathbf{\pp}_2:\la_{r}^*(r),
 \mathbf{\pp}_3:\la_{tan}^0,
 \mathbf{\pp}_4:\la_{tangent}(r)\}
 \eea
and
\bea\nonumber\{\pp_{\sharp}\}=\left\{\frac{(2-r)\sqrt{r_g}}{\sqrt{2-r_g}},\frac{r\sqrt{2-r_g}}{\sqrt{r_g}},\frac{r_g (3r_g-2)-4 r (r_g-1)}{\sqrt{(2-r_g)r_g}},\frac{r_g(1-r)+r}{\sqrt{(2-r_g)r_g}}\right\},
\eea
 in terms of the tangent radius $r_g\in[0,2M]$ on the horizons.
By inverting Eq.\il(\ref{Eq:Mirr-all}), we obtain
$\la_0={2M_{irr}(all)}/{\sqrt{1-M_{irr}(all)^2}}$  and
{\small
\bea&&	\nonumber\{\pp_{\sharp}\}=\left\{-\frac{M_{irr}(all) (r-2)}{\sqrt{1-M_{irr}(all)^2}},\frac{\sqrt{1-M_{irr}(all)^2} r}{M_{irr}(all)},\right.\\
&&\nonumber\left.\frac{2\left[3M_{irr}(all)^4-M_{irr}(all)^2 (2 r+1)+r \right]}{M_{irr}(all)\sqrt{1-M_{irr}(all)^2}},
\right.\\\nonumber
&&\left.\frac{4M_{irr}(all)^4-2M_{irr}(all)^2 (r+1)-4\sqrt{M_{irr}(all)^2}\left(M_{irr}(all)^2-1\right)M_{irr}(all)+r}{2\sqrt{M_{irr}(all)^2(1-M_{irr}(all)^2)}
}\right\}.\eea}
Using  the irreducible mass
$M_ {irr}^{\pm} $  to express the bundle origin $\la_0$,  we find  $\la_0={2\sqrt{1-(M_{irr}^{\pm})^2}}/{M_{irr}^{\pm}}$ and
{\small
\bea&&\nonumber
\{\pp_{\sharp}\}=\left\{-\frac{\sqrt{1-(M_{irr}^{\pm})^2} (r-2)}{M_{irr}^{\pm}},\frac{M_{irr}^{\pm} r}{\sqrt{1-(M_{irr}^{\pm})^2}},\right.\\&&\nonumber\left.\frac{2\left[(M_{irr}^{\pm})^2 (2 r-5)+3(M_{irr}^{\pm})^4- r+2\right]}{M_{irr}^{\pm}\sqrt{1-(M_{irr}^{\pm})^2}},\right.
\\
&&\nonumber\left.\frac{2M_{irr}^{\pm}
\left\{2(M_{irr}^{\pm})^3-2\sqrt{(M_{irr}^{\pm})^2}\left[(M_{irr}^{\pm})^2-1\right]+M_{irr}^{\pm}
(r-3)\right\}-r+2}{2\sqrt{(M_{irr}^{\pm})^2\left[1-(M_{irr}^{\pm})^2\right]}}\right\}.
\eea}
It is evident the role of radii $r=2M$, the Schwarzschild  horizon,  and $r=3M$, the ergosurface on the equatorial plane. The limit $M_{irr}=1$ and $r=2/5$,  which is the tangent point to an inner horizon of the spaceme $a/M=4/5$, is related to the horizon confinement\cite{bundle-EPJC-complete}.
\section{Notable areas in the extended plane}\label{Appendix:mainbundleproperties}
%\textbf{}
%
Table\il(\ref{Table:pub-priAreas}) contains  the evaluation of the areas in the extended plane,\footnote{
	It is
 $
\int \ell \delta \omega=\arcsin\left(M_{irr(\nu)}\right)+M_{irr(\mu)} M_{irr(\nu)},\;\mbox{where},\; 2 M_{irr(\mu)} M_{irr(\nu)}=a_{\pm}, \; \left(\int \ell \delta r=({r-\log r})/{2}\right).$}  according to the regions of  Figs\il(\ref{Fig:zongiaaransboc}), which constitute the  integral form of  the relations in Sec.\il(\ref{Sec:mass-termo-smarr}).
Therefore, considering also the results of \cite{bundle-EPJC-complete}, we obtain
\bea&&\label{Eq:det-traf-vua-sol-aatt} \int_{0}^{2M}a_0(r) dr=2\pi, \quad\int a_{\epsilon}^{\pm} d\sigma dr=\pi,
\\\label{Eq:inse-elem-nov-contro}
&&\int_{0}^M \omega_H^{\pm}\delta a=\frac{\left(1-r_{\pm }+\ln\left[r_{\pm }\right]\right)}{2},\quad\mbox{where}\\&&\nonumber \int_{0}^M  \omega_H^+ da=-\int_{\omega_H^+} a_g(\omega) d\omega+\frac{1}{2}=\frac{1-\log (2)}{2},
\eea
where $a_0$ is the bundle spin origin, $a_{\epsilon}^{\pm} $ are the ergosurfaces curves in the extended plane, relating  characteristic quantities of the  bundles in the extended plane.
%
%\end{table*}}
%%
\begin{table*}
\caption{$A_{areas}$ represent the \textbf{BH} areas in the extended plane, $a_g$ is the tangent spin to the horizon curve, $r_g$ is the tangent radius of the bundle curve to the horizon in the extended plane, $\omega$ is the horizon frequency and bundle characteristic frequency, $a_{\pm}$ are the horizon curves in the extended plane, $r_{\pm}$ are the outer and inner \textbf{BH} horizon, respectively, $\la_0=a_0\sqrt{\sigma}$ is the origin spin of the bundle, where $\sigma\equiv\sin^2\theta$, $M_{irr}$ is the irreducible mass of the \textbf{BH}, $M_{irr(\mu)}$ is defined in Eq.\il(\ref{Eq:xtautau}).}
\label{Table:pub-priAreas}
\centering
\resizebox{.71\textwidth}{!}{%
\begin{tabular}{l|l}
\textbf{Tangent radius}: & $ \int r_g(\omega) d\omega=\arctan(2 \omega );\quad
\int_{\omega_H^+} r_g(\omega) d\omega=\int_{\omega_H^-} r_g(\omega) d\omega=\frac{\pi}{4}$
\\
\hline
\\
\textbf{Horizons curves:}&$
\int a_{\pm} \delta r=\frac{(r-1) a_{\pm}}{2} -\arcsin\left[\frac{M_{irr(\mu)}}{{M}}\right];\quad
\int_{r_+} a_{\pm} \delta r=\int_{r_-} a_{\pm} \delta r=\frac{\pi}{4}$
\\
\hline
\\
\textbf{Tangent spin $a_g$:}&$\int a_g(\omega) d\omega=\frac{\log \left(4 \omega ^2+1\right)}{2};\quad \int_{\omega_H^+} a_g(\omega) d\omega=\frac{\log (2)}{2};\quad
\int a_{g}(r_g) dr_g=\frac{2 \pi }{\sqrt{\sigma }}$
 \\
 &$ \int a_g(\la_0)d \la_0=2 \log \left(\la_0^2+4\right);\quad
\int\limits_{0}^2 a_g(\la_0)d \la_0=\log(4)$
\\
 \hline
 \textbf{Areas:}&$ \int A_{area} (a_{\pm}) \delta r=4\pi r^2;
\quad
\int_0^M A_{area} (a_{\pm}) \delta r=4\pi;\quad \int_M^{2M }A_{area} (a_{\pm}) \delta r=12\pi$
\\
&$\int A_{area}(a_g,r_g)\delta\omega=8\pi \int r_g(\omega) d\omega=8 \pi  \arctan(2 \omega)$
 \\
&$\int_{\omega_H^+} A_{area}(a_g,r_g)\delta\omega=
\int_{\omega_H^-} A_{area}(a_g,r_g)\delta\omega=2\pi^2$
\\
&$\int A_{area}(a_g,r_g)\delta\la_0=16 \pi  \left[\la_0-2 \arctan\left(\frac{\la_0}{2}\right)\right];\quad
\int_0^{2M} A_{area}(a_g,r_g)\delta\la_0=8 (4-\pi) \pi$
\\
&$\int A_{area}(r_{\pm})\delta a=\pi  \pm 4 \left[a \left(\sqrt{1-a^2}\pm 2\right)+\arcsin(a)\right];\quad
\int_{0}^M A_{area}(r_{\pm})\delta a=\pi  \pm 2 (\pi \pm 4)$
 \\
 \hline\hline
\end{tabular}}
\end{table*}

\end{document}